\documentclass[%
 reprint,
 amsmath,amssymb,
 aps,
]{revtex4-2}

\usepackage{graphicx}
\usepackage{dcolumn}
\usepackage{bm}

\usepackage[utf8]{inputenc}
\usepackage[T1]{fontenc}
\usepackage[english]{babel}
\usepackage{amsfonts}
\usepackage{amssymb}
\usepackage{makeidx}
\usepackage{pgf,tikz}
\usepackage{listings}
\usepackage{amsmath, amsthm}
\usepackage{placeins}
\usepackage{graphicx}
\usepackage{dsfont}
\usepackage{wrapfig}
\usetikzlibrary{arrows}
\usepackage{stmaryrd}

\usepackage{url,csquotes}
\usepackage[linktocpage=true]{hyperref}
\hypersetup{colorlinks=true, citecolor=black, filecolor=black, linkcolor=black, urlcolor=blue} 
\renewcommand{\thesection}{\Roman{section}}

\def \N {\mathbb{N}}

\def \i {\mathrm{i}}
\def \d {\mathrm{d}}

\def \lb {\llbracket}
\def \rb {\rrbracket}
\def \ua {\uparrow}
\def \da {\downarrow}

\def \la {\langle}
\def \ra {\rangle}

\def \cal {\mathcal}

\def \le { \leqslant}
\def \ge {\geqslant}
\def \rm {\mathrm}

\def \eq {\Leftrightarrow}

\begin{document}


\title{Quantum Measurement Spintronic Engine powered by Quantum Fluctuations} 

\author{Mathieu Lamblin}
    \affiliation{Institut de Physique et Chimie des Mat\'eriaux de Strasbourg, UMR 7504 CNRS, Universit\'e de Strasbourg, 23 Rue du L\oe ss, BP 43, 67034 Strasbourg, France}
    \email{mathieu.lamblin@ipcms.unistra.fr}
\author{Martin Bowen}
    \affiliation{Institut de Physique et Chimie des Mat\'eriaux de Strasbourg, UMR 7504 CNRS, Universit\'e de Strasbourg, 23 Rue du L\oe ss, BP 43, 67034 Strasbourg, France}
    \email{bowen@unistra.fr}

\date{\today}

\begin{abstract}
Quantum fluctuations, which result from the Heisenberg uncertainty principle, explain a number of physical observations, from the finite mass of elementary particles to the Lamb shift in hydrogen and the Casimir effect. The local violation of the conservation of energy raises the question of whether the energy of quantum fluctuations can sustain the cycle of a quantum engine \cite{klatzow_experimental_2019, bresque_two-qubit_2021}. So far, a proposal has hinted that this is possible, but contains important caveats \cite{jussiau_many-body_nodate}. In this Letter, we predict that quantum fluctuations can power an autonomous spintronic quantum information engine by converting entanglement energy into useful electrical work. Our two-stroke engine operates on two entangled spin quantum dots (QDs) that are connected in series with two fully spin-polarized baths. The ultrafast measurement stroke breaks the entanglement, thereby energizing the system on average. This energy is released into the leads as electrical current when the thermalizing stroke equilibrates the QDs with the electrode baths. Using a master equation approach, we analytically demonstrate the efficiency of the quantum fluctuation-driven engine, and we study the cycle numerically to gain insight into the relevant parameters to maximize power. Our results suggest that quantum fluctuations and the measurement back-action alone cannot explain prior experimental results \cite{katcko_spin-driven_2019, chowrira_quantum_2022}. Measuring the spin dynamics of the engine's ferromagnetic electrodes should help determine its efficiency \cite{wright_quantum_2018}. This electronically driven feedback on quantum entanglement should also boost quantum chemistry \cite{Li2019}, biology \cite{Kim2021} and cognition \cite{Adams2020}.\\
\end{abstract}


\maketitle


In the emerging field of quantum thermodynamics \cite{Vinjanampathy_quantum_2016} and specifically of quantum energetics \cite{rogers_post-selection_2022}, much effort has been dedicated to the study of quantum thermal machines and quantum batteries \cite{bhattacharjee_quantum_2021}, in the hope of finding new ways of producing energy at the nanoscale. 
A first approach describes quantum analogues to a classical engines that rely either on a Maxwell demon \cite{ptaszynski_autonomous_2018, seah_maxwells_2020}, a cycle between different baths \cite{piccitto_ising_2022, manzano_entropy_2016, molitor_stroboscopic_2020} or an external drive \cite{donvil_thermodynamics_2018, zhao_quantum_2021, santos_efficiency_2022}. A second approach focuses on systems that specifically rely on quantum features such as coherence \cite{francica_quantum_2020, shi_quantum_2020, aimet_engineering_2023, monsel_energetic_2020} and entanglement \cite{buffoni_quantum_2019, bresque_two-qubit_2021, jussiau_many-body_nodate} in order to extract energy from the environment using the singular properties of quantum superposition and quantum measurements. These new kinds of quantum engines which have been demonstrated experimentally \cite{katcko_spin-driven_2019, chowrira_quantum_2022, klatzow_experimental_2019, ji_spin_2022, micadei_reversing_2019} 
redefine the notion of temperature when examining engine efficiencies against the Carnot limit \cite{niedenzu_quantum_2018, nanoscale_2014, manzano_entropy_2016, klaers_squeezed_2017}, notably when harvesting energy from single heat baths \cite{scully_extracting_2003, yi_single-temperature_2017}.

Quantum measurements constitute a critical process that could lead to active devices that use entanglement as a fuel \cite{elouard_efficient_2018, francica_daemonic_2017}. Indeed, the resulting projection that such a measurement performs on a quantum state involves an irreversible energy exchange between that state and the environment \cite{rogers_post-selection_2022} that can be viewed as a form of quantum heat \cite{elouard_extracting_2017}. The information obtained from the measurement can either be used by a Maxwell demon \cite{elouard_extracting_2017, bresque_two-qubit_2021} that can extract energy by applying some unitary transformation to the working substance (WS), or the measurement back-action in itself can result in an energy increase that can be harvested into usable work \cite{yi_single-temperature_2017}.

In these engine models \cite{piccitto_ising_2022, jussiau_many-body_nodate, klaers_squeezed_2017, molitor_stroboscopic_2020, ptaszynski_autonomous_2018, seah_maxwells_2020, francica_quantum_2020, shi_quantum_2020, aimet_engineering_2023, niedenzu_quantum_2018, nanoscale_2014, manzano_entropy_2016, klaers_squeezed_2017, henriet_electrical_2015}, the proposed cycle can be difficult to implement experimentally, while the cost of turning on/off interactions between the WS and the baths is ignored. In this Letter, we model a quantum electronic engine that mostly lifts these limitations, that is explicitly powered by quantum fluctuations in contrast to recent proposals \cite{jussiau_many-body_nodate, xiao_thermodynamics_nodate}, and that advantageously mimics a recent experimental implementation \cite{katcko_spin-driven_2019, chowrira_quantum_2022}. 


This quantum spintronic engine's WS is composed of two spin-split quantum dots (QDs) that exhibit a tunnel coupling $\gamma$, a magnetic exchange interaction $J$, and coulombic repulsion $U$ to prevent excessive charging. This open system is connected in series with two ferromagnetic, fully spin-polarized leads. Since the engine is a solid-state device, electronic interactions are inherently always on and time-independent, such that the engine cycle is only driven by periodic quantum measurements through two strokes : an instantaneous measurement stroke that partially projects the QD system (i.e. WS), followed by a thermalizing stroke of duration $\tau$ during which the WS relaxes towards the steady-state. 

Our initial formalism follows prior literature \cite{fransson_pauli_2006, weymann_transport_2007, fransson_electrical_2014}, and was used to successfully model magnetotransport across a quantum spintronic engine under the assumption of effective work \cite{katcko_spin-driven_2019}.

\textit{Model.}\textemdash In line with experimental input \cite{katcko_spin-driven_2019, chowrira_quantum_2022}, we make the following assumptions: $J < 0$, $\gamma$ is spin-independent, only one electronic band is involved, the spintronic anisotropy remains constant (no external bias voltage), the $\ua$ spin energy level on the right side is inaccessible, and the tunnelling coefficients $\gamma_L$ and $\gamma_R$ are asymmetric. We can thus simplify the system (see schematic in Fig.~\ref{model_2} and SI note 1), so that the system Hamiltonian reads: 
\begin{multline}
    H_S = \epsilon_{\ua} n_{\ua} + \epsilon_{\da} n_{\da} + \epsilon_{R} n_{R} + \gamma\, c^{\dagger}_{\da}c_{R} + \gamma^*\, c_{\da}c^{\dagger}_{R} \\ + J\, n_{\da}n_{R} + U\,n_{\ua}n_{\da}\ ,
\end{multline}
where $c_{\da}$, $c_{\ua}$ and $c_R$ are the annihilation operators of the left $\da$ spin, the left $\ua$ spin and the right $\da$ spin respectively ; $n = c^{\dagger}c$ are the number operators and $\epsilon$ are the effective bare energy of each electronic level. Note the absence of an explicit spin channel between the electrodes, and of a spin-flip mechanism ; hence our model can lead to electron transport through engine operation that can only arise from so-called quantum fluctuations via two-electron processes.

\begin{figure}[t]
    \includegraphics[width=0.48\textwidth]{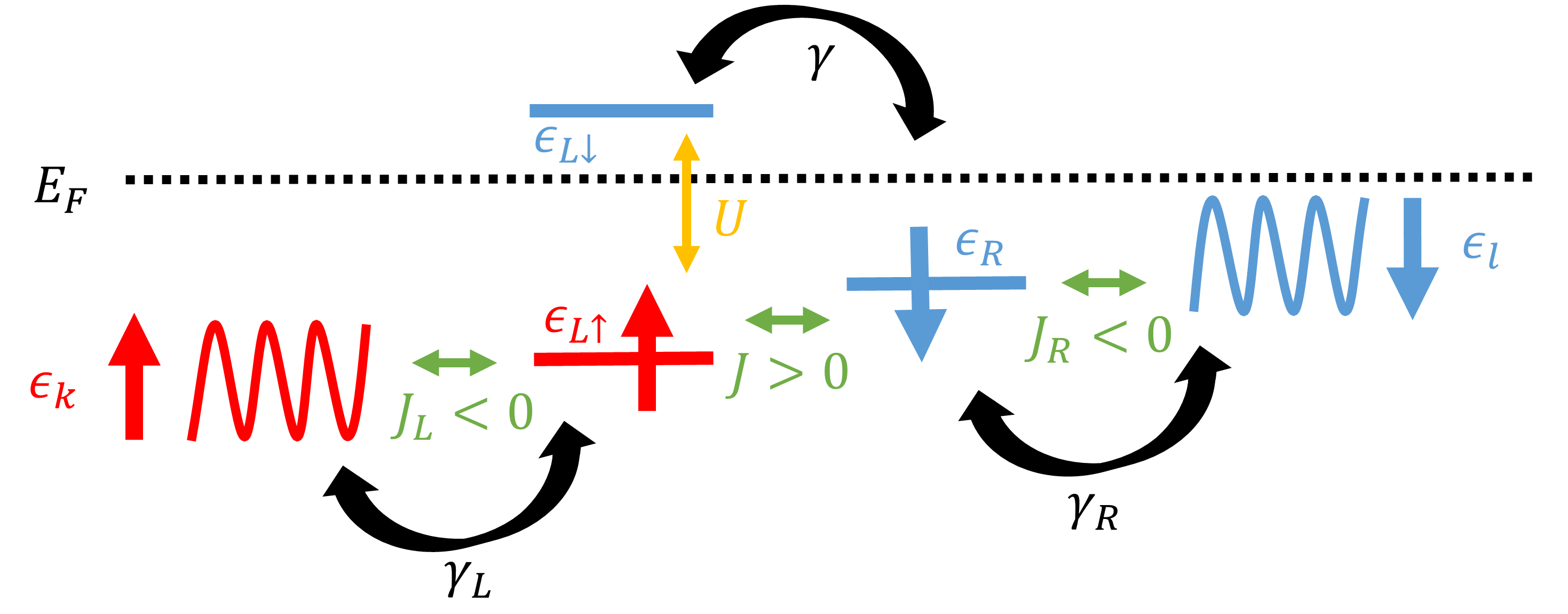}
    \caption{Schematic of the quantum spintronic engine, featuring two quantum dots coupled in series to ferromagnetic leads through fully spin-polarized interactions. Spin $\da$/$\ua$ energy levels (blue/red), magnetic couplings (green double arrows), capacitive couplings (yellow double arrows) and tunnel couplings (black arrows) are shown.}  
    \label{model_2}
\end{figure}

In SI Note. 1, we 
establish the master equation describing the evolution of this open quantum system:
\begin{multline}\label{me}
    \frac{\d\rho_S}{\d t} = -i[H_S,\rho_S] + \cal{T}^-_L\cal{D}[c^{\dagger}_{\ua}](\rho_S) + \cal{T}^+_L\cal{D}[c_{\ua}](\rho_S) \\+ \cal{T}^-_R\cal{D}[c^{\dagger}_{R}](\rho_S) + \cal{T}^+_R\cal{D}[c_{R}](\rho_S),
\end{multline}
where $\rho_S = \rm{Tr}_E\, \rho$ is the QD system's reduced density matrix w.r.t the environment degrees of freedom, $\cal{T}^-_L$ and $\cal{T}^-_R$ represent the electron hopping intensity beteen the QD and the left/right lead respectively, while $\cal{T}^+_L$ and $\cal{T}^+_R$ are the hole counterparts; and with the superoperator $\cal{D}[c](\rho) \equiv  c\rho c^{\dagger}-\frac{1}{2} \{c^{\dagger}c,\rho \}$.

A key element of our quantum spintronic engine is the ferromagnetic metal/molecule interface. This so-called spinterface exhibits a low density of spatiospectrally confined states with high spin polarization (89\% \cite{chowrira_quantum_2022}) at the Fermi level \cite{djeghloul_high_2016, delprat_molecular_2018, katcko_spin-driven_2019, chowrira_quantum_2022}. The proposal that a Maxwell demon can operate electronically at the molecular level \cite{bergfield_probing_2013}, along with recent thermodynamical theory on quantum measurements \cite{manzano_optimal_2018, erez_thermodynamic_2008, erez_thermodynamics_2012}, indicate that the spinterface could \cite{chowrira_quantum_2022} act as an autonomous quantum measurement apparatus by performing frequent projective measurements on the nearby QD, thereby collapsing the WS's quantum state \cite{lindblad_entropy_1973, jacobs_quantum_2012, ban_state_1999}. Following related studies that all postulate an external non-thermal input quantum resource \cite{scully_extracting_2003, bresque_two-qubit_2021, jussiau_many-body_nodate, elouard_extracting_2017, yi_single-temperature_2017}, we assume that the spinterface can perform these quantum measurements without an energy cost. Here, however, this assumption is backed by the ability of the ferromagnetic electrode to maintain a constant spin polarization, thus allowing it to behave as an entropy sink. Indeed, information erasure would require a much lower entropy cost than the Landauer bound by involving the transfer of spin angular momentum into a large spin reservoir, rather than energy \cite{wright_quantum_2018, bormashenko_entropy_2020, vaccaro_information_2011}. 

\begin{figure}[t]
    \includegraphics[width=0.48\textwidth]{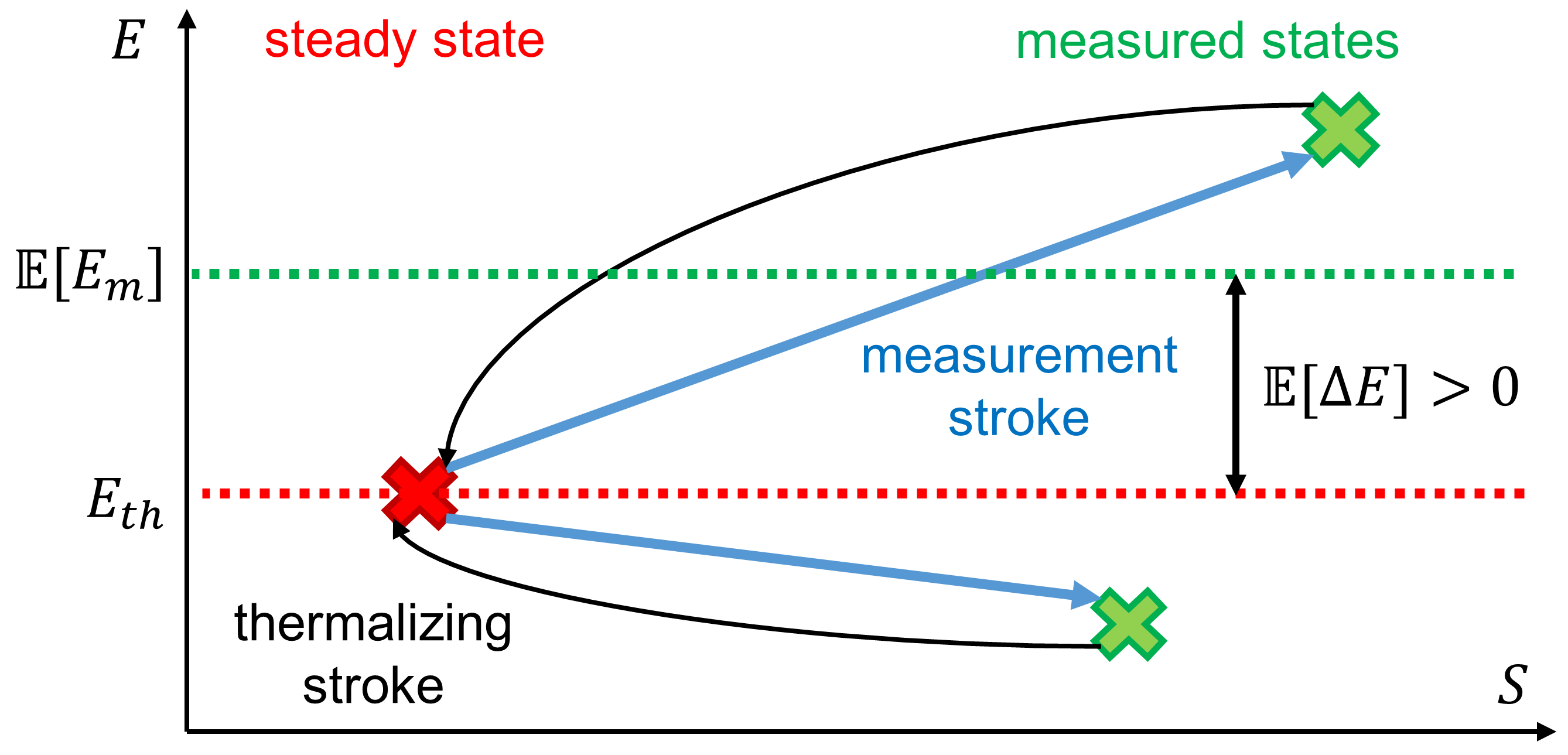}
    \caption{Energy-entropy schematic of the engine cycle. The unselective measurement stroke (straight blue arrows) instantly projects the thermalized steady-state (red) onto several possible separated states (green) with higher entropy and potentially higher energy. Thermalizing strokes (curved black arrows) reset the system to the steady-state while allowing for work extraction on average.}  
    \label{cycle}
\end{figure}

\textit{Engine.}\textemdash Using our Hamiltonian, we consider a two-phased engine cycle (see Fig.~\ref{cycle}). A first 'thermalizing' stroke places the QDs into equilibrium with the electrode baths. The relaxation of the QD sytems during this time-dependent evolution transfers energy from the system to the baths, some of which is harvested to produce useful electrical work. Then, once the system reaches its entangled steady-state, a 'measurement' stroke on a single QD splits the WS into two separated subsystems, thereby killing the entanglement. As we will show, this projective partial unselective measurement acting on a superposed mixed stated with indefinite energy is mathematically described by a quantum channel that leads to a projected system with a higher mean energy than the previous steady-state. The excess energy that results from this back-action \cite{bresque_two-qubit_2021, seah_maxwells_2020, yi_single-temperature_2017} of the measurement is then dissipated into the baths during the next 'thermalizing' stroke. We will show that it can be used to produce electrical work.


To find the equilibrium state, we first derive the density matrix $\rho \equiv \rho_{ss}$ such as $\frac{\d\rho}{\d t}=0$ so that $\rho$ nullify the right hand side of Eq.\ref{me}. To obtain an approximate analytical solution, we use a perturbation approach on $\gamma$ (see SI Note 2), i.e. assume that other interaction energies dominate \cite{chowrira_quantum_2022} the hybridization between the QDs. The solution is given by an affine space, parameterized by the initial population $\mu = \la n_{\da}(0) \ra$ of the spin $\da$ energy level on the left QD. Upon choosing the basis:
\begin{equation}
\begin{array}{c}
     |0\ra \equiv |00\ra,\ |1\ra \equiv |0\da\ra,\ |2\ra \equiv |\ua 0\ra,\ |3\ra \equiv |\ua\da\ra,\\ |4\ra \equiv |\da 0\ra,\ |5\ra \equiv |\da\da\ra,\ |6\ra \equiv |20\ra,\ |7\ra \equiv |2\da\ra\ 
\end{array} ,
\end{equation}
we obtain the full density matrix $\rho^{\mu}$ after thermalization:
\begin{equation}
    \left\{\begin{array}{l}
        \rho_{00} = \frac{1-\mu}{\mu}\rho_{44} = \alpha(1-\mu)\cal{T}_L^+\cal{T}_R^+\\
        \rho_{11} = \frac{1-\mu}{\mu}\rho_{55} = \alpha(1-\mu)\cal{T}_L^+\cal{T}_R^-\\
        \rho_{22} = \frac{1-\mu}{\mu}\rho_{66} = \alpha(1-\mu)\cal{T}_L^-\cal{T}_R^+\\
        \rho_{33} = \frac{1-\mu}{\mu}\rho_{77} = \alpha(1-\mu)\cal{T}_L^-\cal{T}_R^-\\
        \rho_{41} = \rho_{14}^* = \frac{\i\gamma\alpha\beta^{\mu}\cal{T}_L^+}{\rm{det}\,B^*}\Big(\frac{2\cal{T}_L^+ + 2\cal{T}_L^- +\cal{T}_R^-+\cal{T}_R^+}{2} +\i(\Delta + U)\Big) \\
        \rho_{63} = \rho_{36}^* = \frac{\i\gamma\alpha\beta^{\mu}\cal{T}_L^-}{\rm{det}\,B^*}\Big( \frac{2\cal{T}_L^+ + 2\cal{T}_L^- +\cal{T}_R^-+\cal{T}_R^+}{2} +\i\Delta\Big)
    \end{array}\right.
\end{equation}
where $\beta^{\mu} \equiv \mu\cal{T}_R^+-(1-\mu)\cal{T}_R^-$, $1/\alpha \equiv  (\cal{T}_L^++\cal{T}_L^-)(\cal{T}_R^++\cal{T}_R^-)$, $\Delta \equiv \epsilon_{\da}-\epsilon_R$, $B$ is referenced in SI Note. 2 and the other terms are null.


Let us initialize our engine at $t_0=0$ in some state $\rho(0)$ such that $\mu = \la n_{\da}(0) \ra$. After completing a first thermalization process, the electrode performs a partial projective measurement of the entangled QDs at time $t_1=\tau_{m} \equiv \tau$, which represents the duration of one cycle. This measurement projects the system from the steady state $\rho(\tau^-) = \rho^{\mu} \equiv \rho$ to a projected state $\rho(\tau^+)$ that depends on the measurement outcome. Supposing that the right electrode operates frequent unselective measurements of the occupation of the right QD at times $t = t_n \equiv n\tau$, the associated observable is simply $n_R$. The measurement yields either the presence ($n_R$ = 1) or the absence ($n_R$ = 0) of one electron on the right-hand QD. The two possible projectors on the respective eigenspaces are $\Pi_0 = 1-n_R$ and $\Pi_1 = n_R$, leading to the projected state:
\begin{equation}
    \rho(\tau^+) = \Pi_0\rho\Pi_0 + \Pi_1\rho\Pi_1 = \rho + 2\cal{D}[n_R](\rho).
\end{equation}
Using the following expression for the density matrix,
\begin{equation}
    \rho = \sum_{i=0}^7 \rho_{ii}|i\ra\la i| + \rho_{14}|1\ra\la 4| + \rho_{41}|4\ra\la 1| + \rho_{36}|3\ra\la 6| + \rho_{63}|6\ra\la 3|\ ,
\end{equation}
we immediately witness that the off-diagonal terms do not contribute in the projected state because they encode the tunnelling of one electron from one site to the next, leaving either the initial state or the final state with no electron on the right side. Hence we calculate $\rho(\tau^+) = \sum_{i=0}^7 \rho_{ii}|i\ra\la i|$ so that $\rho(\tau^+)$ is the diagonal part of $\rho$, while $-2\cal{D}[n_R](\rho)$ is the off-diagonal part.

We can now calculate the energy and entropy impact of the measurement. The average energy of the system changes by an amount $\Delta E$:
\begin{equation}
    \Delta E = \rm{Tr}[H_S\rho(\tau^+)] - \rm{Tr}[H_S\rho(\tau^-)] = 2\rm{Tr}\Big[H_S\cal{D}[n_R](\rho)\Big]\ .
\end{equation}
which thus represents the energy of the off-diagonal part:
\begin{equation}
    \Delta E = -2\Re[\gamma(\rho_{14}+\rho_{36})] = -\la T(\tau^-) \ra = -\rm{Tr}[T\rho]\ .
\end{equation}
where $T \equiv \gamma c_{\da}^{\dagger}c_R + \gamma^*c_R^{\dagger}c_{\da}$ is the inter-dot tunnel operator.
Thus, the measurement induces a back-action on the system by disentangling it, leading to an energy change $\Delta E$ compared to the thermalized state.

To study the thermalized state of the next cycle, we consider the particle number with spin $\da$ in the left QD. Since the projected state is diagonal, we directly obtain:
\begin{equation}
    \rm{Tr}[n_{\da}\rho(\tau^+)] = \rho_{44}+\rho_{55}+ \rho_{66}+\rho_{77} = \mu 
\end{equation}
So, for both measurement outcomes, the spin $\da$ occupation number remains unchanged after both the thermalizing and measurement strokes. Therefore, the second cycle starts again with $\la n_{\da}(\tau^+) \ra = \mu$, and so yields the same thermalized state just before the second measurement 
, such that $\rho(2\tau^-) = \rho(\tau^-) = \rho$ and thus an instant recursion yields the system state after each cycle $n$:  
\begin{equation}
\rho(n\tau^+) = \rho(n\tau^-) + 2D[n_R](\rho) = \rho + 2D[n_R](\rho).
\end{equation}
This shows that at time $t = n\tau^+$, the system has received a total average energy $n\Delta E$ from quantum measurements. Therefore, if $\Delta E > 0$, then the measurement on average energizes the system, and a fraction of that energy upon deexciting to the thermalized state can be harvested in the form of electron transport.

In SI Note. 3, we show that the engine operation/output is unchanged when measuring an observable that acts on only one QD (e.g. when the occupation of the left QD, or the charge (or spin) of the right QD, is measured). Measuring an observable that operates on both QDs, such as the total charge, does not yield this energy increment: the thermalized and measured states both present the same average energy. Work extraction from these cycles is possible only when the measurement separates the two QDs. Then, in SI Note. 4, we study the case of selective quantum measurements and show that the energy increment relation holds by linearity while entropic considerations differ.



We can express the energy $\Delta E = -2\Re[\gamma(\rho_{14}+\rho_{36})]$ associated with the measurement back-action as:
\begin{equation}
    \Delta E = \frac{\beta^{\mu}|\gamma|^2 }{|\rm{det}\,B|^2}\frac{(s+r)^2}{sr}(s\Delta + mU) \ ,
\end{equation}
with $s \equiv \cal{T}_L^++\cal{T}_L^-$, $m \equiv \cal{T}_L^-$, $r \equiv \frac{\cal{T}_R^++\cal{T}_R^-}{2}$.
To study $\Delta E$, we first assume (see SI. Note 1) that all the energies involved in the system $\cal{T}_L^+$, $\cal{T}_L^-$, $\cal{T}_R^+$, $\cal{T}_R^-$, $\Delta$ and $U$ are strictly positive. This immediately leads to $s,m,r > 0$ and then to $\Delta E$ having the same sign as $\beta^{\mu}$. Therefore, the measurement energizes the two QDs whenever $\mu < \frac{\cal{T}_R^-}{\cal{T}_R^+ + \cal{T}_R^-} \equiv \mu^{c}$. 


We emphasize that the engine is still bounded by a Carnot limit. Indeed, in SI Note. 5, we define the engine efficiency $\eta \equiv \frac{W_{el}}{\Delta E}$, where $W_{el}$ is the electronic work obtained during the thermalization process, and we show that
$\eta \le 1 - \frac{T}{T_c}$. Here, the critical temperature $T_c \equiv \frac{\Delta E}{\Delta S}$ defines a threshold above which this engine cannot operate, and $\Delta S > 0$ is the difference of Von-Neumann entropy between the measured state and the steady-state.




\begin{figure}[t]
    \includegraphics[width=0.48\textwidth]{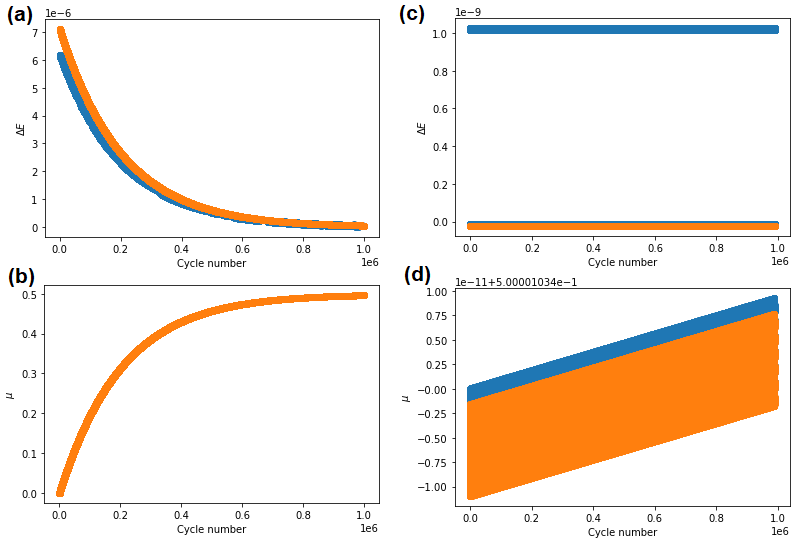}
    \caption{Simulation results of $\Delta E$ in (a) and (c), and $\mu$ in (b) and (d) for $10^5$ cycles when measuring $n_R$. The corrected perturbative results (orange) and the numerically calculated solution at 4 ps (blue) are shown. The parameters used are $\epsilon_{\da} = 8,\ \epsilon_{\ua} = -3,\ \epsilon_R = 1,\ J = 8,\ U = 100,\ \gamma = 0.1,\ \cal{T}_L^+ = \cal{T}_R^+=\cal{T}_L^- = \cal{T}_R^-=1$ (all units may be taken in meV). For (a) and (b), 
    $\rho_0 = |\ua\da\ra\la\ua\da|$. For (c) and (d), 
    $\rho_0 = \frac{1}{2}|\ua\da\ra\la\ua\da| + \frac{1}{2}|\da\da\ra\la\da\da|$.}
    \label{drift}
\end{figure}

\textit{Simulations.}\textemdash To evaluate these analytical results, we simulated the engine operation in order to witness the drift from the perturbative solution, and to show that the measurement reliably provides energy over many cycles. We present in Fig.~\ref{drift} the results of measuring $n_R$ across $10^6$ cycles (see also SI Note. 6). Starting from the pure state $\rho_0 = |\ua\da\ra\la\ua\da|$, we first see that the identity $\Delta E = - \la T \ra$ is well respected (see Fig.SI.1), and that $\Delta E$ decreases after each cycle (see Fig.~\ref{drift}(a)). This is because the population of the down spin $\mu = \la n_{\da} \ra$ gradually increases up to $\mu^{c}=1/2$ in this case, drifting away from the approximate solution that considers it constant to first order in $\gamma$. This equilibration therefore reduces the magnitude of the off-diagonal terms that code for the energizing process, as they are proportional to $\beta^{\mu}$ (see SI Note. 2). Choosing different values for $\cal{T}_R^+$ and $\cal{T}_R^-$ should therefore favor a higher energizing capability. The perturbative solution we derived to approximate the state after each thermalizing stroke deviates from the numerical solution but it correctly mimicks how observables evolve, even after many cycles. Note that the solution should be used each time with a different parameter $\mu$ obtained from the numerically measured state (see Fig.~\ref{drift}(c)).

The simulation strongly depends on the the initial state $\rho_0$. In Fig.~\ref{drift}(c) and (d), we show that, starting from a mixed state $\rho_0 = \frac{1}{2}|\ua\da\ra\la\ua\da| + \frac{1}{2}|\da\da\ra\la\da\da|$, we instead find that $\Delta E$ oscillates randomly between two values. We conjecture that, after a long time, the engine initialized with $\rho_0 = |\ua\da\ra\la\ua\da|$ will approach this fluctuating behavior around 0, as the information on the initial condition is progressively lost through the non-unitary system evolution during thermalization. This eventually leads to nearly vanishing power output due to excessive entropy: the temporal average energy increment falls from $\mathbb{E}[\Delta E] = 10^{-6}$ meV to $\mathbb{E}[\Delta E] = 10^{-9}$ meV. See SI Note 5 for additional data exploring different parameters, measurement protocols, as well as the case of selective measurements. These results show similar behaviors, i.e. all exhibit the ability to extract energy. 


We now consider cases outside the perturbative regime detailed above. First, in SI Note. 7, we showed that 
for a wide range of parameters, it is unreasonable to approximate the state at the end of the thermalizing stroke as the steady-state solution.
Thus, in this general case, we can only hope to reach a partial thermalization, though it may be beneficial to the power output of the device. Indeed, statistics (not shown) on the steady-state entanglement energy reveal that, at infinite time $|\la T \ra| \approx 10^{-9}-10^{-16}$ meV for standard parameter ranges, while after $t = 1$ meV$^{-1} \approx 4$ ps, we can reach up to $|\la T \ra| \approx 10^3$ meV for the same parameter ranges and for specific initial conditions. This suggests a higher energy increment. Indeed, the average energy increment per cycle is still given by $\Delta E = -\la T\ra$ in this general case, so that the previous energetic description of the cycle remain valid here. A proof of this assertion is given in SI Note. 8 along with a numerical justification based on measurement statistics.

\begin{figure}[b]
    \includegraphics[width=0.48\textwidth]{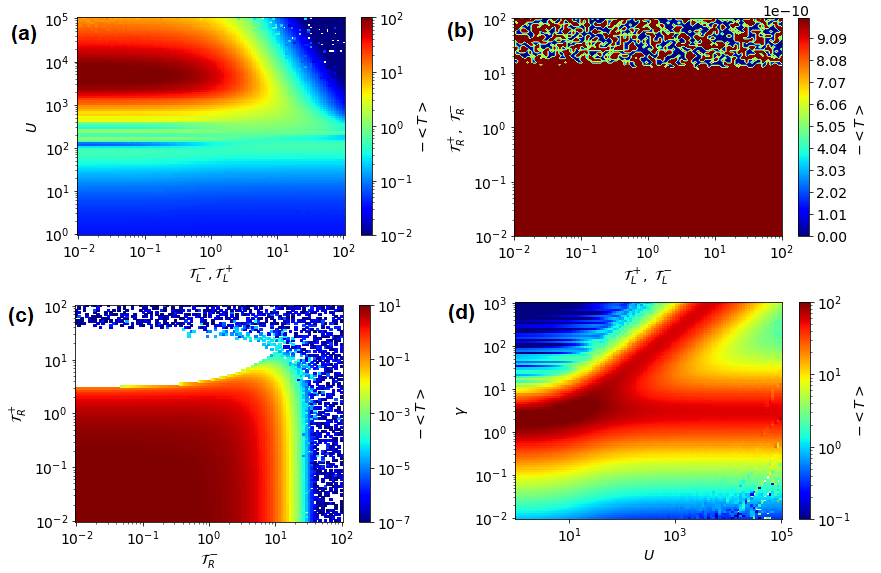}
    \caption{Parameter pair dependence of $-\la T \ra$, calculated at 4 ps starting from the pure state $\rho_0 = |\ua\da\ra\la\ua\da|$, with $\epsilon_{\da} = 8,\ \epsilon_{\ua} = -3,\ \epsilon_R = 1,\ J= 8,\ U=10,\ \gamma = 1000,\ \cal{T}_L^+=\cal{T}_L^-=1,\ \cal{T}_R^+=\cal{T}_R^-=5$.}
    \label{color}
\end{figure}

To study how the parameters that impact the steady-state $\rho$ affect the entanglement energy $-\la T\ra$, we present In Fig.~\ref{color} several color plots of $-\la T\ra$ calculated after partial thermalization starting from the pure state $\rho_0 = |\ua\da\ra\la\ua\da|$ as a function of the most relevant different pairs of parameters (see SI Note. 9 for other plots), while keeping other parameters fixed, and with $\gamma \gg U \sim \epsilon \sim \cal{T}$. In Fig~\ref{color}(a), we notice that $-\la T\ra$ is maximized when $U \approx 10^4$ and $\cal{T}_L^+=\cal{T}_L^- < 1$. Indeed, a higher $U$ could lead to a bigger entanglement energy that is related to the charging energy of a QD, while a lower $\cal{T}_L$ favors the tunneling between the QDs over the tunneling from/to the electrodes. In Fig~\ref{color}(b), we observe that the asymmetry between $\cal{T}_L$ and $\cal{T}_R$ is completely irrelevant for this set of parameters for low $\cal{T}_R$: the engine generates power and energy harvesting may be possible. Above a phase transition around $\cal{T}_R \approx 10$ (see also Fig~\ref{color}(d)), a chaotic phase ensues, the value of $\la T \ra$ almost vanishes and its sign strongly depends on small fluctuations of the parameters: we cannot hope to extract energy in this configuration. 
In Fig~\ref{color}(c), we examine the electron/hole asymmetry on the right electrode. The data reveals a third phase in white in which $-\la T \ra < 0$: a dissipative phase with no energy extraction. 
Finally, in Fig~\ref{color}(d), the $U$ / $\gamma$ dependence reveals two branches that maximize $-\la T \ra$: one for $\gamma \approx 1$ that weakly depends on $U$, and a second for $\gamma \approx U$. This may help tune experimental device parameters to maximize energy harvesting.

\begin{figure}[t]
    \includegraphics[width=0.48\textwidth]{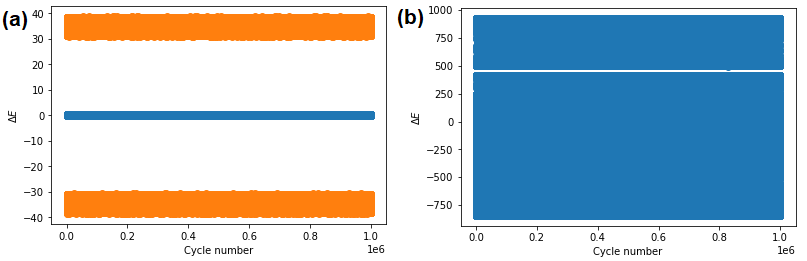}
    \caption{Simulation results of $\Delta E$ for $10^6$ cycles when measuring the charge $Q$ of the left QD using (a) unselective and (b) selective measurements. The numerical calculation (orange) and the perturbative solution (blue) are shown. $\epsilon_{\da} = 8,\ \epsilon_{\ua} = -3,\ \epsilon_R = 1,\ J = 8,\ U = 1000,\ \gamma = 1000,\ \cal{T}_L^+ = \cal{T}_R^+=\cal{T}_L^- = \cal{T}_R^-=0.1$ (all units may be taken in meV), and with the initial condition $\rho_0 = |\ua\da\ra\la\ua\da|$.}
    \label{fluctuations}
\end{figure}

Using the data of SI Note. 9, we infer a regime wherein the entanglement energy is highest when $\cal{T} \ll \epsilon \ll \gamma \approx U \approx 1000\epsilon$. To confirm the high power output $P = \frac{\mathbb{E}[\Delta E]}{\tau}$ within this parameter space, we simulated $10^6$ engine cycles (see plots in Fig~\ref{fluctuations}). Strikingly, we observe strong fluctuations of $\Delta E$ that ultimately kill the temporal average of the energy increment. This shows that maximizing $-\la T \ra$ also yields a strong dependence of the tunneling energy after partial thermalization on the initial conditions. This increases the fluctuations and negatively impacts $P$: maximizing $P$ requires balancing energy and fluctuations. Indeed, from the Heisenberg uncertainty relations, when selectively measuring $n_R$, we may write:
\begin{equation}\label{HIR}
    \Delta T \Delta n_R \ge \frac{1}{2}|\la [T, n_R] \ra | = \frac{1}{2}|\la [H_S, n_R] \ra |\approx \frac{1}{2}\bigg| \frac{\d \la n_R\ra }{\d t}\bigg| .
\end{equation}
Here, at the end of each cycle, $\Delta n_R \lesssim 1$ is known and fixed by the statistical outcomes of the measurements and should be of order unity since the measurement alternatively projects the system into a $n_R =$ 0 or 1 state. Moreover, the right-hand side describes the $n_R$ oscillation rate, which is strongly driven by the $|\la T \ra|$ energy scale. Thus, Eq.~\ref{HIR} indeed justifies that $\Delta T \gtrsim |\la T \ra| $.


\textit{Conclusions.}\textemdash We studied a quantum information engine built around autonomous solid-state spintronic interactions that can promote the energy harvesting of quantum fluctuations. Our model considered a pair of entangled spin quantum dots that electronically interact with spin-selecting electrodes. By using physical assumptions from prior experiments\cite{katcko_spin-driven_2019,chowrira_quantum_2022} to reduce parameter space, we derived a master equation that describes a two-cycle engine featuring: a thermalizing stroke that entangles the two quantum dots and generates electron transport, then a quantum measurement stroke that extracts energy from their mutual information, i.e. through quantum separation. This changes their entropy by separating and projecting the system in a higher energy state on average. In the limiting case that quantum fluctuations
constitute the only energy source, numerical simulations predicted a finite appreciable power output in some cases. The far lower magnitude compared with experiments \cite{chowrira_quantum_2022} is either due to a faster thermalization, or to another energy source such as phonons. 
Our work opens fruitful research into spintronic interaction dynamics between ferromagnets and paramagnetic centers \cite{bowen_atom_2023}, e.g. using scanning tunnelling and ferromagnetic resonance techniques \cite{harder_electrical_2016}, to elucidate the thermodynamic role \cite{vaccaro_information_2011,wright_quantum_2018} of the ferromagnetic metal / molecule quantum measurement apparatus\cite{katcko_spin-driven_2019,chowrira_quantum_2022}. Our model's autonomous interlocking electronic strokes to generate feedback on quantum entanglement should inspire the fields of quantum chemistry\cite{Li2019}, biology \cite{Kim2021} and cognition \cite{Adams2020}.\\

We thank C. Elouard, J. Monsel, K. Singer, and R. Whitney for stimulating discussions. We gratefully acnkowledge PhD funding for M.L. from Ecole Polytechnique. We acknowledge financial support from the ANR (ANR-21-CE50-0039), the Contrat de Plan Etat-Region grants in 2006 and 2008, by “NanoTérahertz”, a project co-funded by the ERDF 2014–2020 in Alsace (European Union fund) and by the Region Grand Est through its FRCR call, by the impact project LUE-N4S part of the French PIA project “Lorraine Université d’Excellence”, reference ANR-15IDEX-04-LUE and by the FEDER-FSE “Lorraine et Massif Vosges 2014–2020”, a European Union Program. This ‘SpinDrive’ work of the Interdisciplinary Thematic Institute QMat, as part of the ITI 2021-2028 program of the University of Strasbourg, CNRS and Inserm, was supported by IdEx Unistra (ANR 10 IDEX 0002), and by SFRI STRAT’US project (ANR 20 SFRI 0012) and EUR QMAT ANR-17-EURE-0024 under the framework of the French Investments for the Future Program.

\bibliography{ref} 

\begin{thebibliography}{61}%
\makeatletter
\providecommand \@ifxundefined [1]{%
 \@ifx{#1\undefined}
}%
\providecommand \@ifnum [1]{%
 \ifnum #1\expandafter \@firstoftwo
 \else \expandafter \@secondoftwo
 \fi
}%
\providecommand \@ifx [1]{%
 \ifx #1\expandafter \@firstoftwo
 \else \expandafter \@secondoftwo
 \fi
}%
\providecommand \natexlab [1]{#1}%
\providecommand \enquote  [1]{``#1''}%
\providecommand \bibnamefont  [1]{#1}%
\providecommand \bibfnamefont [1]{#1}%
\providecommand \citenamefont [1]{#1}%
\providecommand \href@noop [0]{\@secondoftwo}%
\providecommand \href [0]{\begingroup \@sanitize@url \@href}%
\providecommand \@href[1]{\@@startlink{#1}\@@href}%
\providecommand \@@href[1]{\endgroup#1\@@endlink}%
\providecommand \@sanitize@url [0]{\catcode `\\12\catcode `\$12\catcode
  `\&12\catcode `\#12\catcode `\^12\catcode `\_12\catcode `\%12\relax}%
\providecommand \@@startlink[1]{}%
\providecommand \@@endlink[0]{}%
\providecommand \url  [0]{\begingroup\@sanitize@url \@url }%
\providecommand \@url [1]{\endgroup\@href {#1}{\urlprefix }}%
\providecommand \urlprefix  [0]{URL }%
\providecommand \Eprint [0]{\href }%
\providecommand \doibase [0]{https://doi.org/}%
\providecommand \selectlanguage [0]{\@gobble}%
\providecommand \bibinfo  [0]{\@secondoftwo}%
\providecommand \bibfield  [0]{\@secondoftwo}%
\providecommand \translation [1]{[#1]}%
\providecommand \BibitemOpen [0]{}%
\providecommand \bibitemStop [0]{}%
\providecommand \bibitemNoStop [0]{.\EOS\space}%
\providecommand \EOS [0]{\spacefactor3000\relax}%
\providecommand \BibitemShut  [1]{\csname bibitem#1\endcsname}%
\let\auto@bib@innerbib\@empty
\bibitem [{\citenamefont {Klatzow}\ \emph {et~al.}(2019)\citenamefont
  {Klatzow}, \citenamefont {Becker}, \citenamefont {Ledingham}, \citenamefont
  {Weinzetl}, \citenamefont {Kaczmarek}, \citenamefont {Saunders},
  \citenamefont {Nunn}, \citenamefont {Walmsley}, \citenamefont {Uzdin},\ and\
  \citenamefont {Poem}}]{klatzow_experimental_2019}%
  \BibitemOpen
  \bibfield  {author} {\bibinfo {author} {\bibfnamefont {J.}~\bibnamefont
  {Klatzow}}, \bibinfo {author} {\bibfnamefont {J.~N.}\ \bibnamefont {Becker}},
  \bibinfo {author} {\bibfnamefont {P.~M.}\ \bibnamefont {Ledingham}}, \bibinfo
  {author} {\bibfnamefont {C.}~\bibnamefont {Weinzetl}}, \bibinfo {author}
  {\bibfnamefont {K.~T.}\ \bibnamefont {Kaczmarek}}, \bibinfo {author}
  {\bibfnamefont {D.~J.}\ \bibnamefont {Saunders}}, \bibinfo {author}
  {\bibfnamefont {J.}~\bibnamefont {Nunn}}, \bibinfo {author} {\bibfnamefont
  {I.~A.}\ \bibnamefont {Walmsley}}, \bibinfo {author} {\bibfnamefont
  {R.}~\bibnamefont {Uzdin}},\ and\ \bibinfo {author} {\bibfnamefont
  {E.}~\bibnamefont {Poem}},\ }\bibfield  {title} {\bibinfo {title}
  {Experimental {Demonstration} of {Quantum} {Effects} in the {Operation} of
  {Microscopic} {Heat} {Engines}},\ }\bibfield  {journal} {\bibinfo  {journal}
  {Physical Review Letters}\ }\textbf {\bibinfo {volume} {122}},\ \href
  {https://doi.org/10.1103/PhysRevLett.122.110601}
  {10.1103/PhysRevLett.122.110601} (\bibinfo {year} {2019})\BibitemShut
  {NoStop}%
\bibitem [{\citenamefont {Bresque}\ \emph {et~al.}(2021)\citenamefont
  {Bresque}, \citenamefont {Camati}, \citenamefont {Rogers}, \citenamefont
  {Murch}, \citenamefont {Jordan},\ and\ \citenamefont
  {Auffèves}}]{bresque_two-qubit_2021}%
  \BibitemOpen
  \bibfield  {author} {\bibinfo {author} {\bibfnamefont {L.}~\bibnamefont
  {Bresque}}, \bibinfo {author} {\bibfnamefont {P.~A.}\ \bibnamefont {Camati}},
  \bibinfo {author} {\bibfnamefont {S.}~\bibnamefont {Rogers}}, \bibinfo
  {author} {\bibfnamefont {K.}~\bibnamefont {Murch}}, \bibinfo {author}
  {\bibfnamefont {A.~N.}\ \bibnamefont {Jordan}},\ and\ \bibinfo {author}
  {\bibfnamefont {A.}~\bibnamefont {Auffèves}},\ }\bibfield  {title} {\bibinfo
  {title} {Two-{Qubit} {Engine} {Fueled} by {Entanglement} and {Local}
  {Measurements}},\ }\bibfield  {journal} {\bibinfo  {journal} {Physical Review
  Letters}\ }\textbf {\bibinfo {volume} {126}},\ \href
  {https://doi.org/10.1103/PhysRevLett.126.120605}
  {10.1103/PhysRevLett.126.120605} (\bibinfo {year} {2021})\BibitemShut
  {NoStop}%
\bibitem [{\citenamefont {Jussiau}\ \emph {et~al.}()\citenamefont {Jussiau},
  \citenamefont {Bresque}, \citenamefont {Auﬀeves}, \citenamefont {Murch},\
  and\ \citenamefont {Jordan}}]{jussiau_many-body_nodate}%
  \BibitemOpen
  \bibfield  {author} {\bibinfo {author} {\bibfnamefont {E.}~\bibnamefont
  {Jussiau}}, \bibinfo {author} {\bibfnamefont {L.}~\bibnamefont {Bresque}},
  \bibinfo {author} {\bibfnamefont {A.}~\bibnamefont {Auﬀeves}}, \bibinfo
  {author} {\bibfnamefont {K.~W.}\ \bibnamefont {Murch}},\ and\ \bibinfo
  {author} {\bibfnamefont {A.~N.}\ \bibnamefont {Jordan}},\ }\bibfield  {title}
  {\bibinfo {title} {Many-body quantum vacuum ﬂuctuation engines},\
  }\href@noop {} {\ ,\ \bibinfo {pages} {19}}\BibitemShut {NoStop}%
\bibitem [{\citenamefont {Katcko}\ \emph {et~al.}(2019)\citenamefont {Katcko},
  \citenamefont {Urbain}, \citenamefont {Taudul}, \citenamefont {Schleicher},
  \citenamefont {Arabski}, \citenamefont {Beaurepaire}, \citenamefont {Vileno},
  \citenamefont {Spor}, \citenamefont {Weber}, \citenamefont {Lacour},
  \citenamefont {Boukari}, \citenamefont {Hehn}, \citenamefont {Alouani},
  \citenamefont {Fransson},\ and\ \citenamefont
  {Bowen}}]{katcko_spin-driven_2019}%
  \BibitemOpen
  \bibfield  {author} {\bibinfo {author} {\bibfnamefont {K.}~\bibnamefont
  {Katcko}}, \bibinfo {author} {\bibfnamefont {E.}~\bibnamefont {Urbain}},
  \bibinfo {author} {\bibfnamefont {B.}~\bibnamefont {Taudul}}, \bibinfo
  {author} {\bibfnamefont {F.}~\bibnamefont {Schleicher}}, \bibinfo {author}
  {\bibfnamefont {J.}~\bibnamefont {Arabski}}, \bibinfo {author} {\bibfnamefont
  {E.}~\bibnamefont {Beaurepaire}}, \bibinfo {author} {\bibfnamefont
  {B.}~\bibnamefont {Vileno}}, \bibinfo {author} {\bibfnamefont
  {D.}~\bibnamefont {Spor}}, \bibinfo {author} {\bibfnamefont {W.}~\bibnamefont
  {Weber}}, \bibinfo {author} {\bibfnamefont {D.}~\bibnamefont {Lacour}},
  \bibinfo {author} {\bibfnamefont {S.}~\bibnamefont {Boukari}}, \bibinfo
  {author} {\bibfnamefont {M.}~\bibnamefont {Hehn}}, \bibinfo {author}
  {\bibfnamefont {M.}~\bibnamefont {Alouani}}, \bibinfo {author} {\bibfnamefont
  {J.}~\bibnamefont {Fransson}},\ and\ \bibinfo {author} {\bibfnamefont
  {M.}~\bibnamefont {Bowen}},\ }\bibfield  {title} {\bibinfo {title}
  {Spin-driven electrical power generation at room temperature},\ }\bibfield
  {journal} {\bibinfo  {journal} {Communications Physics}\ }\textbf {\bibinfo
  {volume} {2}},\ \href {https://doi.org/10.1038/s42005-019-0207-8}
  {10.1038/s42005-019-0207-8} (\bibinfo {year} {2019})\BibitemShut {NoStop}%
\bibitem [{\citenamefont {Chowrira}\ \emph {et~al.}(2022)\citenamefont
  {Chowrira}, \citenamefont {Kandpal}, \citenamefont {Lamblin}, \citenamefont
  {Ngassam}, \citenamefont {Kouakou}, \citenamefont {Zafar}, \citenamefont
  {Mertz}, \citenamefont {Vileno}, \citenamefont {Kieber}, \citenamefont
  {Versini}, \citenamefont {Gobaut}, \citenamefont {Joly}, \citenamefont
  {Ferté}, \citenamefont {Monteblanco}, \citenamefont {Bahouka}, \citenamefont
  {Bernard}, \citenamefont {Mohapatra}, \citenamefont {Prima~Garcia},
  \citenamefont {Elidrissi}, \citenamefont {Gavara}, \citenamefont
  {Sternitzky}, \citenamefont {Da~Costa}, \citenamefont {Hehn}, \citenamefont
  {Montaigne}, \citenamefont {Choueikani}, \citenamefont {Ohresser},
  \citenamefont {Lacour}, \citenamefont {Weber}, \citenamefont {Boukari},
  \citenamefont {Alouani},\ and\ \citenamefont
  {Bowen}}]{chowrira_quantum_2022}%
  \BibitemOpen
  \bibfield  {author} {\bibinfo {author} {\bibfnamefont {B.}~\bibnamefont
  {Chowrira}}, \bibinfo {author} {\bibfnamefont {L.}~\bibnamefont {Kandpal}},
  \bibinfo {author} {\bibfnamefont {M.}~\bibnamefont {Lamblin}}, \bibinfo
  {author} {\bibfnamefont {F.}~\bibnamefont {Ngassam}}, \bibinfo {author}
  {\bibfnamefont {C.}~\bibnamefont {Kouakou}}, \bibinfo {author} {\bibfnamefont
  {T.}~\bibnamefont {Zafar}}, \bibinfo {author} {\bibfnamefont
  {D.}~\bibnamefont {Mertz}}, \bibinfo {author} {\bibfnamefont
  {B.}~\bibnamefont {Vileno}}, \bibinfo {author} {\bibfnamefont
  {C.}~\bibnamefont {Kieber}}, \bibinfo {author} {\bibfnamefont
  {G.}~\bibnamefont {Versini}}, \bibinfo {author} {\bibfnamefont
  {B.}~\bibnamefont {Gobaut}}, \bibinfo {author} {\bibfnamefont
  {L.}~\bibnamefont {Joly}}, \bibinfo {author} {\bibfnamefont {T.}~\bibnamefont
  {Ferté}}, \bibinfo {author} {\bibfnamefont {E.}~\bibnamefont {Monteblanco}},
  \bibinfo {author} {\bibfnamefont {A.}~\bibnamefont {Bahouka}}, \bibinfo
  {author} {\bibfnamefont {R.}~\bibnamefont {Bernard}}, \bibinfo {author}
  {\bibfnamefont {S.}~\bibnamefont {Mohapatra}}, \bibinfo {author}
  {\bibfnamefont {H.}~\bibnamefont {Prima~Garcia}}, \bibinfo {author}
  {\bibfnamefont {S.}~\bibnamefont {Elidrissi}}, \bibinfo {author}
  {\bibfnamefont {M.}~\bibnamefont {Gavara}}, \bibinfo {author} {\bibfnamefont
  {E.}~\bibnamefont {Sternitzky}}, \bibinfo {author} {\bibfnamefont
  {V.}~\bibnamefont {Da~Costa}}, \bibinfo {author} {\bibfnamefont
  {M.}~\bibnamefont {Hehn}}, \bibinfo {author} {\bibfnamefont {F.}~\bibnamefont
  {Montaigne}}, \bibinfo {author} {\bibfnamefont {F.}~\bibnamefont
  {Choueikani}}, \bibinfo {author} {\bibfnamefont {P.}~\bibnamefont
  {Ohresser}}, \bibinfo {author} {\bibfnamefont {D.}~\bibnamefont {Lacour}},
  \bibinfo {author} {\bibfnamefont {W.}~\bibnamefont {Weber}}, \bibinfo
  {author} {\bibfnamefont {S.}~\bibnamefont {Boukari}}, \bibinfo {author}
  {\bibfnamefont {M.}~\bibnamefont {Alouani}},\ and\ \bibinfo {author}
  {\bibfnamefont {M.}~\bibnamefont {Bowen}},\ }\bibfield  {title} {\bibinfo
  {title} {Quantum {Advantage} in a {Molecular} {Spintronic} {Engine} that
  {Harvests} {Thermal} {Fluctuation} {Energy}},\ }\href
  {https://doi.org/10.1002/adma.202206688} {\bibfield  {journal} {\bibinfo
  {journal} {Advanced Materials}\ }\textbf {\bibinfo {volume} {34}},\ \bibinfo
  {pages} {2206688} (\bibinfo {year} {2022})}\BibitemShut {NoStop}%
\bibitem [{\citenamefont {Wright}\ \emph {et~al.}(2018)\citenamefont {Wright},
  \citenamefont {Gould}, \citenamefont {Carvalho}, \citenamefont {Bedkihal},\
  and\ \citenamefont {Vaccaro}}]{wright_quantum_2018}%
  \BibitemOpen
  \bibfield  {author} {\bibinfo {author} {\bibfnamefont {J.~S. S.~T.}\
  \bibnamefont {Wright}}, \bibinfo {author} {\bibfnamefont {T.}~\bibnamefont
  {Gould}}, \bibinfo {author} {\bibfnamefont {A.~R.~R.}\ \bibnamefont
  {Carvalho}}, \bibinfo {author} {\bibfnamefont {S.}~\bibnamefont {Bedkihal}},\
  and\ \bibinfo {author} {\bibfnamefont {J.~A.}\ \bibnamefont {Vaccaro}},\
  }\bibfield  {title} {\bibinfo {title} {Quantum heat engine operating between
  thermal and spin reservoirs},\ }\bibfield  {journal} {\bibinfo  {journal}
  {Physical Review A}\ }\textbf {\bibinfo {volume} {97}},\ \href
  {https://doi.org/10.1103/PhysRevA.97.052104} {10.1103/PhysRevA.97.052104}
  (\bibinfo {year} {2018})\BibitemShut {NoStop}%
\bibitem [{\citenamefont {Li}\ and\ \citenamefont {Kais}(2019)}]{Li2019}%
  \BibitemOpen
  \bibfield  {author} {\bibinfo {author} {\bibfnamefont {J.}~\bibnamefont
  {Li}}\ and\ \bibinfo {author} {\bibfnamefont {S.}~\bibnamefont {Kais}},\
  }\bibfield  {title} {\bibinfo {title} {Entanglement classifier in chemical
  reactions},\ }\bibfield  {journal} {\bibinfo  {journal} {Science Advances}\
  }\textbf {\bibinfo {volume} {5}},\ \href
  {https://doi.org/10.1126/sciadv.aax5283} {10.1126/sciadv.aax5283} (\bibinfo
  {year} {2019})\BibitemShut {NoStop}%
\bibitem [{\citenamefont {Kim}\ \emph {et~al.}(2021)\citenamefont {Kim},
  \citenamefont {Bertagna}, \citenamefont {D'Souza}, \citenamefont {Heyes},
  \citenamefont {Johannissen}, \citenamefont {Nery}, \citenamefont {Pantelias},
  \citenamefont {Jimenez}, \citenamefont {Slocombe}, \citenamefont {Spencer},
  \citenamefont {Al-Khalili}, \citenamefont {Engel}, \citenamefont {Hay},
  \citenamefont {Hingley-Wilson}, \citenamefont {Jeevaratnam}, \citenamefont
  {Jones}, \citenamefont {Kattnig}, \citenamefont {Lewis}, \citenamefont
  {Sacchi}, \citenamefont {Scrutton}, \citenamefont {Silva},\ and\
  \citenamefont {McFadden}}]{Kim2021}%
  \BibitemOpen
  \bibfield  {author} {\bibinfo {author} {\bibfnamefont {Y.}~\bibnamefont
  {Kim}}, \bibinfo {author} {\bibfnamefont {F.}~\bibnamefont {Bertagna}},
  \bibinfo {author} {\bibfnamefont {E.~M.}\ \bibnamefont {D'Souza}}, \bibinfo
  {author} {\bibfnamefont {D.~J.}\ \bibnamefont {Heyes}}, \bibinfo {author}
  {\bibfnamefont {L.~O.}\ \bibnamefont {Johannissen}}, \bibinfo {author}
  {\bibfnamefont {E.~T.}\ \bibnamefont {Nery}}, \bibinfo {author}
  {\bibfnamefont {A.}~\bibnamefont {Pantelias}}, \bibinfo {author}
  {\bibfnamefont {A.~S.-P.}\ \bibnamefont {Jimenez}}, \bibinfo {author}
  {\bibfnamefont {L.}~\bibnamefont {Slocombe}}, \bibinfo {author}
  {\bibfnamefont {M.~G.}\ \bibnamefont {Spencer}}, \bibinfo {author}
  {\bibfnamefont {J.}~\bibnamefont {Al-Khalili}}, \bibinfo {author}
  {\bibfnamefont {G.~S.}\ \bibnamefont {Engel}}, \bibinfo {author}
  {\bibfnamefont {S.}~\bibnamefont {Hay}}, \bibinfo {author} {\bibfnamefont
  {S.~M.}\ \bibnamefont {Hingley-Wilson}}, \bibinfo {author} {\bibfnamefont
  {K.}~\bibnamefont {Jeevaratnam}}, \bibinfo {author} {\bibfnamefont {A.~R.}\
  \bibnamefont {Jones}}, \bibinfo {author} {\bibfnamefont {D.~R.}\ \bibnamefont
  {Kattnig}}, \bibinfo {author} {\bibfnamefont {R.}~\bibnamefont {Lewis}},
  \bibinfo {author} {\bibfnamefont {M.}~\bibnamefont {Sacchi}}, \bibinfo
  {author} {\bibfnamefont {N.~S.}\ \bibnamefont {Scrutton}}, \bibinfo {author}
  {\bibfnamefont {S.~R.~P.}\ \bibnamefont {Silva}},\ and\ \bibinfo {author}
  {\bibfnamefont {J.}~\bibnamefont {McFadden}},\ }\bibfield  {title} {\bibinfo
  {title} {Quantum biology: An update and perspective},\ }\href
  {https://doi.org/10.3390/quantum3010006} {\bibfield  {journal} {\bibinfo
  {journal} {Quantum Reports}\ }\textbf {\bibinfo {volume} {3}},\ \bibinfo
  {pages} {80} (\bibinfo {year} {2021})}\BibitemShut {NoStop}%
\bibitem [{\citenamefont {Adams}\ and\ \citenamefont
  {Petruccione}(2020)}]{Adams2020}%
  \BibitemOpen
  \bibfield  {author} {\bibinfo {author} {\bibfnamefont {B.}~\bibnamefont
  {Adams}}\ and\ \bibinfo {author} {\bibfnamefont {F.}~\bibnamefont
  {Petruccione}},\ }\bibfield  {title} {\bibinfo {title} {Quantum effects in
  the brain: A review},\ }\href {https://doi.org/10.1116/1.5135170} {\bibfield
  {journal} {\bibinfo  {journal} {{AVS} Quantum Science}\ }\textbf {\bibinfo
  {volume} {2}},\ \bibinfo {pages} {022901} (\bibinfo {year}
  {2020})}\BibitemShut {NoStop}%
\bibitem [{\citenamefont {Vinjanampathy}\ and\ \citenamefont
  {Anders}(2016)}]{Vinjanampathy_quantum_2016}%
  \BibitemOpen
  \bibfield  {author} {\bibinfo {author} {\bibfnamefont {S.}~\bibnamefont
  {Vinjanampathy}}\ and\ \bibinfo {author} {\bibfnamefont {J.}~\bibnamefont
  {Anders}},\ }\bibfield  {title} {\bibinfo {title} {Quantum thermodynamics},\
  }\href {https://doi.org/10.1080/00107514.2016.1201896} {\bibfield  {journal}
  {\bibinfo  {journal} {CONTEMPORARY PHYSICS}\ }\textbf {\bibinfo {volume}
  {57}},\ \bibinfo {pages} {545} (\bibinfo {year} {2016})}\BibitemShut
  {NoStop}%
\bibitem [{\citenamefont {Rogers}\ and\ \citenamefont
  {Jordan}(2022)}]{rogers_post-selection_2022}%
  \BibitemOpen
  \bibfield  {author} {\bibinfo {author} {\bibfnamefont {S.}~\bibnamefont
  {Rogers}}\ and\ \bibinfo {author} {\bibfnamefont {A.~N.}\ \bibnamefont
  {Jordan}},\ }\bibfield  {title} {\bibinfo {title} {Postselection and quantum
  energetics},\ }\bibfield  {journal} {\bibinfo  {journal} {Physical Review A}\
  }\textbf {\bibinfo {volume} {106}},\ \href
  {https://doi.org/10.1103/PhysRevA.106.052214} {10.1103/PhysRevA.106.052214}
  (\bibinfo {year} {2022})\BibitemShut {NoStop}%
\bibitem [{\citenamefont {Bhattacharjee}\ and\ \citenamefont
  {Dutta}(2021)}]{bhattacharjee_quantum_2021}%
  \BibitemOpen
  \bibfield  {author} {\bibinfo {author} {\bibfnamefont {S.}~\bibnamefont
  {Bhattacharjee}}\ and\ \bibinfo {author} {\bibfnamefont {A.}~\bibnamefont
  {Dutta}},\ }\bibfield  {title} {\bibinfo {title} {Quantum thermal machines
  and batteries},\ }\bibfield  {journal} {\bibinfo  {journal} {The European
  Physical Journal B}\ }\textbf {\bibinfo {volume} {94}},\ \href
  {https://doi.org/10.1140/epjb/s10051-021-00235-3}
  {10.1140/epjb/s10051-021-00235-3} (\bibinfo {year} {2021})\BibitemShut
  {NoStop}%
\bibitem [{\citenamefont {Ptaszyński}(2018)}]{ptaszynski_autonomous_2018}%
  \BibitemOpen
  \bibfield  {author} {\bibinfo {author} {\bibfnamefont {K.}~\bibnamefont
  {Ptaszyński}},\ }\bibfield  {title} {\bibinfo {title} {Autonomous quantum
  {Maxwell}'s demon based on two exchange-coupled quantum dots},\ }\bibfield
  {journal} {\bibinfo  {journal} {Physical Review E}\ }\textbf {\bibinfo
  {volume} {97}},\ \href {https://doi.org/10.1103/PhysRevE.97.012116}
  {10.1103/PhysRevE.97.012116} (\bibinfo {year} {2018})\BibitemShut {NoStop}%
\bibitem [{\citenamefont {Seah}\ \emph {et~al.}(2020)\citenamefont {Seah},
  \citenamefont {Nimmrichter},\ and\ \citenamefont
  {Scarani}}]{seah_maxwells_2020}%
  \BibitemOpen
  \bibfield  {author} {\bibinfo {author} {\bibfnamefont {S.}~\bibnamefont
  {Seah}}, \bibinfo {author} {\bibfnamefont {S.}~\bibnamefont {Nimmrichter}},\
  and\ \bibinfo {author} {\bibfnamefont {V.}~\bibnamefont {Scarani}},\
  }\bibfield  {title} {\bibinfo {title} {Maxwell’s {Lesser} {Demon}: {A}
  {Quantum} {Engine} {Driven} by {Pointer} {Measurements}},\ }\bibfield
  {journal} {\bibinfo  {journal} {Physical Review Letters}\ }\textbf {\bibinfo
  {volume} {124}},\ \href {https://doi.org/10.1103/PhysRevLett.124.100603}
  {10.1103/PhysRevLett.124.100603} (\bibinfo {year} {2020})\BibitemShut
  {NoStop}%
\bibitem [{\citenamefont {Piccitto}\ \emph {et~al.}(2022)\citenamefont
  {Piccitto}, \citenamefont {Campisi},\ and\ \citenamefont
  {Rossini}}]{piccitto_ising_2022}%
  \BibitemOpen
  \bibfield  {author} {\bibinfo {author} {\bibfnamefont {G.}~\bibnamefont
  {Piccitto}}, \bibinfo {author} {\bibfnamefont {M.}~\bibnamefont {Campisi}},\
  and\ \bibinfo {author} {\bibfnamefont {D.}~\bibnamefont {Rossini}},\
  }\bibfield  {title} {\bibinfo {title} {The ising critical quantum otto
  engine},\ }\bibfield  {journal} {\bibinfo  {journal} {New Journal of
  Physics}\ }\textbf {\bibinfo {volume} {24}},\ \href
  {https://doi.org/10.1088/1367-2630/ac963b} {10.1088/1367-2630/ac963b}
  (\bibinfo {year} {2022})\BibitemShut {NoStop}%
\bibitem [{\citenamefont {Manzano}\ \emph {et~al.}(2016)\citenamefont
  {Manzano}, \citenamefont {Galve}, \citenamefont {Zambrini},\ and\
  \citenamefont {Parrondo}}]{manzano_entropy_2016}%
  \BibitemOpen
  \bibfield  {author} {\bibinfo {author} {\bibfnamefont {G.}~\bibnamefont
  {Manzano}}, \bibinfo {author} {\bibfnamefont {F.}~\bibnamefont {Galve}},
  \bibinfo {author} {\bibfnamefont {R.}~\bibnamefont {Zambrini}},\ and\
  \bibinfo {author} {\bibfnamefont {J.~M.~R.}\ \bibnamefont {Parrondo}},\
  }\bibfield  {title} {\bibinfo {title} {Entropy production and thermodynamic
  power of the squeezed thermal reservoir},\ }\bibfield  {journal} {\bibinfo
  {journal} {Physical Review E}\ }\textbf {\bibinfo {volume} {93}},\ \href
  {https://doi.org/10.1103/PhysRevE.93.052120} {10.1103/PhysRevE.93.052120}
  (\bibinfo {year} {2016})\BibitemShut {NoStop}%
\bibitem [{\citenamefont {Molitor}\ and\ \citenamefont
  {Landi}(2020)}]{molitor_stroboscopic_2020}%
  \BibitemOpen
  \bibfield  {author} {\bibinfo {author} {\bibfnamefont {O.~A.~D.}\
  \bibnamefont {Molitor}}\ and\ \bibinfo {author} {\bibfnamefont {G.~T.}\
  \bibnamefont {Landi}},\ }\bibfield  {title} {\bibinfo {title} {Stroboscopic
  two-stroke quantum heat engines},\ }\bibfield  {journal} {\bibinfo  {journal}
  {Physical Review A}\ }\textbf {\bibinfo {volume} {102}},\ \href
  {https://doi.org/10.1103/PhysRevA.102.042217} {10.1103/PhysRevA.102.042217}
  (\bibinfo {year} {2020})\BibitemShut {NoStop}%
\bibitem [{\citenamefont {Donvil}(2018)}]{donvil_thermodynamics_2018}%
  \BibitemOpen
  \bibfield  {author} {\bibinfo {author} {\bibfnamefont {B.}~\bibnamefont
  {Donvil}},\ }\bibfield  {title} {\bibinfo {title} {Thermodynamics of a
  periodically driven qubit},\ }\href
  {https://doi.org/10.1088/1742-5468/aab857} {\bibfield  {journal} {\bibinfo
  {journal} {Journal of Statistical Mechanics: Theory and Experiment}\ }\textbf
  {\bibinfo {volume} {2018}},\ \bibinfo {pages} {043104} (\bibinfo {year}
  {2018})}\BibitemShut {NoStop}%
\bibitem [{\citenamefont {Zhao}\ \emph {et~al.}(2021)\citenamefont {Zhao},
  \citenamefont {Dou},\ and\ \citenamefont {Zhao}}]{zhao_quantum_2021}%
  \BibitemOpen
  \bibfield  {author} {\bibinfo {author} {\bibfnamefont {F.}~\bibnamefont
  {Zhao}}, \bibinfo {author} {\bibfnamefont {F.-Q.}\ \bibnamefont {Dou}},\ and\
  \bibinfo {author} {\bibfnamefont {Q.}~\bibnamefont {Zhao}},\ }\bibfield
  {title} {\bibinfo {title} {Quantum battery of interacting spins with
  environmental noise},\ }\bibfield  {journal} {\bibinfo  {journal} {Physical
  Review A}\ }\textbf {\bibinfo {volume} {103}},\ \href
  {https://doi.org/10.1103/PhysRevA.103.033715} {10.1103/PhysRevA.103.033715}
  (\bibinfo {year} {2021})\BibitemShut {NoStop}%
\bibitem [{\citenamefont {Santos}\ and\ \citenamefont
  {Santos}(2022)}]{santos_efficiency_2022}%
  \BibitemOpen
  \bibfield  {author} {\bibinfo {author} {\bibfnamefont {T.~F.~F.}\
  \bibnamefont {Santos}}\ and\ \bibinfo {author} {\bibfnamefont {M.~F.}\
  \bibnamefont {Santos}},\ }\bibfield  {title} {\bibinfo {title} {Efficiency of
  optically pumping a quantum battery and a two-stroke heat engine},\ }\href
  {https://doi.org/10.1103/PhysRevA.106.052203} {\bibfield  {journal} {\bibinfo
   {journal} {Phys. Rev. A}\ }\textbf {\bibinfo {volume} {106}},\ \bibinfo
  {pages} {052203} (\bibinfo {year} {2022})}\BibitemShut {NoStop}%
\bibitem [{\citenamefont {Francica}\ \emph {et~al.}(2020)\citenamefont
  {Francica}, \citenamefont {Binder}, \citenamefont {Guarnieri}, \citenamefont
  {Mitchison}, \citenamefont {Goold},\ and\ \citenamefont
  {Plastina}}]{francica_quantum_2020}%
  \BibitemOpen
  \bibfield  {author} {\bibinfo {author} {\bibfnamefont {G.}~\bibnamefont
  {Francica}}, \bibinfo {author} {\bibfnamefont {F.}~\bibnamefont {Binder}},
  \bibinfo {author} {\bibfnamefont {G.}~\bibnamefont {Guarnieri}}, \bibinfo
  {author} {\bibfnamefont {M.}~\bibnamefont {Mitchison}}, \bibinfo {author}
  {\bibfnamefont {J.}~\bibnamefont {Goold}},\ and\ \bibinfo {author}
  {\bibfnamefont {F.}~\bibnamefont {Plastina}},\ }\bibfield  {title} {\bibinfo
  {title} {Quantum {Coherence} and {Ergotropy}},\ }\bibfield  {journal}
  {\bibinfo  {journal} {Physical Review Letters}\ }\textbf {\bibinfo {volume}
  {125}},\ \href {https://doi.org/10.1103/PhysRevLett.125.180603}
  {10.1103/PhysRevLett.125.180603} (\bibinfo {year} {2020})\BibitemShut
  {NoStop}%
\bibitem [{\citenamefont {Shi}\ \emph {et~al.}(2020)\citenamefont {Shi},
  \citenamefont {Shi}, \citenamefont {Wang}, \citenamefont {Hu}, \citenamefont
  {Liu}, \citenamefont {Yang},\ and\ \citenamefont {Fan}}]{shi_quantum_2020}%
  \BibitemOpen
  \bibfield  {author} {\bibinfo {author} {\bibfnamefont {Y.-H.}\ \bibnamefont
  {Shi}}, \bibinfo {author} {\bibfnamefont {H.-L.}\ \bibnamefont {Shi}},
  \bibinfo {author} {\bibfnamefont {X.-H.}\ \bibnamefont {Wang}}, \bibinfo
  {author} {\bibfnamefont {M.-L.}\ \bibnamefont {Hu}}, \bibinfo {author}
  {\bibfnamefont {S.-Y.}\ \bibnamefont {Liu}}, \bibinfo {author} {\bibfnamefont
  {W.-L.}\ \bibnamefont {Yang}},\ and\ \bibinfo {author} {\bibfnamefont
  {H.}~\bibnamefont {Fan}},\ }\bibfield  {title} {\bibinfo {title} {Quantum
  coherence in a quantum heat engine},\ }\bibfield  {journal} {\bibinfo
  {journal} {Journal of Physics A-Mathematical and Theoretical}\ }\textbf
  {\bibinfo {volume} {53}},\ \href {https://doi.org/10.1088/1751-8121/ab6a6b}
  {10.1088/1751-8121/ab6a6b} (\bibinfo {year} {2020})\BibitemShut {NoStop}%
\bibitem [{\citenamefont {Aimet}\ and\ \citenamefont
  {Kwon}(2023)}]{aimet_engineering_2023}%
  \BibitemOpen
  \bibfield  {author} {\bibinfo {author} {\bibfnamefont {S.}~\bibnamefont
  {Aimet}}\ and\ \bibinfo {author} {\bibfnamefont {H.}~\bibnamefont {Kwon}},\
  }\bibfield  {title} {\bibinfo {title} {Engineering a heat engine purely
  driven by quantum coherence},\ }\href
  {https://doi.org/10.1103/PhysRevA.107.012221} {\bibfield  {journal} {\bibinfo
   {journal} {Phys. Rev. A}\ }\textbf {\bibinfo {volume} {107}},\ \bibinfo
  {pages} {012221} (\bibinfo {year} {2023})}\BibitemShut {NoStop}%
\bibitem [{\citenamefont {Monsel}\ \emph {et~al.}(2020)\citenamefont {Monsel},
  \citenamefont {Fellous-Asiani}, \citenamefont {Huard},\ and\ \citenamefont
  {Auffèves}}]{monsel_energetic_2020}%
  \BibitemOpen
  \bibfield  {author} {\bibinfo {author} {\bibfnamefont {J.}~\bibnamefont
  {Monsel}}, \bibinfo {author} {\bibfnamefont {M.}~\bibnamefont
  {Fellous-Asiani}}, \bibinfo {author} {\bibfnamefont {B.}~\bibnamefont
  {Huard}},\ and\ \bibinfo {author} {\bibfnamefont {A.}~\bibnamefont
  {Auffèves}},\ }\bibfield  {title} {\bibinfo {title} {The {Energetic} {Cost}
  of {Work} {Extraction}},\ }\bibfield  {journal} {\bibinfo  {journal}
  {Physical Review Letters}\ }\textbf {\bibinfo {volume} {124}},\ \href
  {https://doi.org/10.1103/PhysRevLett.124.130601}
  {10.1103/PhysRevLett.124.130601} (\bibinfo {year} {2020})\BibitemShut
  {NoStop}%
\bibitem [{\citenamefont {Buffoni}\ \emph {et~al.}(2019)\citenamefont
  {Buffoni}, \citenamefont {Solfanelli}, \citenamefont {Verrucchi},
  \citenamefont {Cuccoli},\ and\ \citenamefont
  {Campisi}}]{buffoni_quantum_2019}%
  \BibitemOpen
  \bibfield  {author} {\bibinfo {author} {\bibfnamefont {L.}~\bibnamefont
  {Buffoni}}, \bibinfo {author} {\bibfnamefont {A.}~\bibnamefont {Solfanelli}},
  \bibinfo {author} {\bibfnamefont {P.}~\bibnamefont {Verrucchi}}, \bibinfo
  {author} {\bibfnamefont {A.}~\bibnamefont {Cuccoli}},\ and\ \bibinfo {author}
  {\bibfnamefont {M.}~\bibnamefont {Campisi}},\ }\bibfield  {title} {\bibinfo
  {title} {Quantum {Measurement} {Cooling}},\ }\bibfield  {journal} {\bibinfo
  {journal} {Physical Review Letters}\ }\textbf {\bibinfo {volume} {122}},\
  \href {https://doi.org/10.1103/PhysRevLett.122.070603}
  {10.1103/PhysRevLett.122.070603} (\bibinfo {year} {2019})\BibitemShut
  {NoStop}%
\bibitem [{\citenamefont {Ji}\ \emph {et~al.}(2022)\citenamefont {Ji},
  \citenamefont {Chai}, \citenamefont {Wang}, \citenamefont {Guo},
  \citenamefont {Rong}, \citenamefont {Shi}, \citenamefont {Ren}, \citenamefont
  {Wang},\ and\ \citenamefont {Du}}]{ji_spin_2022}%
  \BibitemOpen
  \bibfield  {author} {\bibinfo {author} {\bibfnamefont {W.}~\bibnamefont
  {Ji}}, \bibinfo {author} {\bibfnamefont {Z.}~\bibnamefont {Chai}}, \bibinfo
  {author} {\bibfnamefont {M.}~\bibnamefont {Wang}}, \bibinfo {author}
  {\bibfnamefont {Y.}~\bibnamefont {Guo}}, \bibinfo {author} {\bibfnamefont
  {X.}~\bibnamefont {Rong}}, \bibinfo {author} {\bibfnamefont {F.}~\bibnamefont
  {Shi}}, \bibinfo {author} {\bibfnamefont {C.}~\bibnamefont {Ren}}, \bibinfo
  {author} {\bibfnamefont {Y.}~\bibnamefont {Wang}},\ and\ \bibinfo {author}
  {\bibfnamefont {J.}~\bibnamefont {Du}},\ }\bibfield  {title} {\bibinfo
  {title} {Spin {Quantum} {Heat} {Engine} {Quantified} by {Quantum}
  {Steering}},\ }\bibfield  {journal} {\bibinfo  {journal} {Physical Review
  Letters}\ }\textbf {\bibinfo {volume} {128}},\ \href
  {https://doi.org/10.1103/PhysRevLett.128.090602}
  {10.1103/PhysRevLett.128.090602} (\bibinfo {year} {2022})\BibitemShut
  {NoStop}%
\bibitem [{\citenamefont {Micadei}\ \emph {et~al.}(2019)\citenamefont
  {Micadei}, \citenamefont {Peterson}, \citenamefont {Souza}, \citenamefont
  {Sarthour}, \citenamefont {Oliveira}, \citenamefont {Landi}, \citenamefont
  {Batalhão}, \citenamefont {Serra},\ and\ \citenamefont
  {Lutz}}]{micadei_reversing_2019}%
  \BibitemOpen
  \bibfield  {author} {\bibinfo {author} {\bibfnamefont {K.}~\bibnamefont
  {Micadei}}, \bibinfo {author} {\bibfnamefont {J.~P.~S.}\ \bibnamefont
  {Peterson}}, \bibinfo {author} {\bibfnamefont {A.~M.}\ \bibnamefont {Souza}},
  \bibinfo {author} {\bibfnamefont {R.~S.}\ \bibnamefont {Sarthour}}, \bibinfo
  {author} {\bibfnamefont {I.~S.}\ \bibnamefont {Oliveira}}, \bibinfo {author}
  {\bibfnamefont {G.~T.}\ \bibnamefont {Landi}}, \bibinfo {author}
  {\bibfnamefont {T.~B.}\ \bibnamefont {Batalhão}}, \bibinfo {author}
  {\bibfnamefont {R.~M.}\ \bibnamefont {Serra}},\ and\ \bibinfo {author}
  {\bibfnamefont {E.}~\bibnamefont {Lutz}},\ }\bibfield  {title} {\bibinfo
  {title} {Reversing the direction of heat flow using quantum correlations},\
  }\bibfield  {journal} {\bibinfo  {journal} {Nature Communications}\ }\textbf
  {\bibinfo {volume} {10}},\ \href {https://doi.org/10.1038/s41467-019-10333-7}
  {10.1038/s41467-019-10333-7} (\bibinfo {year} {2019})\BibitemShut {NoStop}%
\bibitem [{\citenamefont {Niedenzu}\ \emph {et~al.}(2018)\citenamefont
  {Niedenzu}, \citenamefont {Mukherjee}, \citenamefont {Ghosh}, \citenamefont
  {Kofman},\ and\ \citenamefont {Kurizki}}]{niedenzu_quantum_2018}%
  \BibitemOpen
  \bibfield  {author} {\bibinfo {author} {\bibfnamefont {W.}~\bibnamefont
  {Niedenzu}}, \bibinfo {author} {\bibfnamefont {V.}~\bibnamefont {Mukherjee}},
  \bibinfo {author} {\bibfnamefont {A.}~\bibnamefont {Ghosh}}, \bibinfo
  {author} {\bibfnamefont {A.~G.}\ \bibnamefont {Kofman}},\ and\ \bibinfo
  {author} {\bibfnamefont {G.}~\bibnamefont {Kurizki}},\ }\bibfield  {title}
  {\bibinfo {title} {Quantum engine efficiency bound beyond the second law of
  thermodynamics},\ }\bibfield  {journal} {\bibinfo  {journal} {Nature
  Communications}\ }\textbf {\bibinfo {volume} {9}},\ \href
  {https://doi.org/10.1038/s41467-017-01991-6} {10.1038/s41467-017-01991-6}
  (\bibinfo {year} {2018})\BibitemShut {NoStop}%
\bibitem [{\citenamefont {Roßnagel}\ \emph {et~al.}(2014)\citenamefont
  {Roßnagel}, \citenamefont {Abah}, \citenamefont {Schmidt-Kaler},
  \citenamefont {Singer},\ and\ \citenamefont {Lutz}}]{nanoscale_2014}%
  \BibitemOpen
  \bibfield  {author} {\bibinfo {author} {\bibfnamefont {J.}~\bibnamefont
  {Roßnagel}}, \bibinfo {author} {\bibfnamefont {O.}~\bibnamefont {Abah}},
  \bibinfo {author} {\bibfnamefont {F.}~\bibnamefont {Schmidt-Kaler}}, \bibinfo
  {author} {\bibfnamefont {K.}~\bibnamefont {Singer}},\ and\ \bibinfo {author}
  {\bibfnamefont {E.}~\bibnamefont {Lutz}},\ }\bibfield  {title} {\bibinfo
  {title} {Nanoscale heat engine beyond the carnot limit},\ }\href
  {https://doi.org/10.1103/PhysRevLett.112.030602} {\bibfield  {journal}
  {\bibinfo  {journal} {Physical review letters}\ }\textbf {\bibinfo {volume}
  {112}},\ \bibinfo {pages} {030602} (\bibinfo {year} {2014})}\BibitemShut
  {NoStop}%
\bibitem [{\citenamefont {Klaers}\ \emph {et~al.}(2017)\citenamefont {Klaers},
  \citenamefont {Faelt}, \citenamefont {Imamoglu},\ and\ \citenamefont
  {Togan}}]{klaers_squeezed_2017}%
  \BibitemOpen
  \bibfield  {author} {\bibinfo {author} {\bibfnamefont {J.}~\bibnamefont
  {Klaers}}, \bibinfo {author} {\bibfnamefont {S.}~\bibnamefont {Faelt}},
  \bibinfo {author} {\bibfnamefont {A.}~\bibnamefont {Imamoglu}},\ and\
  \bibinfo {author} {\bibfnamefont {E.}~\bibnamefont {Togan}},\ }\bibfield
  {title} {\bibinfo {title} {Squeezed {Thermal} {Reservoirs} as a {Resource}
  for a {Nanomechanical} {Engine} beyond the {Carnot} {Limit}},\ }\bibfield
  {journal} {\bibinfo  {journal} {Physical Review X}\ }\textbf {\bibinfo
  {volume} {7}},\ \href {https://doi.org/10.1103/PhysRevX.7.031044}
  {10.1103/PhysRevX.7.031044} (\bibinfo {year} {2017})\BibitemShut {NoStop}%
\bibitem [{\citenamefont {Scully}(2003)}]{scully_extracting_2003}%
  \BibitemOpen
  \bibfield  {author} {\bibinfo {author} {\bibfnamefont {M.~O.}\ \bibnamefont
  {Scully}},\ }\bibfield  {title} {\bibinfo {title} {Extracting {Work} from a
  {Single} {Heat} {Bath} via {Vanishing} {Quantum} {Coherence}},\ }\href
  {https://doi.org/10.1126/science.1078955} {\bibfield  {journal} {\bibinfo
  {journal} {Science}\ }\textbf {\bibinfo {volume} {299}},\ \bibinfo {pages}
  {862} (\bibinfo {year} {2003})}\BibitemShut {NoStop}%
\bibitem [{\citenamefont {Yi}\ \emph {et~al.}(2017)\citenamefont {Yi},
  \citenamefont {Talkner},\ and\ \citenamefont
  {Kim}}]{yi_single-temperature_2017}%
  \BibitemOpen
  \bibfield  {author} {\bibinfo {author} {\bibfnamefont {J.}~\bibnamefont
  {Yi}}, \bibinfo {author} {\bibfnamefont {P.}~\bibnamefont {Talkner}},\ and\
  \bibinfo {author} {\bibfnamefont {Y.~W.}\ \bibnamefont {Kim}},\ }\bibfield
  {title} {\bibinfo {title} {Single-temperature quantum engine without feedback
  control},\ }\bibfield  {journal} {\bibinfo  {journal} {Physical Review E}\
  }\textbf {\bibinfo {volume} {96}},\ \href
  {https://doi.org/10.1103/PhysRevE.96.022108} {10.1103/PhysRevE.96.022108}
  (\bibinfo {year} {2017})\BibitemShut {NoStop}%
\bibitem [{\citenamefont {Elouard}\ and\ \citenamefont
  {Jordan}(2018)}]{elouard_efficient_2018}%
  \BibitemOpen
  \bibfield  {author} {\bibinfo {author} {\bibfnamefont {C.}~\bibnamefont
  {Elouard}}\ and\ \bibinfo {author} {\bibfnamefont {A.~N.}\ \bibnamefont
  {Jordan}},\ }\bibfield  {title} {\bibinfo {title} {Efficient {Quantum}
  {Measurement} {Engines}},\ }\bibfield  {journal} {\bibinfo  {journal}
  {Physical Review Letters}\ }\textbf {\bibinfo {volume} {120}},\ \href
  {https://doi.org/10.1103/PhysRevLett.120.260601}
  {10.1103/PhysRevLett.120.260601} (\bibinfo {year} {2018})\BibitemShut
  {NoStop}%
\bibitem [{\citenamefont {Francica}\ \emph {et~al.}(2017)\citenamefont
  {Francica}, \citenamefont {Goold}, \citenamefont {Plastina},\ and\
  \citenamefont {Paternostro}}]{francica_daemonic_2017}%
  \BibitemOpen
  \bibfield  {author} {\bibinfo {author} {\bibfnamefont {G.}~\bibnamefont
  {Francica}}, \bibinfo {author} {\bibfnamefont {J.}~\bibnamefont {Goold}},
  \bibinfo {author} {\bibfnamefont {F.}~\bibnamefont {Plastina}},\ and\
  \bibinfo {author} {\bibfnamefont {M.}~\bibnamefont {Paternostro}},\
  }\bibfield  {title} {\bibinfo {title} {Daemonic ergotropy: enhanced work
  extraction from quantum correlations},\ }\bibfield  {journal} {\bibinfo
  {journal} {npj Quantum Information}\ }\textbf {\bibinfo {volume} {3}},\ \href
  {https://doi.org/10.1038/s41534-017-0012-8} {10.1038/s41534-017-0012-8}
  (\bibinfo {year} {2017})\BibitemShut {NoStop}%
\bibitem [{\citenamefont {Elouard}\ \emph {et~al.}(2017)\citenamefont
  {Elouard}, \citenamefont {Herrera-Martí}, \citenamefont {Huard},\ and\
  \citenamefont {Auffèves}}]{elouard_extracting_2017}%
  \BibitemOpen
  \bibfield  {author} {\bibinfo {author} {\bibfnamefont {C.}~\bibnamefont
  {Elouard}}, \bibinfo {author} {\bibfnamefont {D.}~\bibnamefont
  {Herrera-Martí}}, \bibinfo {author} {\bibfnamefont {B.}~\bibnamefont
  {Huard}},\ and\ \bibinfo {author} {\bibfnamefont {A.}~\bibnamefont
  {Auffèves}},\ }\bibfield  {title} {\bibinfo {title} {Extracting {Work} from
  {Quantum} {Measurement} in {Maxwell}’s {Demon} {Engines}},\ }\bibfield
  {journal} {\bibinfo  {journal} {Physical Review Letters}\ }\textbf {\bibinfo
  {volume} {118}},\ \href {https://doi.org/10.1103/PhysRevLett.118.260603}
  {10.1103/PhysRevLett.118.260603} (\bibinfo {year} {2017})\BibitemShut
  {NoStop}%
\bibitem [{\citenamefont {Henriet}\ \emph {et~al.}(2015)\citenamefont
  {Henriet}, \citenamefont {Jordan},\ and\ \citenamefont
  {Le~Hur}}]{henriet_electrical_2015}%
  \BibitemOpen
  \bibfield  {author} {\bibinfo {author} {\bibfnamefont {L.}~\bibnamefont
  {Henriet}}, \bibinfo {author} {\bibfnamefont {A.~N.}\ \bibnamefont
  {Jordan}},\ and\ \bibinfo {author} {\bibfnamefont {K.}~\bibnamefont
  {Le~Hur}},\ }\bibfield  {title} {\bibinfo {title} {Electrical current from
  quantum vacuum fluctuations in nanoengines},\ }\href
  {https://doi.org/10.1103/PhysRevB.92.125306} {\bibfield  {journal} {\bibinfo
  {journal} {Physical Review B}\ }\textbf {\bibinfo {volume} {92}},\ \bibinfo
  {pages} {125306} (\bibinfo {year} {2015})}\BibitemShut {NoStop}%
\bibitem [{\citenamefont {Xiao}\ \emph {et~al.}()\citenamefont {Xiao},
  \citenamefont {Liu}, \citenamefont {He},\ and\ \citenamefont
  {Liu}}]{xiao_thermodynamics_nodate}%
  \BibitemOpen
  \bibfield  {author} {\bibinfo {author} {\bibfnamefont {Y.}~\bibnamefont
  {Xiao}}, \bibinfo {author} {\bibfnamefont {D.}~\bibnamefont {Liu}}, \bibinfo
  {author} {\bibfnamefont {J.}~\bibnamefont {He}},\ and\ \bibinfo {author}
  {\bibfnamefont {W.-M.}\ \bibnamefont {Liu}},\ }\bibfield  {title} {\bibinfo
  {title} {Thermodynamics and {Fluctuations} in {Quantum} {Heat} {Engines}
  under {Reservoir} {Squeezing}},\ }\href@noop {} {\ ,\ \bibinfo {pages}
  {13}}\BibitemShut {NoStop}%
\bibitem [{\citenamefont {Fransson}\ and\ \citenamefont
  {Råsander}(2006)}]{fransson_pauli_2006}%
  \BibitemOpen
  \bibfield  {author} {\bibinfo {author} {\bibfnamefont {J.}~\bibnamefont
  {Fransson}}\ and\ \bibinfo {author} {\bibfnamefont {M.}~\bibnamefont
  {Råsander}},\ }\bibfield  {title} {\bibinfo {title} {Pauli spin blockade in
  weakly coupled double quantum dots},\ }\bibfield  {journal} {\bibinfo
  {journal} {Physical Review B}\ }\textbf {\bibinfo {volume} {73}},\ \href
  {https://doi.org/10.1103/PhysRevB.73.205333} {10.1103/PhysRevB.73.205333}
  (\bibinfo {year} {2006})\BibitemShut {NoStop}%
\bibitem [{\citenamefont {Weymann}\ and\ \citenamefont
  {Barnaś}(2007)}]{weymann_transport_2007}%
  \BibitemOpen
  \bibfield  {author} {\bibinfo {author} {\bibfnamefont {I.}~\bibnamefont
  {Weymann}}\ and\ \bibinfo {author} {\bibfnamefont {J.}~\bibnamefont
  {Barnaś}},\ }\bibfield  {title} {\bibinfo {title} {Transport through
  two-level quantum dots weakly coupled to ferromagnetic leads},\ }\href
  {https://doi.org/10.1088/0953-8984/19/9/096208} {\bibfield  {journal}
  {\bibinfo  {journal} {Journal of Physics: Condensed Matter}\ }\textbf
  {\bibinfo {volume} {19}},\ \bibinfo {pages} {096208} (\bibinfo {year}
  {2007})}\BibitemShut {NoStop}%
\bibitem [{\citenamefont {Fransson}\ \emph {et~al.}(2014)\citenamefont
  {Fransson}, \citenamefont {Ren},\ and\ \citenamefont
  {Zhu}}]{fransson_electrical_2014}%
  \BibitemOpen
  \bibfield  {author} {\bibinfo {author} {\bibfnamefont {J.}~\bibnamefont
  {Fransson}}, \bibinfo {author} {\bibfnamefont {J.}~\bibnamefont {Ren}},\ and\
  \bibinfo {author} {\bibfnamefont {J.-X.}\ \bibnamefont {Zhu}},\ }\bibfield
  {title} {\bibinfo {title} {Electrical and {Thermal} {Control} of {Magnetic}
  {Exchange} {Interactions}},\ }\bibfield  {journal} {\bibinfo  {journal}
  {Physical Review Letters}\ }\textbf {\bibinfo {volume} {113}},\ \href
  {https://doi.org/10.1103/PhysRevLett.113.257201}
  {10.1103/PhysRevLett.113.257201} (\bibinfo {year} {2014})\BibitemShut
  {NoStop}%
\bibitem [{\citenamefont {Djeghloul}\ \emph {et~al.}(2016)\citenamefont
  {Djeghloul}, \citenamefont {Gruber}, \citenamefont {Urbain}, \citenamefont
  {Xenioti}, \citenamefont {Joly}, \citenamefont {Boukari}, \citenamefont
  {Arabski}, \citenamefont {Bulou}, \citenamefont {Scheurer}, \citenamefont
  {Bertran}, \citenamefont {Le~F{\`e}vre}, \citenamefont {Taleb-Ibrahimi},
  \citenamefont {Wulfhekel}, \citenamefont {Garreau}, \citenamefont
  {Hajjar-Garreau}, \citenamefont {Wetzel}, \citenamefont {Alouani},
  \citenamefont {Beaurepaire}, \citenamefont {Bowen},\ and\ \citenamefont
  {Weber}}]{djeghloul_high_2016}%
  \BibitemOpen
  \bibfield  {author} {\bibinfo {author} {\bibfnamefont {F.}~\bibnamefont
  {Djeghloul}}, \bibinfo {author} {\bibfnamefont {M.}~\bibnamefont {Gruber}},
  \bibinfo {author} {\bibfnamefont {E.}~\bibnamefont {Urbain}}, \bibinfo
  {author} {\bibfnamefont {D.}~\bibnamefont {Xenioti}}, \bibinfo {author}
  {\bibfnamefont {L.}~\bibnamefont {Joly}}, \bibinfo {author} {\bibfnamefont
  {S.}~\bibnamefont {Boukari}}, \bibinfo {author} {\bibfnamefont
  {J.}~\bibnamefont {Arabski}}, \bibinfo {author} {\bibfnamefont
  {H.}~\bibnamefont {Bulou}}, \bibinfo {author} {\bibfnamefont
  {F.}~\bibnamefont {Scheurer}}, \bibinfo {author} {\bibfnamefont
  {F.}~\bibnamefont {Bertran}}, \bibinfo {author} {\bibfnamefont
  {P.}~\bibnamefont {Le~F{\`e}vre}}, \bibinfo {author} {\bibfnamefont
  {A.}~\bibnamefont {Taleb-Ibrahimi}}, \bibinfo {author} {\bibfnamefont
  {W.}~\bibnamefont {Wulfhekel}}, \bibinfo {author} {\bibfnamefont
  {G.}~\bibnamefont {Garreau}}, \bibinfo {author} {\bibfnamefont
  {S.}~\bibnamefont {Hajjar-Garreau}}, \bibinfo {author} {\bibfnamefont
  {P.}~\bibnamefont {Wetzel}}, \bibinfo {author} {\bibfnamefont
  {M.}~\bibnamefont {Alouani}}, \bibinfo {author} {\bibfnamefont
  {E.}~\bibnamefont {Beaurepaire}}, \bibinfo {author} {\bibfnamefont
  {M.}~\bibnamefont {Bowen}},\ and\ \bibinfo {author} {\bibfnamefont
  {W.}~\bibnamefont {Weber}},\ }\bibfield  {title} {\bibinfo {title} {High spin
  polarization at ferromagnetic metal--organic interfaces: A generic
  property},\ }\href@noop {} {\bibfield  {journal} {\bibinfo  {journal} {J.
  Phys. Chem. Lett.}\ }\textbf {\bibinfo {volume} {7}},\ \bibinfo {pages}
  {2310} (\bibinfo {year} {2016})}\BibitemShut {NoStop}%
\bibitem [{\citenamefont {Delprat}\ \emph {et~al.}(2018)\citenamefont
  {Delprat}, \citenamefont {Galbiati}, \citenamefont {Tatay}, \citenamefont
  {Quinard}, \citenamefont {Barraud}, \citenamefont {Petroff}, \citenamefont
  {Seneor},\ and\ \citenamefont {Mattana}}]{delprat_molecular_2018}%
  \BibitemOpen
  \bibfield  {author} {\bibinfo {author} {\bibfnamefont {S.}~\bibnamefont
  {Delprat}}, \bibinfo {author} {\bibfnamefont {M.}~\bibnamefont {Galbiati}},
  \bibinfo {author} {\bibfnamefont {S.}~\bibnamefont {Tatay}}, \bibinfo
  {author} {\bibfnamefont {B.}~\bibnamefont {Quinard}}, \bibinfo {author}
  {\bibfnamefont {C.}~\bibnamefont {Barraud}}, \bibinfo {author} {\bibfnamefont
  {F.}~\bibnamefont {Petroff}}, \bibinfo {author} {\bibfnamefont
  {P.}~\bibnamefont {Seneor}},\ and\ \bibinfo {author} {\bibfnamefont
  {R.}~\bibnamefont {Mattana}},\ }\bibfield  {title} {\bibinfo {title}
  {Molecular spintronics: the role of spin-dependent hybridization},\
  }\href@noop {} {\bibfield  {journal} {\bibinfo  {journal} {J. Phys. D Appl.
  Phys.}\ }\textbf {\bibinfo {volume} {51}},\ \bibinfo {pages} {473001}
  (\bibinfo {year} {2018})}\BibitemShut {NoStop}%
\bibitem [{\citenamefont {Bergfield}\ \emph {et~al.}(2013)\citenamefont
  {Bergfield}, \citenamefont {Story}, \citenamefont {Stafford},\ and\
  \citenamefont {Stafford}}]{bergfield_probing_2013}%
  \BibitemOpen
  \bibfield  {author} {\bibinfo {author} {\bibfnamefont {J.~P.}\ \bibnamefont
  {Bergfield}}, \bibinfo {author} {\bibfnamefont {S.~M.}\ \bibnamefont
  {Story}}, \bibinfo {author} {\bibfnamefont {R.~C.}\ \bibnamefont
  {Stafford}},\ and\ \bibinfo {author} {\bibfnamefont {C.~A.}\ \bibnamefont
  {Stafford}},\ }\bibfield  {title} {\bibinfo {title} {Probing {Maxwell}’s
  {Demon} with a {Nanoscale} {Thermometer}},\ }\href
  {https://doi.org/10.1021/nn401027u} {\bibfield  {journal} {\bibinfo
  {journal} {ACS Nano}\ }\textbf {\bibinfo {volume} {7}},\ \bibinfo {pages}
  {4429} (\bibinfo {year} {2013})}\BibitemShut {NoStop}%
\bibitem [{\citenamefont {Manzano}\ \emph {et~al.}(2018)\citenamefont
  {Manzano}, \citenamefont {Plastina},\ and\ \citenamefont
  {Zambrini}}]{manzano_optimal_2018}%
  \BibitemOpen
  \bibfield  {author} {\bibinfo {author} {\bibfnamefont {G.}~\bibnamefont
  {Manzano}}, \bibinfo {author} {\bibfnamefont {F.}~\bibnamefont {Plastina}},\
  and\ \bibinfo {author} {\bibfnamefont {R.}~\bibnamefont {Zambrini}},\
  }\bibfield  {title} {\bibinfo {title} {Optimal {Work} {Extraction} and
  {Thermodynamics} of {Quantum} {Measurements} and {Correlations}},\ }\bibfield
   {journal} {\bibinfo  {journal} {Physical Review Letters}\ }\textbf {\bibinfo
  {volume} {121}},\ \href {https://doi.org/10.1103/PhysRevLett.121.120602}
  {10.1103/PhysRevLett.121.120602} (\bibinfo {year} {2018})\BibitemShut
  {NoStop}%
\bibitem [{\citenamefont {Erez}\ \emph {et~al.}(2008)\citenamefont {Erez},
  \citenamefont {Gordon}, \citenamefont {Nest},\ and\ \citenamefont
  {Kurizki}}]{erez_thermodynamic_2008}%
  \BibitemOpen
  \bibfield  {author} {\bibinfo {author} {\bibfnamefont {N.}~\bibnamefont
  {Erez}}, \bibinfo {author} {\bibfnamefont {G.}~\bibnamefont {Gordon}},
  \bibinfo {author} {\bibfnamefont {M.}~\bibnamefont {Nest}},\ and\ \bibinfo
  {author} {\bibfnamefont {G.}~\bibnamefont {Kurizki}},\ }\bibfield  {title}
  {\bibinfo {title} {Thermodynamic control by frequent quantum measurements},\
  }\href {https://doi.org/10.1038/nature06873} {\bibfield  {journal} {\bibinfo
  {journal} {Nature}\ }\textbf {\bibinfo {volume} {452}},\ \bibinfo {pages}
  {724} (\bibinfo {year} {2008})}\BibitemShut {NoStop}%
\bibitem [{\citenamefont {Erez}(2012)}]{erez_thermodynamics_2012}%
  \BibitemOpen
  \bibfield  {author} {\bibinfo {author} {\bibfnamefont {N.}~\bibnamefont
  {Erez}},\ }\bibfield  {title} {\bibinfo {title} {Thermodynamics of projective
  quantum measurements},\ }\href
  {https://doi.org/10.1088/0031-8949/2012/T151/014028} {\bibfield  {journal}
  {\bibinfo  {journal} {Physica Scripta}\ }\textbf {\bibinfo {volume} {T151}},\
  \bibinfo {pages} {014028} (\bibinfo {year} {2012})}\BibitemShut {NoStop}%
\bibitem [{\citenamefont {Lindblad}(1973)}]{lindblad_entropy_1973}%
  \BibitemOpen
  \bibfield  {author} {\bibinfo {author} {\bibfnamefont {G.}~\bibnamefont
  {Lindblad}},\ }\bibfield  {title} {\bibinfo {title} {Entropy, information and
  quantum measurements},\ }\href {https://doi.org/10.1007/BF01646743}
  {\bibfield  {journal} {\bibinfo  {journal} {Communications in Mathematical
  Physics}\ }\textbf {\bibinfo {volume} {33}},\ \bibinfo {pages} {305}
  (\bibinfo {year} {1973})}\BibitemShut {NoStop}%
\bibitem [{\citenamefont {Jacobs}(2012)}]{jacobs_quantum_2012}%
  \BibitemOpen
  \bibfield  {author} {\bibinfo {author} {\bibfnamefont {K.}~\bibnamefont
  {Jacobs}},\ }\bibfield  {title} {\bibinfo {title} {Quantum measurement and
  the first law of thermodynamics: {The} energy cost of measurement is the work
  value of the acquired information},\ }\href
  {https://doi.org/10.1103/PhysRevE.86.040106} {\bibfield  {journal} {\bibinfo
  {journal} {Physical Review E}\ }\textbf {\bibinfo {volume} {86}},\ \bibinfo
  {pages} {040106} (\bibinfo {year} {2012})}\BibitemShut {NoStop}%
\bibitem [{\citenamefont {Ban}(1999)}]{ban_state_1999}%
  \BibitemOpen
  \bibfield  {author} {\bibinfo {author} {\bibfnamefont {M.}~\bibnamefont
  {Ban}},\ }\bibfield  {title} {\bibinfo {title} {State reduction, information
  and entropy in quantum measurement processes},\ }\href
  {https://doi.org/10.1088/0305-4470/32/9/012} {\bibfield  {journal} {\bibinfo
  {journal} {Journal of Physics A: Mathematical and General}\ }\textbf
  {\bibinfo {volume} {32}},\ \bibinfo {pages} {1643} (\bibinfo {year}
  {1999})}\BibitemShut {NoStop}%
\bibitem [{\citenamefont {Bormashenko}(2020)}]{bormashenko_entropy_2020}%
  \BibitemOpen
  \bibfield  {author} {\bibinfo {author} {\bibfnamefont {E.}~\bibnamefont
  {Bormashenko}},\ }\bibfield  {title} {\bibinfo {title} {Entropy,
  {Information}, and {Symmetry}; {Ordered} {Is} {Symmetrical}, {II}: {System}
  of {Spins} in the {Magnetic} {Field}},\ }\href
  {https://doi.org/10.3390/e22020235} {\bibfield  {journal} {\bibinfo
  {journal} {Entropy}\ }\textbf {\bibinfo {volume} {22}},\ \bibinfo {pages}
  {235} (\bibinfo {year} {2020})}\BibitemShut {NoStop}%
\bibitem [{\citenamefont {Vaccaro}\ and\ \citenamefont
  {Barnett}(2011)}]{vaccaro_information_2011}%
  \BibitemOpen
  \bibfield  {author} {\bibinfo {author} {\bibfnamefont {J.~A.}\ \bibnamefont
  {Vaccaro}}\ and\ \bibinfo {author} {\bibfnamefont {S.~M.}\ \bibnamefont
  {Barnett}},\ }\bibfield  {title} {\bibinfo {title} {Information erasure
  without an energy cost},\ }\href {https://doi.org/10.1098/rspa.2010.0577}
  {\bibfield  {journal} {\bibinfo  {journal} {Proceedings of the Royal Society
  A: Mathematical, Physical and Engineering Sciences}\ }\textbf {\bibinfo
  {volume} {467}},\ \bibinfo {pages} {1770} (\bibinfo {year}
  {2011})}\BibitemShut {NoStop}%
\bibitem [{\citenamefont {Bowen}(2023)}]{bowen_atom_2023}%
  \BibitemOpen
  \bibfield  {author} {\bibinfo {author} {\bibfnamefont {M.}~\bibnamefont
  {Bowen}},\ }\bibfield  {title} {\bibinfo {title} {Atom-level electronic
  physicists are needed to develop practical engines with a quantum
  advantage},\ }\href@noop {} {\bibfield  {journal} {\bibinfo  {journal} {Npj
  Quantum Inf.}\ }\textbf {\bibinfo {volume} {9}} (\bibinfo {year}
  {2023})}\BibitemShut {NoStop}%
\bibitem [{\citenamefont {Harder}\ \emph {et~al.}(2016)\citenamefont {Harder},
  \citenamefont {Gui},\ and\ \citenamefont {Hu}}]{harder_electrical_2016}%
  \BibitemOpen
  \bibfield  {author} {\bibinfo {author} {\bibfnamefont {M.}~\bibnamefont
  {Harder}}, \bibinfo {author} {\bibfnamefont {Y.}~\bibnamefont {Gui}},\ and\
  \bibinfo {author} {\bibfnamefont {C.-M.}\ \bibnamefont {Hu}},\ }\bibfield
  {title} {\bibinfo {title} {Electrical detection of magnetization dynamics via
  spin rectification effects},\ }\href
  {https://doi.org/10.1016/j.physrep.2016.10.002} {\bibfield  {journal}
  {\bibinfo  {journal} {Physics Reports}\ }\textbf {\bibinfo {volume} {661}},\
  \bibinfo {pages} {1} (\bibinfo {year} {2016})}\BibitemShut {NoStop}%
\bibitem [{\citenamefont {Bowen}\ \emph {et~al.}(2006)\citenamefont {Bowen},
  \citenamefont {Barthélémy}, \citenamefont {Bellini}, \citenamefont {Bibes},
  \citenamefont {Seneor}, \citenamefont {Jacquet}, \citenamefont {Contour},\
  and\ \citenamefont {Dederichs}}]{bowen_observation_2006}%
  \BibitemOpen
  \bibfield  {author} {\bibinfo {author} {\bibfnamefont {M.}~\bibnamefont
  {Bowen}}, \bibinfo {author} {\bibfnamefont {A.}~\bibnamefont {Barthélémy}},
  \bibinfo {author} {\bibfnamefont {V.}~\bibnamefont {Bellini}}, \bibinfo
  {author} {\bibfnamefont {M.}~\bibnamefont {Bibes}}, \bibinfo {author}
  {\bibfnamefont {P.}~\bibnamefont {Seneor}}, \bibinfo {author} {\bibfnamefont
  {E.}~\bibnamefont {Jacquet}}, \bibinfo {author} {\bibfnamefont {J.-P.}\
  \bibnamefont {Contour}},\ and\ \bibinfo {author} {\bibfnamefont {P.~H.}\
  \bibnamefont {Dederichs}},\ }\bibfield  {title} {\bibinfo {title}
  {Observation of {Fowler}–{Nordheim} hole tunneling across an electron
  tunnel junction due to total symmetry filtering},\ }\bibfield  {journal}
  {\bibinfo  {journal} {Physical Review B}\ }\textbf {\bibinfo {volume} {73}},\
  \href {https://doi.org/10.1103/PhysRevB.73.140408}
  {10.1103/PhysRevB.73.140408} (\bibinfo {year} {2006})\BibitemShut {NoStop}%
\bibitem [{\citenamefont {Miao}\ \emph {et~al.}(2014)\citenamefont {Miao},
  \citenamefont {Chang}, \citenamefont {Assaf}, \citenamefont {Heiman},\ and\
  \citenamefont {Moodera}}]{miao_spin_2014}%
  \BibitemOpen
  \bibfield  {author} {\bibinfo {author} {\bibfnamefont {G.-X.}\ \bibnamefont
  {Miao}}, \bibinfo {author} {\bibfnamefont {J.}~\bibnamefont {Chang}},
  \bibinfo {author} {\bibfnamefont {B.~A.}\ \bibnamefont {Assaf}}, \bibinfo
  {author} {\bibfnamefont {D.}~\bibnamefont {Heiman}},\ and\ \bibinfo {author}
  {\bibfnamefont {J.~S.}\ \bibnamefont {Moodera}},\ }\bibfield  {title}
  {\bibinfo {title} {Spin regulation in composite spin-filter barrier
  devices},\ }\bibfield  {journal} {\bibinfo  {journal} {Nature
  Communications}\ }\textbf {\bibinfo {volume} {5}},\ \href
  {https://doi.org/10.1038/ncomms4682} {10.1038/ncomms4682} (\bibinfo {year}
  {2014})\BibitemShut {NoStop}%
\bibitem [{\citenamefont {Nielsen}\ \emph {et~al.}(2002)\citenamefont
  {Nielsen}, \citenamefont {Chuang},\ and\ \citenamefont
  {Grover}}]{nielsen_quantum_2002}%
  \BibitemOpen
  \bibfield  {author} {\bibinfo {author} {\bibfnamefont {M.~A.}\ \bibnamefont
  {Nielsen}}, \bibinfo {author} {\bibfnamefont {I.}~\bibnamefont {Chuang}},\
  and\ \bibinfo {author} {\bibfnamefont {L.~K.}\ \bibnamefont {Grover}},\
  }\bibfield  {title} {\bibinfo {title} {{Quantum Computation and Quantum
  Information}},\ }\href {https://doi.org/10.1119/1.1463744} {\bibfield
  {journal} {\bibinfo  {journal} {American Journal of Physics}\ }\textbf
  {\bibinfo {volume} {70}},\ \bibinfo {pages} {558} (\bibinfo {year} {2002})},\
  \Eprint
  {https://arxiv.org/abs/https://pubs.aip.org/aapt/ajp/article-pdf/70/5/558/7529938/558\_2\_online.pdf}
  {https://pubs.aip.org/aapt/ajp/article-pdf/70/5/558/7529938/558\_2\_online.pdf}
  \BibitemShut {NoStop}%
\bibitem [{\citenamefont {Kammerlander}\ and\ \citenamefont
  {Anders}(2016)}]{kammerlander_coherence_2016}%
  \BibitemOpen
  \bibfield  {author} {\bibinfo {author} {\bibfnamefont {P.}~\bibnamefont
  {Kammerlander}}\ and\ \bibinfo {author} {\bibfnamefont {J.}~\bibnamefont
  {Anders}},\ }\bibfield  {title} {\bibinfo {title} {Coherence and measurement
  in quantum thermodynamics},\ }\bibfield  {journal} {\bibinfo  {journal}
  {Scientific Reports}\ }\textbf {\bibinfo {volume} {6}},\ \href
  {https://doi.org/10.1038/srep22174} {10.1038/srep22174} (\bibinfo {year}
  {2016})\BibitemShut {NoStop}%
\bibitem [{\citenamefont {Kumar}\ and\ \citenamefont
  {Staﬀord}()}]{kumar_first_nodate}%
  \BibitemOpen
  \bibfield  {author} {\bibinfo {author} {\bibfnamefont {P.}~\bibnamefont
  {Kumar}}\ and\ \bibinfo {author} {\bibfnamefont {C.~A.}\ \bibnamefont
  {Staﬀord}},\ }\bibfield  {title} {\bibinfo {title} {On the {First} {Law} of
  {Thermodynamics} in {Time}-{Dependent} {Open} {Quantum} {Systems}},\
  }\href@noop {} {\ ,\ \bibinfo {pages} {19}}\BibitemShut {NoStop}%
\bibitem [{\citenamefont {Hormoz}(2013)}]{hormoz_quantum_2013}%
  \BibitemOpen
  \bibfield  {author} {\bibinfo {author} {\bibfnamefont {S.}~\bibnamefont
  {Hormoz}},\ }\bibfield  {title} {\bibinfo {title} {Quantum collapse and the
  second law of thermodynamics},\ }\href
  {https://doi.org/10.1103/PhysRevE.87.022129} {\bibfield  {journal} {\bibinfo
  {journal} {Physical Review E}\ }\textbf {\bibinfo {volume} {87}},\ \bibinfo
  {pages} {022129} (\bibinfo {year} {2013})}\BibitemShut {NoStop}%
\bibitem [{\citenamefont {Lloyd}(1989)}]{lloyd_use_1989}%
  \BibitemOpen
  \bibfield  {author} {\bibinfo {author} {\bibfnamefont {S.}~\bibnamefont
  {Lloyd}},\ }\bibfield  {title} {\bibinfo {title} {Use of mutual information
  to decrease entropy: {Implications} for the second law of thermodynamics},\
  }\href {https://doi.org/10.1103/PhysRevA.39.5378} {\bibfield  {journal}
  {\bibinfo  {journal} {Physical Review A}\ }\textbf {\bibinfo {volume} {39}},\
  \bibinfo {pages} {5378} (\bibinfo {year} {1989})}\BibitemShut {NoStop}%
\bibitem [{\citenamefont {D'Abramo}(2012)}]{dabramo_peculiar_2012}%
  \BibitemOpen
  \bibfield  {author} {\bibinfo {author} {\bibfnamefont {G.}~\bibnamefont
  {D'Abramo}},\ }\bibfield  {title} {\bibinfo {title} {The peculiar status of
  the second law of thermodynamics and the quest for its violation},\ }\href
  {https://doi.org/10.1016/j.shpsb.2012.05.004} {\bibfield  {journal} {\bibinfo
   {journal} {Studies in History and Philosophy of Science Part B: Studies in
  History and Philosophy of Modern Physics}\ }\textbf {\bibinfo {volume}
  {43}},\ \bibinfo {pages} {226} (\bibinfo {year} {2012})}\BibitemShut
  {NoStop}%
\end{thebibliography}%

\newpage
\onecolumngrid
\appendix 

\renewcommand{\appendixname}{SI}
\renewcommand\thefigure{SI.\arabic{figure}}
\renewcommand\thesection{Note \arabic{section}}
\renewcommand{\theequation}{SI-\arabic{equation}} 

\begin{center}
    {\large \textbf{SUPPLEMENTARY INFORMATION\\Quantum Measurement Spintronic Engine powered by Quantum Fluctuations}}
\end{center}

\begin{center}
    Mathieu Lamblin and Martin Bowen
\end{center}

\title{SUPPLEMENTARY INFORMATION\\Quantum Measurement Spintronic Engine powered by Quantum Fluctuations} 

\author{Mathieu Lamblin}
    \affiliation{Institut de Physique et Chimie des Matériaux de Strasbourg, UMR 7504 CNRS, Université de Strasbourg, 23 Rue du Lœss, BP 43, 67034 Strasbourg, France}
    \email{mathieu.lamblin@ipcms.unistra.fr}
\author{Martin Bowen}
    \affiliation{Institut de Physique et Chimie des Matériaux de Strasbourg, UMR 7504 CNRS, Université de Strasbourg, 23 Rue du Lœss, BP 43, 67034 Strasbourg, France}
    \email{bowen@unistra.fr}

\date{\today}



\maketitle


\section{\label{SI1} The Model}

\subsection{\label{1A} The general Hamiltonian}

In this paper, we consider two QDs. Each QD, or atomic dot, consists in two non-degenerate electronic energy levels that code for two opposite spins on the Bloch sphere. The two QDs are coupled with one another by a tunneling interaction of magnitude $\gamma$ and a magnetic exchange interaction of magnitude $J$. A coulombic repulsion term $U$ is also considered so as to prevent excessive charging on each dot. The environment is composed of two ferromagnetic leads, the left one $L$ and the right one $R$, each of them respectively coupled to the left and right QDs. From these elements, the total Hamiltonian $H$ can be separated as:
\begin{equation}
    H= H_S +  H_E + H_{SE}.
\end{equation}

The first term $H_S$, called the Hamiltonian of the system, represents the two spin qubits and can be written as:
\begin{multline}
    H_S = \epsilon_{L\ua} n_{L\ua} + \epsilon_{L\da} n_{L\da} + \epsilon_{R\ua} n_{R\ua} + \epsilon_{R\da} n_{R\da}  + (\gamma c^{\dagger}_{L\ua}c_{R\ua} + \gamma^*c_{L\ua}c^{\dagger}_{R\ua} + \gamma c^{\dagger}_{L\da}c_{R\da} + \gamma^*c_{L\da}c^{\dagger}_{R\da}) \\
     - U((1-n_{L\ua})n_{L\da} + n_{L\ua}(1-n_{L\da})) -U((1-n_{R\ua})n_{R\da} + n_{R\ua}(1-n_{R\da})) - J (n_{L\ua}n_{R\ua} + n_{L\da}n_{R\da}).
\end{multline}
Where we have defined $c^{\dagger}$ and $c$ the raising and lowering operators with the left index identifying the left or right quantum dot and the right index identifying the spin. The $n$ correspond to the number operators defined as $n = c^{\dagger}c$. Let us explain the physical meaning of those terms. The first terms in $\epsilon$ correspond to the bare energy of each of the four electrons that can occupy the two quantum dots. We assume that the energies $\epsilon$ are different for each level and we will see below that their relative values can be tuned through the couplings.\\
The terms in $\gamma$ code for the hopping electron transmission between the QDs. The spin is preserved during this transfer as no spin flip is possible during the hopping to leading order. The electron hopping argument $\gamma$ is taken as independent of the tunnelling spin for simplicity. Although this tunneling argument should strongly depend on the considered spin channel given the spin-splitting of the energy level, this assumption is not critical here given the approximations we make later.\\
The terms in $J$ represents the magnetic coupling between the two quantum dots. Since prior literature indicates antiferromagnetic coupling and spontaneous current flow at $V = 0$ \cite{katcko_spin-driven_2019, chowrira_quantum_2022}, we therefore assume that $J < 0$. As we can see, this contribution adds an energy penalty of $-J$  when an electron of same spin is present on both QDs. This repulsion term $J$ is considered independent of the spin orientation for simplicity, and we will see in the following that this approximation holds given the weak relevance of $J$ in the next results.\\
Finally, the terms in $U$ correspond to the Coulombic repulsion which lowers the energy when a quantum dot is singly occupied. The term has been included to avoid excessive charge being retained on the system. We assume that this Coulomb repulsion energy is identical on the two sites.

This Hamiltonian can be simplified a little by rescaling the energies. Redefining $\epsilon \equiv \epsilon - U$ and $U \equiv U/2$, we get:
\begin{multline}
    H_S = \epsilon_{L\ua} n_{L\ua} + \epsilon_{L\da} n_{L\da} + \epsilon_{R\ua} n_{R\ua} + \epsilon_{R\da} n_{R\da} + (\gamma c^{\dagger}_{L\ua}c_{R\ua} + \gamma^*c_{L\ua}c^{\dagger}_{R\ua} + \gamma c^{\dagger}_{L\da}c_{R\da} + \gamma^*c_{L\da}c^{\dagger}_{R\da}) \\
    - J (n_{L\ua}n_{R\ua} + n_{L\da}n_{R\da}) + U(n_{L\ua}n_{L\da} + n_{R\ua}n_{R\da}) 
\end{multline}

The second contribution to the Hamiltonian describes the energy of the ferromagnetic reservoirs, i.e. the environment of the QDs. It can be split into two terms $H_E = H_L + H_R$ describing each electrode:
\begin{equation}
    H_L = \sum_{k\sigma} \epsilon_{k\sigma}\, c^{\dagger}_{k\sigma}c_{k\sigma} ,\ H_R = \sum_{p\sigma} \epsilon_{p\sigma}\, c^{\dagger}_{p\sigma}c_{p\sigma} \ ,\ 
\end{equation}
where the index $\sigma$ accounts for the spin degrees of freedom while the indexes $k$ and $p$ are used for the left and right leads respectively, such that $\epsilon_{k,\sigma}$ and $\epsilon_{p,\sigma}$ are the energies of each fermionic mode of the field while $c_{k,\sigma}^{\dagger}$, $c_{p,\sigma}^{\dagger}$, $c_{k,\sigma}$ and $c_{p,\sigma}$ are the creation and annihilation operators. Only a single band is considered on each lead. This hypothesis is in line with a description of dominant transmission from a specific wavefunction in most tunneling spintronic devices \cite{bowen_observation_2006}. It is especially valid given prior experiments \cite{katcko_spin-driven_2019,chowrira_quantum_2022} on the quantum spintronic engine that utilize the ferromagnetic metal/molecule interface (aka the spinterface \cite{djeghloul_high_2016, delprat_molecular_2018} to generate electrodes with a spectrally narrow band of conduction states with full spin-polarization.\\

The final term $H_{SE}$ describes the tunnelling interaction between the system (i.e. the QDs) and the environment (i.e. the leads). This term can also be split into two parts $H_{SE} = H_{SL} + H_{SR}$. For each lead, we consider two contributions. The first contribution describes the exchange of electrons between the lead and the system, more precisely the adjacent QD since we initially considered a series geometry. This allows for a current to emerge in the model. The second contribution describes the magnetic pinning exerted by the lead on the nearby site to model the effective magnetic field generated by spintronic anisotropy\cite{katcko_spin-driven_2019,chowrira_quantum_2022, miao_spin_2014}. Following these assumptions, we write:
\begin{equation}
    \begin{array}{l}
    H_{SL} = \sum_{k\sigma} (\gamma_{k\sigma}\, c^{\dagger}_{L\sigma} c_{k\sigma} + \gamma_{k\sigma}^*\, c_{L\sigma} c^{\dagger}_{k\sigma}) + \sum_{k\sigma} J_{k\sigma}\,  n_{L\sigma}n_{k\sigma}, \\
    H_{SR} = \sum_{p\sigma} (\gamma_{p\sigma}\, c^{\dagger}_{R\sigma} c_{p\sigma} + \gamma_{p\sigma}^*\, c_{R\sigma} c^{\dagger}_{p\sigma}) + \sum_{p\sigma} J_{p\sigma}\,  n_{R\sigma}n_{p\sigma} \ .
    \end{array}
\end{equation}
The coefficients $\gamma$ represent the hopping coefficients between the QDs and the electrodes, while the $J$ represent the magnetic coupling. Note that our Hamiltonian does not describe an external bias voltage applied across the device: we are considering the case of spontaneous current flow.

\subsection{\label{2B} Primary approximations}

The Hamiltonian we are considering is too complex to be tackled as such analytically. We therefore physically justify the three following approximations. 

First, the spinterface present in experimental devices generates conduction electrons of only one spin that, furthermore, are fixed on the Bloch sphere due to the remanent magnetization of the ferromagnetic electrode underscoring this interfacial effect\cite{chowrira_quantum_2022}. Furthermore, experiments indicate better current output when the device's electrode magnetizations are oriented anti-parallel. As a result, assuming identical L \& R interfaces, we will consider only spin $\uparrow$ electrons in the left lead and spin $\downarrow$ electrons in the right lead. This consideration leads to an approximation of the electrodes and the tunnel Hamiltonians such that:
\begin{equation}
    H_L = \sum_{k} \epsilon_{k}\, c^{\dagger}_{k}c_{k} ,\ H_R = \sum_{p} \epsilon_{p}\, c^{\dagger}_{p}c_{p} \ ,\ 
\end{equation}
and
\begin{equation}
    \begin{array}{l}
    H_{SL} = \sum_{k} (\gamma_{k}\, c^{\dagger}_{L\ua} c_{k} + \gamma_{k}^*\, c_{L\ua} c^{\dagger}_{k}) + \sum_{k} J_{k}\,  n_{L\ua}n_{k},\\
    H_{SR} = \sum_{p} (\gamma_{p}\, c^{\dagger}_{R\da} c_{p} + \gamma_{p}^*\, c_{R\da} c^{\dagger}_{p}) + \sum_{p} J_{p}\,  n_{R\da}n_{p}.
    \end{array}
\end{equation}
It should be made clear that $c_{k}^{\dagger}$ creates an excitation with spin $\uparrow$ and momentum $k$ in the left lead, while $c_p^{\dagger}$ creates an excitation of spin $\downarrow$ and momentum $p$ in the right lead.

Our second assumption is that the effective magnetic field generated through spintronic anisotropy by the fully spin-polarized transport from a lead onto the adjacent QD is constant. This holds at constant bias voltage \cite{katcko_spin-driven_2019}, consistently with the absence of an applied bias in our model. This is also reasonable to first order during engine operation given the much lower formation energy of the ferromagnetic state relative to the engine energies, owing in part to a much larger size compared to that of the atomic dots. We therefore rely on a mean field approach which allows to approximate the magnetic couplings as:
\begin{equation}
    \begin{array}{l}
    \sum_k J_kn_k = \Big{\la} \sum_k J_k n_k \Big{\ra} \equiv J_L,\\
    \sum_p J_pn_p = \Big{\la} \sum_p J_p n_p \Big{\ra} \equiv J_R.
    \end{array}
\end{equation}

This approximation allows us to drop the magnetic coupling term in the system-lead interaction and add it to the system Hamiltonian without changing its structure by rescaling the QDs' energy level. Redefining $\epsilon_{L\ua} \equiv \epsilon_{L\ua} + J_L$ and $\epsilon_{R\da} \equiv \epsilon_{R\da} + J_R$, the system Hamiltonian remains unchanged and the tunnel Hamiltonian now reads:
\begin{equation}
    \begin{array}{l}
        H_{SL} = \sum_{k} (\gamma_{k}\, c^{\dagger}_{L\ua} c_{k} + \gamma_{k}^*\, c_{L\ua} c^{\dagger}_{k}),\\
        H_{SR} = \sum_{p} (\gamma_{p}\, c^{\dagger}_{R\da} c_{p} + \gamma_{p}^*\, c_{R\da} c^{\dagger}_{p}).
    \end{array}
\end{equation}

Finally, the quantum spintronic engine concept as proposed \cite{katcko_spin-driven_2019, chowrira_quantum_2022} includes an asymmetry in the tunnelling coefficients $\gamma_L$ and $\gamma_R$. This not only helps to further break detailed balance of transport, but also enables one electrode to set a dominant spin referential on the QDs. As a result, the QD that is adjacent to that electrode will experience a larger spin splitting than the other QD. Therefore, we assume that the right QD is positioned such as $\epsilon_{R\ua} \gg \epsilon_{R\da}, \epsilon_{L\ua}, \epsilon_{L\da}$. Placing this energy level farther above the other ones allows us to discard all the states where a spin $\ua$ occupies the right QD, thereby reducing the dimensionality of the system Hamiltonian from 16 down to 8:
\begin{equation}
    H_S = \epsilon_{\ua} n_{\ua} + \epsilon_{\da} n_{\da} + \epsilon_{R} n_{R} + \gamma\, c^{\dagger}_{\da}c_{R} + \gamma^*\, c_{\da}c^{\dagger}_{R} + J\, n_{\da}n_{R} + U\,n_{\ua}n_{\da}\ ,
\end{equation}
where we have redefined $\epsilon_{\ua} \equiv \epsilon_{L\ua}$, $\epsilon_{\da} \equiv \epsilon_{L\da}$ and $\epsilon_{R} \equiv \epsilon_{R\da}$ for simplicity, now that the ambiguity between the spin and the L/R QD has been lifted. This approximation is therefore leaving only one transport channel, which justifies the previously stated independence of $\gamma$ on the spin.

\begin{figure}[t]
    \includegraphics[width=0.48\textwidth]{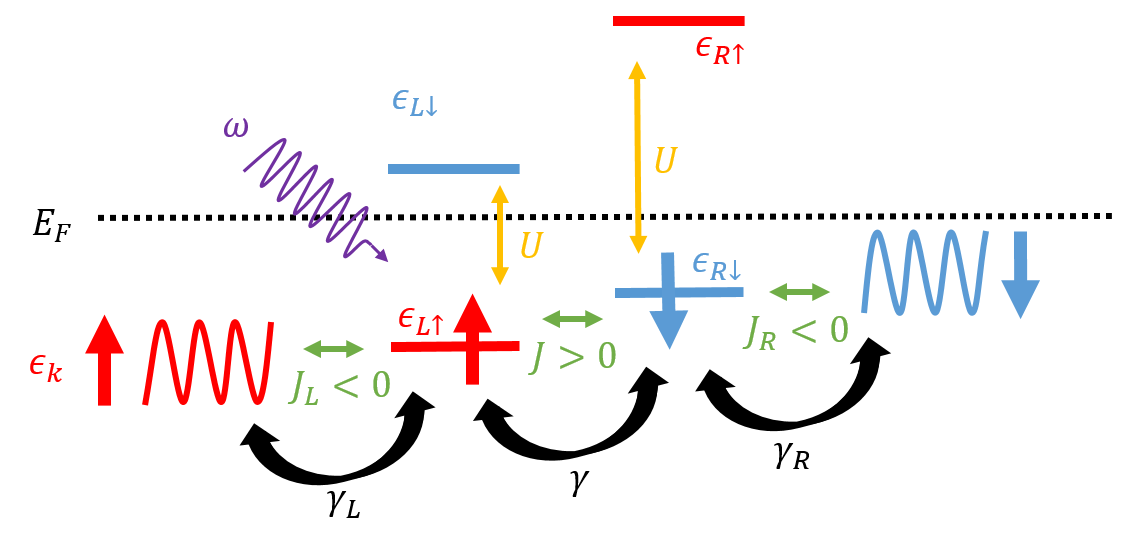}
    \caption{Schematic depiction of the model quantum spintronic engine, featuring two quantum dots trapped in series between two ferromagnetic leads in an antiparallel configuration with fully spin-polarized interactions. Blue/red levels represent spin $\da$/$\ua$ energy levels. Green double arrows represent the magnetic couplings ; yellow double arrows, capacitive couplings and black arrows, tunnel couplings.}  
    \label{model}
\end{figure}

\subsection{The integro-differential master equation}

The quantum system $\rho$ represented in the Schrödinger picture obeys the Von-Neumann equation:
\begin{equation}
    \frac{\d\rho}{\d t} = -\i[H,\rho(t)]\ 
\end{equation}
Switching to the interaction picture where any operator $O$ reads:
\begin{equation}
    \tilde{O}(t) = e^{\i(H_S + H_{E})t}Oe^{-\i(H_S + H_{E})t}\ ,
\end{equation}
the Von-Neumann equation now reads:
\begin{equation}
    \frac{\d\tilde{\rho}}{\d t} = -\i[\tilde{H}_{SE}(t),\tilde{\rho}(t)]\ ,
\end{equation}
which integrates into 
\begin{equation}
    \tilde{\rho}(t) = -\i\int_0^t [\tilde{H}_{SE}(s),\tilde{\rho}(s)]\d s\ .
\end{equation}
Inserting this formula back into the Von-Neumann equation leads to
\begin{equation}
    \frac{\d\tilde{\rho}}{\d t} = -\i[\tilde{H}_{SE}(t),\, \tilde{\rho}(0)]-\int_0^t [\tilde{H}_{SE}(t),\, [\tilde{H}_{SE}(t'),\, \tilde{\rho}(t')]] \d t'\ .
\end{equation}
We now invoke the Born approximation (weak coupling) which states that the system does not influence the environment, so that $\tilde{\rho}_E(t) = \tilde{\rho}_E$ and that the system decomposes as a tensor product at all times, $\tilde{\rho}(t) = \tilde{\rho}_S(t)\otimes\tilde{\rho}_E$. Under this approximation, tracing out the Von-Neumann equation governing the evolution of the composite system leads to
\begin{equation}
    \frac{\d\tilde{\rho}}{\d t} = -\i\rm{Tr}_E[\tilde{H}_{SE}(t),\, \tilde{\rho}_S(0)\otimes \tilde{\rho}_E]-\int_0^t \rm{Tr}_E[\tilde{H}_{SE}(t),\, [\tilde{H}_{SE}(t'),\, \tilde{\rho}_S(t')\otimes \tilde{\rho}_E]] \d t'\ .
\end{equation}
We now write the tunnel interaction as:
\begin{equation}
    H_{SE} = \sum_{i=1}^4 S_i \otimes T_i\ , 
\end{equation}
with
\begin{equation}\begin{array}{ccc}
    S_1\otimes T_1 = c_{\ua}^{\dagger}\otimes \sum_k \gamma_kc_k \equiv S_L^{\dagger}\otimes T_L,\ & S_2\otimes T_2 = c_{\ua}\otimes \sum_k \gamma_k^*c_k^{\dagger} \equiv S_L\otimes T^{\dagger}_L,\\ 
    S_3\otimes T_3 = c_{R}^{\dagger}\otimes \sum_p \gamma_pc_p \equiv S_R^{\dagger}\otimes T_R,\ & S_4\otimes T_4 = c_{R}\otimes \sum_p \gamma_p^*c_p^{\dagger}  \equiv S_R\otimes T^{\dagger}_R 
\end{array}\ .
\end{equation}
Then, in the interaction picture, we obtain the simple form:
\begin{equation}
    \tilde{H}_{SE}(t) = e^{\i H t}H_{SE}e^{-\i H t} = \sum_{i=1}^4 e^{\i H_S t}S_i e^{-\i H_S t} \otimes e^{\i H_{E} t}T_i e^{-\i H_{E} t} \equiv \sum_{i} S_i(t) \otimes T_i(t).
\end{equation}

\subsection{Bath operators in the interaction picture}

We now need to calculate the operators $T_i$ in the interaction picture. From the anti-commutation relations, we have, for all $l \in \N$:
\begin{equation}
    c_kn_k^{l+1} = c_kc_k^{\dagger}c_kn_k^l = (1-c_k^{\dagger}c_k)c_kn_k^l = c_kn_k^l =\ \dots\ = c_k\ ,
\end{equation}
and $n_k^{l+1}c_k = 0$, therefore, expanding the exponential in series, we have:
\begin{equation}
    c_k e^{-\i \epsilon_k n_k t} = \sum_{l=0}^{+\infty} \frac{(-\i\epsilon_k t)^l}{l \text{!}} c_kn_k^l = \sum_{l=0}^{+\infty} \frac{(-\i\epsilon_k t)^l}{l \text{!}} c_k = e^{-\i \epsilon_k t} c_k\ ,
\end{equation}
and 
\begin{equation}
    e^{\i \epsilon_k n_k t}c_k = c_k + \sum_{l=1}^{+\infty} \frac{(\i\epsilon_k t)^l}{l \text{!}} n_k^lc_k = c_k\ ,
\end{equation}
therefore
\begin{equation}
    e^{\i \epsilon_k n_k t}c_ke^{-\i \epsilon_k n_k t} = e^{-\i \epsilon_k t} c_k\ ,
\end{equation}
and because of the commutation relation $[n_k,n_{k'}] = 0$, we obtain:
\begin{equation}
    e^{\i H_{E} t}c_ke^{-\i H_{E} t} = e^{\i H_{L} t}c_ke^{-\i H_{L} t} = e^{\i \sum_{k'}\epsilon_{k'}n_{k'} t}c_ke^{-\i\sum_{k'}\epsilon_{k'}n_{k'} t} = e^{-\i \epsilon_k t} c_k\ .
\end{equation}
An identical relation is obtained for $c^{\dagger}$ by taking the adjoint:
\begin{equation}
    e^{\i H_{E} t}c_k^{\dagger}e^{-\i H_{E} t} = e^{\i \epsilon_k t} c_k^{\dagger}\ ,
\end{equation}
which leads to the particle number by multiplying the two previous results:
\begin{equation}
    e^{\i H_{E} t}n_ke^{-\i H_{E} t} = n_k\ ,
\end{equation}
which is correct because $n_k$ commutes with every term of $H_{E}$. The same relations can be obtained for the right lead by taking $k=p$. We now have all the elements to write the $T_i(t) \equiv e^{\i H_{E} t}T_ie^{-\i H_{E}t}$ in the interaction picture
\begin{equation}\begin{array}{ccc}
    T_L(t) = \sum_k \gamma_ke^{-\i\epsilon_k t} c_k,\ & T_R(t) = \sum_p \gamma_pe^{-\i\epsilon_p t}c_p
\end{array}\ .\end{equation}

\subsection{Averages over the bath}

The average values $\la T_i(t)\ra_{E} = \rm{Tr}(T_i(t)\rho_{E})$ in the $\rho_{E}$ state can now be calculated and yield
\begin{equation}
    \la T_L(t)\ra_{E} = 0 \text{ and } \la T_R(t)\ra_{E} = 0
\end{equation}
Indeed, we recall that the one point functions $\la c \ra_{E}$ all vanish because the number operators are hermitian and commute with $H_{E}$. Therefore, by diagonalizing in a common basis, the $c$ and $c^{\dagger}$ project each eigenspace onto their perpendicular space because they change the particle number, so the trace is null.

The same argument allows us to calculate the averages $\la T_i^{\dagger}(t)T_j(t')\ra $. The only four two-point functions that do not vanish are:
\begin{equation}\begin{array}{l}
    \la T_L^{\dagger}(t)T_L(t')\ra_{E} = \sum_k |\gamma_k|^2e^{\i\epsilon_k(t-t')}\la n_k\ra_{E}  \\
    \la T_L(t)T_L^{\dagger}(t')\ra_{E} = \sum_k |\gamma_k|^2e^{-\i\epsilon_k(t-t')}(1 - \la n_k\ra_{E})  \\
    \la T_R^{\dagger}(t)T_R(t')\ra_{E} = \sum_p |\gamma_p|^2e^{\i\epsilon_p(t-t')}\la n_p\ra_{E}  \\
    \la T_R(t)T_R^{\dagger}(t')\ra_{E} = \sum_p |\gamma_p|^2e^{-\i\epsilon_p(t-t')}(1-\la n_p\ra_{E})  \\
\end{array}\ .\end{equation}

Invoking the rotating wave, or secular approximation, we discard the fast oscillating terms, which lead us to consider that the two-point functions decay rapidly, such that $\la T_i(t)T_j(t')\ra_{E} \propto \delta(t-t')$ up to an additive constant. We now have:
\begin{equation}\begin{array}{l}
    \la T_L^{\dagger}(t)T_L(t')\ra_{E} = \sum_k |\gamma_k|^2\la n_k\ra_{E}\delta(t-t') \equiv \cal{T}^-_L\delta(t-t')/2  \\
    \la T_L(t)T_L^{\dagger}(t')\ra_{E} = \sum_k |\gamma_k|^2(1-\la n_k\ra_{E})\delta(t-t') \equiv \cal{T}^+_L\delta(t-t')/2 \\
    \la T_R^{\dagger}(t)T_R(t')\ra_{E} = \sum_p |\gamma_p|^2\la n_p\ra_{E}\delta(t-t') \equiv \cal{T}^-_R\delta(t-t')/2  \\
    \la T_R(t)T_R^{\dagger}(t')\ra_{E} = \sum_p |\gamma_p|^2(1-\la n_p\ra_{E})\delta(t-t') \equiv \cal{T}^+_R\delta(t-t')/2\\
\end{array}\ .\end{equation}

The constants $\cal{T}^-_L$ and $\cal{T}^-_R$ are real numbers and characterize the electrons tunnelling coefficients of the left and right leads, while $\cal{T}^+_L$ and $\cal{T}^+_R$ characterize the hole tunneling coefficients.

It should be emphasized here that $\cal{T}^-$ and $\cal{T}^+$ are not independent since we should have $\cal{T}^++\cal{T}^-=\sum_k|\gamma_k|^2\equiv \kappa$, such as we should have some detailed balance leading to $\cal{T}^- = \kappa n_F(\mu, T)$ and $\cal{T}^+ = \kappa (1-n_F(\mu, T))$, for a perfectly thermal bath, where $n_F(\mu,T) = \frac{1}{1+e^{-\mu/k_BT}}$ is the Fermi function at temperature $T$ and electrochemical potential $\mu$. In the following, we will discard the link between the two quantities and consider that they can be tuned relatively independently either by acting on the electron filling, the electrochemical potential, the temperature, or by invoking some non-thermal interaction between the spinterface and the QDs, which will be the topic of a future paper.

\subsection{The GKSL master equation}

The first term in the integro-differential equation reads:
\begin{equation}
    \rm{Tr}_{E}[\tilde{H}_{SE}(t),\tilde{\rho}_S(0)\otimes\tilde{\rho}_{E}] = \sum_{i=1}^4 [S_i(t), \tilde{\rho}_S(0)]\la T_i(t) \ra_{E}\ .
\end{equation}
We can always choose a special Hamiltonian by adding a constant to it such that $\la T_i \ra_{E} = 0$. Indeed, the rescaled Hamiltonian $ H = \big( H_S + \sum_i \la T_i \ra_{E} S_i \big) + H_{E} + H_{SE}' $, with $H_{SE}' = \sum_i S_i \otimes (T-\la T_i \ra_{E}) $ is such that $\la T_i' \ra_{E} = 0$, with $T_i' = T_i - \la T_i \ra_{E} $, while having the same dynamics. Therefore, we can discard this term in the equation.

The second term in the Von-Neumann equation reads:
\begin{equation}
    -\int_0^t \sum_{i,j} \rm{Tr}_{E}\Big( [\tilde{H}_{SE}(t),\ [\tilde{H}_T(t'), \tilde{\rho}_S(t)\otimes \tilde{\rho}_E]]\Big)
\end{equation}
which expands into
\begin{multline}
    -\int_0^t \rm{Tr}_{E}\Big(\tilde{H}_{SE}(t)\tilde{H}_{SE}(t')\tilde{\rho}_S(t)\otimes\tilde{\rho}_E - \tilde{H}_{SE}(t') \tilde{\rho}_S(t)\otimes\tilde{\rho}_E \tilde{H}_{SE}(t) \\ - \tilde{H}_{SE}(t)\tilde{\rho}_S(t)\otimes \tilde{\rho}_E \tilde{H}_{SE}(t') + \tilde{\rho}_S(t)\otimes\tilde{\rho}_E\tilde{H}_{SE}(t')\tilde{H}_{SE}(t)\Big) \d t'\ ,
\end{multline}
reading also:
\begin{equation}
    \int_0^t \rm{Tr}_{E}\Big([\tilde{H}_{SE}(t')\tilde{\rho}_S(t)\otimes\tilde{\rho}_E, \tilde{H}_{SE}(t)]\Big) \d t' + \int_0^t \rm{Tr}_{E}\Big([\tilde{H}_{SE}(t), \tilde{\rho}_S(t)\otimes\tilde{\rho}_E \tilde{H}_{SE}(t')]\Big) \d t'\ .
\end{equation}
Invoking now the hermiticity of $\tilde{H}_{SE}$, we replace $\tilde{H}_{SE}(t)$ by $\tilde{H}^{\dagger}_{SE}(t)$ in the first term and $\tilde{H}_{SE}(t')$ by $\tilde{H}^{\dagger}_{SE}(t')$ in the second term, then we express the tunnel Hamiltonians in terms of $S_i$ and $T_i$ and use the cyclicity property of the trace to obtain
\begin{equation}
    \sum_{i,j} \int_0^t \la(T_i^{\dagger}(t)T_j(t')\ra_{E}  [S_j(t')\tilde{\rho}_S(t),S_i^{\dagger}(t)]\d t' + \int_0^t \la T_i^{\dagger}(t')T_j(t)\ra_{E}  [S_j(t),\tilde{\rho}_S(t)S^{\dagger}_i(t')] \d t'
\end{equation}
Using the above expressions for the tunnel two point functions, only four terms remain, leading to the final form of the master equation:
\begin{equation}
    \frac{\d\tilde{\rho}_S}{\d t} = \cal{T}^-_L\cal{D}[S_L^{\dagger}(t)](\tilde{\rho}_S) + \cal{T}^+_L\cal{D}[S_L(t)](\tilde{\rho}_S) + \cal{T}^-_R\cal{D}[S_R^{\dagger}(t)](\tilde{\rho}_S) + \cal{T}^+_R\cal{D}[S_R(t)](\tilde{\rho}_S)
\end{equation}
where the superoperator $\cal{D}$ is given by:
\begin{equation}
    \cal{D}[S](\rho) = S\rho S^{\dagger}-\frac{1}{2} \{S^{\dagger}S,\rho \}\ .
\end{equation}
Going back to the Schrödinger picture, we finally get:
\begin{equation}\label{SIme}
    \frac{\d\rho_S}{\d t} = -i[H_S,\rho_S] + \cal{T}^-_L\cal{D}[S_L^{\dagger}](\rho_S) + \cal{T}^+_L\cal{D}[S_L](\rho_S) + \cal{T}^-_R\cal{D}[S_R^{\dagger}](\rho_S) + \cal{T}^+_R\cal{D}[S_R](\rho_S)
\end{equation}

\section{\label{SI2} Perturbative solution to the steady-state master equation}

\subsection{Reduction of the system}

We will now show how to find the steady-state density matrix $\rho \equiv \rho_{ss}$ such that:
\begin{equation}\label{ss}
    -i[H_S,\rho] + \cal{T}^-_L\cal{D}[c_{\ua}^{\dagger}](\rho) + \cal{T}^+_L\cal{D}[c_{\ua}](\rho) + \cal{T}^-_R\cal{D}[c_R^{\dagger}](\rho) + \cal{T}^+_R\cal{D}[c_R](\rho) = 0.
\end{equation}
This equation is a linear system of 64 equations, which seems hard to solve but can be reduced with a bit of effort, and while no usable analytical solution can be found, it is still possible to derive the approximate steady-state using perturbation theory. 

Let us first numerate the basis states:
\begin{equation}
\begin{array}{c}
     |0\ra \equiv |00\ra,\ |1\ra \equiv |0\da\ra,\ |2\ra \equiv |\ua 0\ra,\ |3\ra \equiv |\ua\da\ra,\\ |4\ra \equiv |\da 0\ra,\ |5\ra \equiv |\da\da\ra,\ |6\ra \equiv |20\ra,\ |7\ra \equiv |2\da\ra\ 
\end{array} .
\end{equation}
Writing the steady-state master equation \eqref{ss} in this basis, we notice that a set of 12 equations are independent of the 52 other and can be used to find the diagonal coefficients and four off-diagonal terms: $\rho_{14} = \la 0\da|\rho|\da 0\ra$, $\rho_{41} = \la \da 0|\rho|0\da\ra$, $\rho_{36} = \la \ua\da|\rho|2 0\ra$ and $\rho_{63} = \la 20|\rho|\ua\da\ra$
\begin{equation}
    \left\{ \begin{array}{l}
        (-\cal{T}_L^--\cal{T}_R^-)\rho_{00} + \cal{T}_R^+\rho_{11} + \cal{T}_L^+\rho_{22} = 0 \\
        \cal{T}_R^-\rho_{00} + (-\cal{T}_L^--\cal{T}_R^+)\rho_{11} + \cal{T}_L^+\rho_{33} + \i\gamma\rho_{14} - \i\gamma^*\rho_{41} = 0\\
        \cal{T}_L^-\rho_{00} + (-\cal{T}_L^+-\cal{T}_R^-)\rho_{22} + \cal{T}_R^+\rho_{33} = 0\\
        \cal{T}_L^-\rho_{11} + \cal{T}_R^-\rho_{22} + (-\cal{T}_L^+-\cal{T}_R^+)\rho_{33} + \i\gamma\rho_{36} - \i\gamma^*\rho_{63} = 0\\
        (-\cal{T}_L^--\cal{T}_R^-)\rho_{44} + \cal{T}_R^+\rho_{55} + \cal{T}_L^+\rho_{66} - \i\gamma\rho_{14} + \i\gamma^*\rho_{41} = 0 \\
        \cal{T}_R^-\rho_{44} + (-\cal{T}_L^--\cal{T}_R^+)\rho_{55} + \cal{T}_L^+\rho_{77} = 0\\
        \cal{T}_L^-\rho_{44} + (-\cal{T}_L^+-\cal{T}_R^-)\rho_{66} + \cal{T}_R^+\rho_{77} - \i\gamma\rho_{36} + \i\gamma^*\rho_{63} = 0\\
        \cal{T}_L^-\rho_{55} + \cal{T}_R^-\rho_{66} + (-\cal{T}_L^+-\cal{T}_R^+)\rho_{77} = 0\\
        \Big(-\cal{T}_L^- - \frac{\cal{T}_R^-+\cal{T}_R^+}{2} + \i\Delta\Big)\rho_{14} + \cal{T}_L^+\rho_{36} - \i\gamma^*\rho_{44} + i \gamma^*\rho_{11} = 0\\
        \cal{T}_L^-\rho_{14} + \Big(-\cal{T}_L^+ - \frac{\cal{T}_R^-+\cal{T}_R^+}{2} + \i(\Delta+U)\Big)\rho_{36} - \i\gamma^*\rho_{66} + \i \gamma^*\rho_{33} = 0\\
        \Big(-\cal{T}_L^- - \frac{\cal{T}_R^-+\cal{T}_R^+}{2} - \i\Delta\Big)\rho_{41} + \cal{T}_L^+\rho_{63} + \i\gamma\rho_{44} - i \gamma\rho_{11} = 0\\
        \cal{T}_L^-\rho_{41} + \Big(-\cal{T}_L^+ - \frac{\cal{T}_R^-+\cal{T}_R^+}{2} - \i(\Delta+U)\Big)\rho_{63} + \i\gamma\rho_{66} - \i \gamma\rho_{33} = 0
    \end{array} \right. ,
\end{equation}
where we have set $\Delta = \epsilon_{\da} - \epsilon_R$.

To study this system, we first vectorize the system and define the vector $\vec{\rho}$ such that:
\begin{align}
    \vec{\rho} &\equiv (\rho_{00},\ \rho_{11},\ \rho_{22},\ \rho_{33},\ \rho_{44},\ \rho_{55},\ \rho_{66},\ \rho_{77}, \rho_{14},\ \rho_{41},\ \rho_{36},\ \rho_{63})\\
    &\equiv  (\rho_0,\ \rho_1,\ \rho_2,\ \rho_3,\ \rho_4,\ \rho_5,\ \rho_6,\ \rho_7,\ \rho_{14},\ \rho_{36},\ \rho_{41},\ \rho_{63}).
\end{align}
What we can now easily see by taking the complex conjugate of the system is that it yields the exact same system but with the following solution:
\begin{equation}
    (\rho_0^*,\ \rho_1^*,\ \rho_2^*,\ \rho_3^*,\ \rho_0^*,\ \rho_1^*,\ \rho_2^*,\ \rho_3^*,\ \rho_{41}^*,\ \rho_{63}^*,\ \rho_{14}^*,\ \rho_{36}^*).
\end{equation}
This checks out with the hermiticity of the density matrix, which is a good confirmation of the correctness of our calculus. Using this property, we obtain the following relations:
\begin{equation}
    \rho_i = \rho_i^*,\ \rho_{14} = \rho_{41}^*, \text{ and } \rho_{36} = \rho_{63}^*.
\end{equation}
This allows us to discard the last two equations from this system, meaning that we can discard the off-diagonal lower terms $\rho_{41}$ and $\rho_{36}$, which we will get from the solved upper terms.

With this reduction, we shall now write this system in matrix format. We start by setting $\gamma \equiv re^{\i\phi}$. Then we choose to separate the real and imaginary parts of the rotated off-diagonal terms by setting $\rho_{\da} \equiv \rho_{\da}^R + \i\rho_{\da}^I \equiv \i e^{\i\phi}\rho_{14}$ and $\rho_{\ua} \equiv \rho_{\ua}^R + \i\rho_{\ua}^I \equiv \i e^{\i\phi}\rho_{36}$. And finally, we redefine the vector $\vec{\rho}$ by deleting the last two redundant components, such that:
\begin{equation}
    \vec{\rho} \equiv (\rho_0,\ \rho_1,\ \rho_2,\ \rho_3,\ \rho_4,\ \rho_5,\ \rho_6,\ \rho_7,\ \rho_{\da},\ \rho_{\ua}).
\end{equation}
Starting from a system with 12 complex parameters, we have now managed to reduce it down to a system with 12 real parameters or 8 real parameters and 2 complex.

This system is too complicated to be solved exactly analytically so we will use perturbation theory in order to find an approximate solution. The small parameter we should use as a perturbation shall the magnitude of the tunnelling coefficient $r = |\gamma|$, which should be an order of magnitude lower than all the energy scales present in this problem. 

We can now write the system in matrix format by defining $\Lambda \equiv \Lambda_0 + \i r \Lambda_1$ such as $\Lambda \vec{\rho} = 0$, with the block matrix
\begin{equation}
    \Lambda_0 = \rm{diag}(A,\ A,\ B),
\end{equation}
filled by
\begin{equation}
    A \equiv \begin{pmatrix}
        -(\cal{T}_L^-+\cal{T}_R^-) & \cal{T}_R^+ & \cal{T}_L^+ & 0 \\
        \cal{T}_R^- & -(\cal{T}_L^- + \cal{T}_R^+) & 0 & \cal{T}_L^+ \\
        \cal{T}_L^- & 0 & -(\cal{T}_R^-+\cal{T}_L^+) & \cal{T}_R^+ \\
        0 & \cal{T}_L^- & \cal{T}_R^- & -(\cal{T}_L^++\cal{T}_R^+)
    \end{pmatrix},
\end{equation}
and
\begin{equation}
    B = \begin{pmatrix}
        -\cal{T}_L^- - \frac{\cal{T}_R^-+\cal{T}_L^+}{2} + \i\Delta & \cal{T}_L^+ \\
        \cal{T}_L^- & -\cal{T}_L^+ - \frac{\cal{T}_R^-+\cal{T}_L^+}{2} + \i(\Delta+U)
    \end{pmatrix}
\end{equation}
and with the perturbation interaction matrix $\Lambda_1$ such as
\begin{equation}
    \Lambda_1\vec{\rho} = (0,\ \rho_{\da}^R,\ 0,\ \rho_{\ua}^R,\ -\rho_{\da}^R,\ 0,\ -\rho_{\ua}^R,\ 0,\ \rho_4-\rho_1,\ \rho_6-\rho_3).
\end{equation}

\subsection{Preliminary calculus}

Let us first start by diagonalizing the matrix $\Lambda_0$, meaning that we should diagonalize $A$ and $B$.

The diagonalisation of $A$ is straightforward and yields the following eigenvalues
\begin{equation}
    \left\{ \begin{array}{l}
        \lambda_0 = 0\\ 
        \lambda_1 = -\cal{T}_L^--\cal{T}_L^+\\
        \lambda_2 = -\cal{T}_R^- - \cal{T}_R^+\\ 
        \lambda_3 = -\cal{T}_L^- - \cal{T}_L^+ - \cal{T}_R^- - \cal{T}_R^+
    \end{array} \right. ,
\end{equation}
with the corresponding eigenvectors
\begin{equation}
\left\{ \begin{array}{l}
    v_0 = (\cal{T}_L^+\cal{T}_R^+,\ \cal{T}_L^+\cal{T}_R^-,\ \cal{T}_L^-\cal{T}_R^+,\ \cal{T}_L^-\cal{T}_R^-)\\
    v_1 = (-\cal{T}_R^+,\ -\cal{T}_R^-,\ \cal{T}_R^+,\ \cal{T}_R^-)\\
    v_2 = (-\cal{T}_L^+,\ \cal{T}_L^+,\ -\cal{T}_L^-,\ \cal{T}_L^-)\\
    v_3 = (1,\ -1,\ -1,\ 1)
\end{array} \right. .
\end{equation}
As we can see, we can have non-unicity issues whenever $\lambda_1$, $\lambda_2$ or $\lambda_3$ vanishes. In the following, we will suppose that the tunnelling parameters $\cal{T}_L^+$, $\cal{T}_L^-$, $\cal{T}_R^+$ and $\cal{T}_R^-$ are chosen such that $\lambda_0$ is the only null eigenvalue.

The diagonalisation of $B$ is also trivial and yields the eigenvalues
\begin{equation}
    \left\{ \begin{array}{l}
        \lambda_- = -\frac{\cal{T}_L^- + \cal{T}_L^++ \cal{T}_R^- + \cal{T}_R^+}{2} + \i(\Delta + \frac{U}{2}) - \frac{1}{2} \sqrt{4\cal{T}_L^-\cal{T}_L^+ + (\cal{T}_L^+-\cal{T}_L^--\i U)^2}\\ 
        \lambda_+ = -\frac{\cal{T}_L^- + \cal{T}_L^++ \cal{T}_R^- + \cal{T}_R^+}{2} + \i(\Delta + \frac{U}{2}) + \frac{1}{2} \sqrt{4\cal{T}_L^-\cal{T}_L^+ + (\cal{T}_L^+-\cal{T}_L^--\i U)^2}
    \end{array} \right. ,
\end{equation}
where the square root of the complex number is chosen such that its real part is positive, and with the following eigenvectors:
\begin{equation}
\left\{ \begin{array}{l}
    v_- = \Big(\cal{T}_L^+-\cal{T}_L^--\i U - \sqrt{4\cal{T}_L^-\cal{T}_L^+ + (\cal{T}_L^+-\cal{T}_L^--\i U)^2},\ 2\cal{T}_L^-\Big)\\
    v_+ = \Big(\cal{T}_L^+-\cal{T}_L^--\i U + \sqrt{4\cal{T}_L^-\cal{T}_L^+ + (\cal{T}_L^+-\cal{T}_L^--\i U)^2},\ 2\cal{T}_L^-\Big)
\end{array} \right. .
\end{equation}
We will also need its inverse, which reads:
\begin{equation}
    B^{-1} = \frac{1}{\rm{det}\,B} \begin{pmatrix}
        -\cal{T}_L^+-\frac{\cal{T}_R^++\cal{T}_R^-}{2}+\i(\Delta+U) & -\cal{T}_L^+ \\
        -\cal{T}_L^- & -\cal{T}_L^--\frac{\cal{T}_R^++\cal{T}_R^-}{2}+\i\Delta
    \end{pmatrix}
\end{equation}
where 
\begin{equation}
    \rm{det}\,B = \bigg(\cal{T}_L^-+\frac{\cal{T}_R^-+\cal{T}_R^+}{2}-\i\Delta\bigg)\bigg(\cal{T}_L^++\frac{\cal{T}_R^-+\cal{T}_R^+}{2}-\i(\Delta+U)\bigg) - \cal{T}_L^-\cal{T}_L^+\ .
\end{equation}

\subsection{Perturbation theory: kernel approach}

Let us look for a perturbed solution $\vec{\rho}$ in the kernel of $\Lambda$, meaning that we are searching $\vec{\rho}$ as an expanded form $\vec{\rho} = \vec{\rho}^{\,(0)} + r\vec{\rho}^{\,(1)} + r^2\vec{\rho}^{\,(2)} + o(r^2)$. Expanding the equation $\Lambda \vec{\rho} = 0$, and identifying each order leads to:
\begin{equation}
    (\Lambda_0 + r\Lambda_1)(\vec{\rho}^{\,(0)} + r\vec{\rho}^{\,(1)} + r^2\vec{\rho}^{\,(2)} + o(r^2)) = 0 \implies \left\{ \begin{array}{l}
        \Lambda_0\vec{\rho}^{\,(0)} = 0\\
        \Lambda_0\vec{\rho}^{\,(1)} = -\Lambda_1\vec{\rho}^{\,(0)}  \\
        \Lambda_0\vec{\rho}^{\,(2)} = -\Lambda_1\vec{\rho}^{\,(1)}  
    \end{array}\right. .
\end{equation}
More generally, it is straightforward to see that we can obtain the $(i+1)$-th order from the $i$-th order by solving the system $\Lambda_0\vec{\rho}^{\,(i+1)} = -\Lambda_1\vec{\rho}^{\,(i)}$, with $\vec{\rho}^{\,(0)}$ in the kernel of $\Lambda_0$.

From the initial diagonalization, we immediately obtain the kernel of $\Lambda_0$, which can be written as:
\begin{equation}
    \vec{\rho}^{\,(0)} = \lambda\, v_0\otimes 0\otimes 0 + \mu\,  0\otimes v_0\otimes 0
\end{equation}
where $\lambda$ and $\mu$ are two real parameters.

Using the properties of the density matrix, we can eliminate one parameter. Because of the conservation of the probabilities, $\rho$ must have a unit trace. This condition leads to:
\begin{equation}
    \rm{Tr}\,\rho^{(0)} = 1 \implies \lambda = \alpha - \mu\ ,
\end{equation}
where we have set $1/\alpha \equiv   (\cal{T}_L^++\cal{T}_L^-)(\cal{T}_R^++\cal{T}_R^-)$. Rewritting $\mu \equiv \alpha\mu$, we thus obtain
\begin{equation}
    \vec{\rho}^{\,(0)} = \alpha\Big((1-\mu)\, v_0\otimes 0\otimes 0 + \mu\,  0\otimes v_0\otimes 0\Big)\ .
\end{equation}
Moreover, the positivity of the density matrix imposes $0 \le \mu \le 1$.

Apparently here, we now have a problem because $\rho^{\,(0)}$ cannot be determined uniquely because of this free $\lambda$ parameter. This means that the steady-state solution to the master equation will depend on its initial condition! Fortunately, in this case, we can deduce the final state corresponding to the initial state quite easily. Indeed, we point out that $\rho^{(0)}$ taken as a density matrix should correspond to a solution of the Liouville-Von Neumann equation with $\gamma = 0$ encoded in the Hamiltonian:
\begin{equation}
    H = \epsilon_{\ua} n_{\ua} + \epsilon_{\da} n_{\da} + \epsilon_R n_R + J\, n_{\da}n_R + U\,n_{\ua}n_{\da} + H_L + H_R + H_{SL} + H_{SR} \ .
\end{equation}
Looking at this operator, it should be clear that we have the commutation relation $[H,\, n_{\da}] = 0$. Therefore, $n_{\da}$ is a conserved quantity during the time evolution of the whole system. This property is transferring directly to $\vec{\rho}^{\,(0)}$, which should therefore verify
\begin{equation}
    \rm{Tr}\, \rho^{(0)}n_{\da} = \rm{Tr}\, \rho(0)n_{\da} \eq \mu = \la n_{\da}(0)\ra\ .
\end{equation}
This initial information gives us the value of $\mu$ that corresponds to the initial occupation number of the down spin energy level of the left qubit. It should be clear that the two extremal values $\mu = 0$ and $\mu = 1$ will be the most interesting.

We can now move to finding the first order $\rho^{(1)}$. We need to solve $\Lambda_0\vec{\rho}^{\,(1)} = \Lambda_1\vec{\rho}^{\,(0)}$. The right-hand side reads:
\begin{align}
    \Lambda_1\vec{\rho}^{\,(0)} &= (0,\ 0,\ 0,\ 0,\ 0,\ 0,\ 0,\ 0,\ \rho_4^{(0)}-\rho_1^{(0)},\ \rho_6^{(0)}-\rho_3^{(0)}) \\
    &= \Big(0,\ 0,\ 0,\ 0,\ 0,\ 0,\ 0,\ 0,\ \alpha\cal{T}_L^+(\mu\cal{T}_R^+-(1-\mu)\cal{T}_R^-),\ \alpha\cal{T}_L^-(\mu\cal{T}_R^+-(1-\mu)\cal{T}_R^-)\Big) \ .
\end{align}
Decomposing $\vec{\rho}^{\,(1)}$ as $\vec{\rho}^{(1)} = \vec{\rho}^{\,(1)}_+\otimes\vec{\rho}^{\,(1)}_-\otimes\begin{pmatrix}
        \rho^{(1)}_{\da} \\ \rho^{(1)}_{\ua}
    \end{pmatrix}$, we immediately obtain the following three equations from the block diagonal expression of $\Lambda_0$:
\begin{equation}
    A\vec{\rho}^{\,(1)}_+ = 0,\ A\vec{\rho}^{\,(1)}_- = 0,\ \text{and}\ B\begin{pmatrix}
        \rho^{(1)}_{\da} \\ \rho^{(1)}_{\ua}
    \end{pmatrix} = -\alpha(\mu\cal{T}_R^+-(1-\mu)\cal{T}_R^-)\begin{pmatrix}
        \cal{T}_L^+ \\ \cal{T}_L^-
    \end{pmatrix},\
\end{equation}
Let us focus first on the last system which gives a unique solution given the inversibility of $B$:
\begin{equation}
    \left\{ \begin{array}{l}
        \rho_{\da}^{(1)} = \frac{\alpha\cal{T}_L^+}{\rm{det}\,B}(\mu\cal{T}_R^+-(1-\mu)\cal{T}_R^-)\Big(\cal{T}_L^+ + \cal{T}_L^- + \frac{\cal{T}_R^-+\cal{T}_R^+}{2} -\i(\Delta + U)\Big)
         \\ \rho_{\ua}^{(1)} = \frac{\alpha\cal{T}_L^-}{\rm{det}\,B}(\mu\cal{T}_R^+-(1-\mu)\cal{T}_R^-)\Big(\cal{T}_L^+ + \cal{T}_L^- + \frac{\cal{T}_R^-+\cal{T}_R^+}{2} -\i\Delta\Big)
    \end{array}\right.\ .
\end{equation}
The general solution to the first-order equation thus leads to the unique off-diagonal terms we just found plus additional diagonal terms in the kernel of $\Lambda_0$. Having non-zero diagonal terms would give first-order corrections to the unit-trace condition of the density matrix and to the mean value of the number operators. Such solutions would therefore be less physical so we should discard them to keep the normalisation of the density matrix intact. 

Unfortunately, this method cannot be used to evaluate the higher orders. Indeed, we can show that the system $\Lambda_0\vec{\rho}^{\,(2)} = -\Lambda_1\vec{\rho}^{\,(1)}$ has no solution because $\Re(\rho_{\da}^{(1)}) \neq \Re(\rho_{\ua}^{(1)})$. We shall thus limit the analysis to the first order we have just derived.

Collecting every piece together thus leads to a steady-state solution $\rho^{\mu}$ calculated up to first order in $\gamma$ with null coefficients everywhere except:
\begin{equation}
    \left\{\begin{array}{l}
        \rho_{00}^{\mu} = \alpha(1-\mu)\cal{T}_L^+\cal{T}_R^+\\
        \rho_{11}^{\mu} = \alpha(1-\mu)\cal{T}_L^+\cal{T}_R^-\\
        \rho_{22}^{\mu} = \alpha(1-\mu)\cal{T}_L^-\cal{T}_R^+\\
        \rho_{33}^{\mu} = \alpha(1-\mu)\cal{T}_L^-\cal{T}_R^-\\
        \rho_{44}^{\mu} = \alpha\mu\cal{T}_L^+\cal{T}_R^+\\
        \rho_{55}^{\mu} = \alpha\mu\cal{T}_L^+\cal{T}_R^-\\
        \rho_{66}^{\mu} = \alpha\mu\cal{T}_L^-\cal{T}_R^+\\
        \rho_{77}^{\mu} = \alpha\mu\cal{T}_L^-\cal{T}_R^-\\
        \rho_{14}^{\mu} = -\i\gamma^*\frac{\alpha\cal{T}_L^+}{\rm{det}\,B}(\mu\cal{T}_R^+-(1-\mu)\cal{T}_R^-)\Big(\cal{T}_L^+ + \cal{T}_L^- + \frac{\cal{T}_R^-+\cal{T}_R^+}{2} -\i(\Delta + U)\Big) \\
        \rho_{36}^{\mu} = -\i\gamma^*\frac{\alpha\cal{T}_L^-}{\rm{det}\,B}(\mu\cal{T}_R^+-(1-\mu)\cal{T}_R^-)\Big(\cal{T}_L^+ + \cal{T}_L^- + \frac{\cal{T}_R^-+\cal{T}_R^+}{2} -\i\Delta\Big) \\
        \rho_{41}^{\mu} = \i\gamma\frac{\alpha\cal{T}_L^+}{\rm{det}\,B^*}(\mu\cal{T}_R^+-(1-\mu)\cal{T}_R^-)\Big(\cal{T}_L^+ + \cal{T}_L^- + \frac{\cal{T}_R^-+\cal{T}_R^+}{2} +\i(\Delta + U)\Big) \\
        \rho_{63}^{\mu} = \i\gamma\frac{\alpha\cal{T}_L^-}{\rm{det}\,B^*}(\mu\cal{T}_R^+-(1-\mu)\cal{T}_R^-)\Big(\cal{T}_L^+ + \cal{T}_L^- + \frac{\cal{T}_R^-+\cal{T}_R^+}{2} +\i\Delta\Big)
    \end{array}\right. \ .
\end{equation}

\section{\label{SI3} Energy for other measurement protocols}

\subsection{Unselective measurement of the spin of the left quantum dot}

Let us now study the case where the environment measures the spin of the left quantum dot at frequent times. The spin operator $S$ reads
\begin{equation}
    S = n_{\ua}-n_{\da} = 1\times n_{\ua}(1-n_{\da})-1\times n_{\da}(1-n_{\ua}) + 0\times(n_{\ua}n_{\da}+(1-n_{\ua})(1-n_{\da}))\ .
\end{equation}
The measurement can give one of the three values 0, 1 or -1 and at time $\tau^+$, the unselected measured state read:
\begin{equation}
    \rho(\tau^+) = (n_{\ua}n_{\da}+(1-n_{\ua})(1-n_{\da})\rho_{\ua}n_{\da}+(1-n_{\ua})(1-n_{\da}) + n_{\ua}(1-n_{\da})\rho n_{\ua}(1-n_{\da}) + n_{\da}(1-n_{\ua})\rho n_{\da}(1-n_{\ua}).
\end{equation}
Once again, we observe that the off-diagonal terms do not contribute to any of the projected states because they involve the tunnelling of one electron between the two sites, leading to 
\begin{equation}
    \rho(\tau^+) = \sum_{i=0}^7 \rho_{ii}|i\ra\la i|.
\end{equation}
The calculation for the average energy increase is then straightforward and identical to the case treated in the main text, and we find the exact previous result $\la \Delta E \ra = -\rm{Tr}[T\rho]$.

\subsection{Unselective measurement of the charge of the left quantum dot}

The very same results hold when measuring the charge of the left quantum dot. In this case, the charge operator $Q$ reads
\begin{equation}
    Q = n_{\da} + n_{\ua} = 0\times (1-n_{\da})(1-n_{\ua}) + 1\times (n_{\ua}(1-n_{\da})+n_{\da}(1-n_{\ua})) + 2\times n_{\da}n_{\ua} \ ,
\end{equation}
which yields the following measured state:
\begin{equation}
    \rho(\tau^+) = (1-n_{\da})(1-n_{\ua})\rho(1-n_{\da})(1-n_{\ua}) + (n_{\ua}(1-n_{\da})+n_{\da}(1-n_{\ua}))\rho (n_{\ua}(1-n_{\da})+n_{\da}(1-n_{\ua})) + n_{\da}n_{\ua}\rho n_{\da}n_{\ua}.
\end{equation}
And again, the off-diagonal terms do not contribute, which leads again to the previous result $\la \Delta E \ra = -\rm{Tr}[T\rho]$.

\subsection{Unselective measurement of the total charge}

As a counterexample, let us now consider an observable that acts on both quantum dots, namely the total charge $C = n_{\ua}+n_{\da}+n_R$, which decomposes as:
\begin{equation}
    C = \begin{array}{l}
         \ \ \, 0\times (1-n_{\ua})(1-n_{\da})(1-n_R)\\
    + 1\times [n_{\ua}(1-n_{\da})(1-n_R)+(1-n_{\ua})n_{\da}(1-n_R)+(1-n_{\ua})(1-n_{\da})n_R]\\
    + 2\times[n_{\ua}n_{\da}(1-n_R)+n_{\ua}(1-n_{\da})n_R+(1-n_{\ua})n_{\da}n_R]\\
    + 3\times n_{\ua}n_{\da}n_R
    \end{array}\ .
\end{equation}
Just as in the previous section, we can calculate the measured state and we find the very simple result $\rho(\tau^+) = \rho$. This time, off-diagonal terms are present in the projected state, which is identical to the thermalized state! Therefore, the average energy increment $\Delta E$ vanishes! In this case, we cannot expect the measurement to energize the system on average because it does not produce the necessary separation and leaves the state unchanged.

\section{\label{SI4} The case of selective quantum measurements}

\subsection{Selective measurement of the occupation of the right quantum dot}

Let us initialize our engine at $t_0=0$ in some state $\rho(0)$ such that $\mu = \la n_{\da}(0) \ra$. After completing a first thermalization process, a selective partial projective measurement of the system is performed by the environment at time $\tau$. This measurement projects the system from the steady state $\rho(\tau^-) = \rho^{\mu} \equiv \rho$ to a projected state $\rho^i(\tau^+)$ that depends on the measurement outcome $i$.

The measurement of the occupation of the right quantum dot $n_R$ can only give one of the two values 1 or 0, which code either the presence or the absence of one electron on the right-hand QD. The two possible projected states read:
\begin{equation}
    \left\{\begin{array}{l}
        \rho^{1}(\tau^+) = \frac{n_R\rho_R}{\rm{Tr}[n_R\rho]}\\
        \rho^{0}(\tau^+) = \frac{(1-n_R)\rho(1-n_R)}{\rm{Tr}[(1-n_R)\rho]}
    \end{array}\right.\ .
\end{equation}
The same argument holds once again: the off-diagonal terms do not contribute in the projected states because they encode the tunnelling of one electron from one site to the next. Hence we calculate
\begin{equation}
    \left\{\begin{array}{l}
    \rho^{1}(\tau^+) = \frac{\rho_{11}|1\ra\la 1|+\rho_{33}|3\ra\la 3| + \rho_{55}|5\ra\la 5|+\rho_{77}|7\ra\la 7|}{\rho_{11} + \rho_{33} + \rho_{55} + \rho_{77}}\\
    \rho^0(\tau^+) = \frac{\rho_{00}|0\ra\la 0|+\rho_{22}|2\ra\la 2| + \rho_{44}|4\ra\la 4|+\rho_{66}|6\ra\la 6|}{\rho_{00} + \rho_{22} + \rho_{44} + \rho_{66}} \end{array}\right.\ .
\end{equation}
The associated probabilities to obtain these two states are given by
\begin{equation}
    \left\{\begin{array}{l}
    p^1(\tau^+) \equiv \rm{Tr}[n_R\rho] = \rho_{11} + \rho_{33} + \rho_{55} + \rho_{77}\\
    p^0(\tau^+) \equiv \rm{Tr}[(1-n_R)\rho] = \rho_{00} + \rho_{22} + \rho_{44} + \rho_{66}
    \end{array}\right.\ .
\end{equation}
From these quantities, we can now calculate the impact of the measurement in terms of energy and entropy. For each of the two measurement outcomes, the energy of the system respectively changes by an amount $\Delta E^1$ and $\Delta E^0$ such that
\begin{equation}
    \Delta E^i = \rm{Tr}[H_S\rho^i(\tau^+)] - \rm{Tr}[H_S\rho(\tau^-)]\ .
\end{equation}
We calculate 
\begin{equation}
    \rm{Tr}[H_S\rho(\tau^-)] = \sum_{i=0}^7 H_{ii}\rho_{ii} + \gamma\rho_{14} + \gamma\rho_{36} + \gamma^*\rho_{41} + \gamma^*\rho_{63}
\end{equation}
then
\begin{equation}
\left\{\begin{array}{l}
    \rm{Tr}[H_S\rho^1(\tau^+)] = \frac{H_{11}\rho_{11}+H_{33}\rho_{33} + H_{55}\rho_{55}+H_{77}\rho_{77}}{\rho_{11} + \rho_{33} + \rho_{55} + \rho_{77}}\\
    \rm{Tr}[H_S\rho^0(\tau^+)] = \frac{H_{00}\rho_{00}+H_{22}\rho_{22} + H_{44}\rho_{44}+H_{66}\rho_{66}}{\rho_{00} + \rho_{22} + \rho_{44} + \rho_{66}} 
 \end{array}\right.\ .
\end{equation}
This allows us to derive the expected value of the energy increment $\overline{\Delta E} \equiv p^1\Delta E^1 + p^0\Delta E^0$. We emphasize the difference here between the average value of a quantum observable over all possible quantum trajectories written between angles $\la \cdot \ra$ and the expectation value of a random variable coming from the measurement of the system, written with an overline $\overline{\cdot}$. We directly notice that the terms coming from the diagonal components all vanish, leaving only
\begin{equation}
    \overline{\Delta E} = - \gamma\rho_{14} - \gamma\rho_{36} - \gamma^*\rho_{41} - \gamma^*\rho_{63} = -2\Re[\gamma(\rho_{14}+\rho_{36})] = -\rm{Tr}[T\rho]\ .
\end{equation}
Therefore, the energy increment is identical to the case of unselective quantum measurement.

Now, for the next cycle, we need to obtain is the particle number with spin $\da$ in the left quantum dot so that we can calculate the next thermalized state. Fortunately, because the two possible projected states are diagonal, we obtain directly
\begin{equation}
\left\{\begin{array}{l}
    \rm{Tr}[n_{\da}\rho^1(\tau^+)] = \frac{\rho_{55}+\rho_{77}}{\rho_{11} + \rho_{33} + \rho_{55} + \rho_{77}} = \mu \\
    \rm{Tr}[n_{\da}\rho^0(\tau^+)] = \frac{\rho_{44}+\rho_{66}}{\rho_{00} + \rho_{22} + \rho_{44} + \rho_{66}} = \mu
\end{array}\right.\ .
\end{equation}
This means that the occupation number with spin $\da$ remains unchanged after both the thermalizing and the measurement processes for both outcomes of the measurement! Therefore, the second cycle starts again with $\la n_{\da}(\tau^+) \ra = \mu$, so it yields the same thermalized state just before the second measurement as in the previous cycle. This means that $\rho(2\tau^-) = \rho(\tau^-) = \rho$ and thus $\rho^i(2\tau^+) = \rho^i(\tau^+)$. 

An instant recursion then allows us to explicitly obtain the state of the system at the end-point of each cycle $n$:  
\begin{equation}
\left\{\begin{array}{l}
    \rho(n\tau^-) = \rho\\
    \rho^1(n\tau^+) = \frac{\rho_{11}|1\ra\la 1|+\rho_{33}|3\ra\la 3| + \rho_{55}|5\ra\la 5|+\rho_{77}|7\ra\la 7|}{\rho_{11} + \rho_{33} + \rho_{55} + \rho_{77}} \\
    \rho^0(n\tau^+) =  \frac{\rho_{00}|0\ra\la 0|+\rho_{22}|2\ra\la 2| + \rho_{44}|4\ra\la 4|+\rho_{66}|6\ra\la 6|}{\rho_{00} + \rho_{22} + \rho_{44} + \rho_{66}}
\end{array}\right.\ .
\end{equation}

\subsection{Selective measurement of the spin of the left quantum dot}

Let us now study the case where the environment operates a selective measurement the spin of the left quantum dot at frequent times. At time $\tau^+$, the three possible projected states read:
\begin{equation}
    \left\{ \begin{array}{l}
         \rho^0(\tau^+) = \frac{n_{\ua}n_{\da}+(1-n_{\ua})(1-n_{\da})\rho_{\ua}n_{\da}+(1-n_{\ua})(1-n_{\da})}{\rm{Tr}n_{\ua}n_{\da}+(1-n_{\ua})(1-n_{\da})\rho]} = \frac{\rho_{00}|0\ra\la0| + \rho_{11}|1\ra\la 1| + \rho_{66}|6\ra\la6| + \rho_{77}|7\ra\la 7|}{\rho_{00} + \rho_{11} + \rho_{66} + \rho_{77}} \\
         \rho^1(\tau^+) = \frac{n_{\ua}(1-n_{\da})\rho n_{\ua}(1-n_{\da})}{\rm{Tr}[n_{\ua}(1-n_{\da})\rho]} =\frac{\rho_{22}|2\ra\la2| + \rho_{33}|3\ra\la 3|}{\rho_{22} + \rho_{33}}\\
         \rho^{-1}(\tau^+) = \frac{n_{\da}(1-n_{\ua})\rho n_{\da}(1-n_{\ua})}{\rm{Tr}[n_{\da}(1-n_{\ua})\rho]} = \frac{\rho_{44}|4\ra\la4| + \rho_{55}|5\ra\la 5|}{\rho_{44} + \rho_{55}}
    \end{array} \right.\ ,
\end{equation}
with the corresponding probabilities
\begin{equation}
    \left\{ \begin{array}{l}
         p^0(\tau^+) =\rm{Tr}[(2n_{\da}n_{\ua}-n_{\ua}-n_{\da})\rho] = \rho_{00} + \rho_{11} + \rho_{66} + \rho_{77} \\
         p^1(\tau^+) = \rm{Tr}[n_{\ua}(1-n_{\da})\rho] = \rho_{22} + \rho_{33}\\
         p^{-1}(\tau^+) = \rm{Tr}[n_{\da}(1-n_{\ua})\rho] = \rho_{44} + \rho_{55}
    \end{array} \right.\ ,
\end{equation}
Once again, we observe that the off-diagonal terms do not contribute to any of the projected states because they involve the tunnelling of one electron between the two sites.

Therefore, we can write $\overline{\Delta E} = p^0\Delta E^0 + p^1\Delta E^1 + p^{-1}\Delta E^{-1}$ with $\Delta E^i = \rm{Tr}[H_S\rho^i(\tau^+)] - \rm{Tr}[H_S\rho]$. Given that we have
\begin{equation}
    \left\{ \begin{array}{l}
         \rm{Tr}[H_S\rho^0(\tau^+)] = \frac{H_{00}\rho_{00} + H_{11}\rho_{11} + H_{66}\rho_{66} + H_{77}\rho_{77}}{\rho_{00} + \rho_{11} + \rho_{66} + \rho_{77}} \\
         \rm{Tr}[H_S\rho^1(\tau^+)] = \frac{H_{22}\rho_{22} + H_{33}\rho_{33}}{\rho_{22} + \rho_{33}}\\
         \rm{Tr}[H_S\rho^{-1}(\tau^+)] = \frac{H_{44}\rho_{44} + H_{55}\rho_{55}}{\rho_{44} + \rho_{55}}
    \end{array} \right.\ ,
\end{equation}
we find the exact previous result $\overline{\Delta E} = -\rm{Tr}[T\rho]$.

Compared to the previous in the main text where the parameter $\mu$ remains unchanged for each cycle, in this case, the measurement has an impact on the spin $\downarrow$ population on the left quantum dot, which can lead to three different values for the initial condition of the next cycle:
\begin{equation}
    \left\{ \begin{array}{l}
        \mu^{0}(\tau^+) = \rm{Tr}[n_{\da}\rho^0(\tau^+)] = \frac{\rho_{66}+\rho_{77}}{\rho_{00} + \rho_{11} + \rho_{66} + \rho_{77}} = \frac{\mu\cal{T}_L^-}{(1-\mu)\cal{T}_L^++\mu\cal{T}_L^-}\\
        \mu^{1}(\tau^+) = \rm{Tr}[n_{\da}\rho^1(\tau^+)] = 0\\
        \mu^{-1}(\tau^+) = \rm{Tr}[n_{\da}\rho^{-1}(\tau^+)] = 1\\
    \end{array}\right. .
\end{equation}
Nonetheless, its average value $\overline{\mu} = p^0\mu^{0} + p^1\mu^1 + p^{-1}\mu^{-1}$ remains unchanged:
\begin{equation}
    \overline{\mu} = \frac{\mu\cal{T}_L^-}{(1-\mu)\cal{T}_L^++\mu\cal{T}_L^-}\times\Big[\alpha(\cal{T}_R^++ \cal{T}_R^-)((1-\mu)\cal{T}_L^++\mu\cal{T}_L^-)\Big] + 1\times \Big[\alpha\mu\cal{T}_L^+(\cal{T}_R^++ \cal{T}_R^-)\Big] = \mu \ .
\end{equation}
This means that the initial value at the beginning of each thermalizing stroke describes a stochastic sequence $\mu_n$ such that:
\begin{equation}
    \mu_{n+1} = \left\{ \begin{array}{lll}
        \frac{\mu^n\cal{T}_L^-}{(1-\mu^n)\cal{T}_L^++\mu^n\cal{T}_L^-} & \text{ with probability} & p^0(n\tau^+)\\
        0 & \text{ with probability}&  p^1(n\tau^+) \\
        1 & \text{ with probability}& p^{-1}(n\tau^+) \\
    \end{array}\right. .
\end{equation}
It is quite straightforward to see that after some time, starting from a random initial value $\mu^0$, the sequence $\mu^n$ can only take one of the two values $0$ or $1$. Indeed, let us consider the first cycle, such as $\mu_{n_0} = 0$ or $\mu_{n_0} = 1$. In the first case $\mu_{n_0} = 0$, we have:
\begin{equation}
    \mu_{n_0+1} = \left\{ \begin{array}{lll}
        0 & \text{ with probability}&  p^1(n\tau^+)+p^0(n\tau^+)=1 \\
        1 & \text{ with probability}& p^{-1}(n\tau^+)=0 \\
    \end{array}\right. .
\end{equation}
Recursively, we then show that the sequence $\mu_n$ stabilizes at 0. And in the second case $\mu_{n_0} = 1$, we have:
\begin{equation}
    \mu_{n_1+1} = \left\{ \begin{array}{lll}
        0 & \text{ with probability}&  p^1(n\tau^+)=0 \\
        1 & \text{ with probability}&  p^{-1}(n\tau^+)+p^0(n\tau^+)=1 \\
    \end{array}\right. .
\end{equation}
A trivial recursion hence shows the stabilization at either $\mu_n = 0$ or $\mu_n = 1$. This shows that when starting with $\mu_0 = 0$, the value of $\mu$ remains constant and is stabilized by the measurements, which guarantees that the energy increment $\overline{ \Delta E}$ remains positive, i.e. that energy can potentially be extracted during each cycle.

\subsection{Selective measurement of the charge of the left quantum dot}

The very same results hold when measuring the charge of the left quantum dot. It yields the following projected states:
\begin{equation}
    \left\{ \begin{array}{l}
         \rho^0(\tau^+) = \frac{(1-n_{\da})(1-n_{\ua})\rho(1-n_{\da})(1-n_{\ua})}{\rm{Tr}[(1-n_{\da})(1-n_{\ua})\rho]} = \frac{\rho_{00}|0\ra\la 0| + \rho_{11}|1\ra\la 1|}{\rho_{00} + \rho_{11}} \\
         \rho^1(\tau^+) = \frac{(n_{\ua}(1-n_{\da})+n_{\da}(1-n_{\ua}))\rho (n_{\ua}(1-n_{\da})+n_{\da}(1-n_{\ua}))}{\rm{Tr}[(n_{\ua}(1-n_{\da})+n_{\da}(1-n_{\ua}))\rho]} = \frac{\rho_{22}|2\ra\la 2| + \rho_{33}|3\ra\la 3| + \rho_{44}|4\ra\la 4| + \rho_{55}|5\ra\la 5|}{\rho_{22} + \rho_{33} + \rho_{44} + \rho_{55}}\\
         \rho^{2}(\tau^+) = \frac{n_{\da}n_{\ua}\rho n_{\da}n_{\ua}}{\rm{Tr}[n_{\da}n_{\ua}\rho]} = \frac{\rho_{66}|6\ra\la 6| + \rho_{77}|7\ra\la 7|}{\rho_{66} + \rho_{77}}
    \end{array} \right.\ ,
\end{equation}
with the associated probabilities
\begin{equation}
    \left\{ \begin{array}{l}
         p^0(\tau^+) =\rm{Tr}[(1-n_{\da})(1-n_{\ua})\rho] = \rho_{00} + \rho_{11} \\
         p^1(\tau^+) = \rm{Tr}[(n_{\ua}(1-n_{\da})+n_{\da}(1-n_{\ua}))\rho] = \rho_{22} + \rho_{33} + \rho_{44} + \rho_{55}\\
         p^2(\tau^+) = \rm{Tr}[n_{\da}n_{\ua}\rho] = \rho_{66} + \rho_{77}
    \end{array} \right.\ .
\end{equation}
Once again, the off-diagonal terms do not contribute, which leads to the following mean energies for each outcome:
\begin{equation}
    \left\{ \begin{array}{l}
         \rm{Tr}[H_S\rho^0(\tau^+)] = \frac{H_{00}\rho_{00}+ \rho_{11}H_{11}}{\rho_{00}+\rho_{11}}\\
         \rm{Tr}[H_S\rho^1(\tau^+)] = \frac{H_{22}\rho_{22}+ \rho_{33}H_{33} + H_{44}\rho_{44}+ \rho_{55}H_{55}}{\rho_{22} + \rho_{33} + \rho_{44} + \rho_{55}}\\
         \rm{Tr}[H_S\rho^2(\tau^+)] = \frac{H_{66}\rho_{66}+ \rho_{77}H_{77}}{\rho_{66}+\rho_{77}}
    \end{array} \right.\ ,
\end{equation}
This leads again to the previous result $\overline{\Delta E } = -\rm{Tr}[T\rho]$.

Similarly to the previous case, we find that the inital values follow a stochastic sequence $\mu_n$ such that:
\begin{equation}
    \mu_{n+1} = \left\{ \begin{array}{lll}
        0 & \text{ with probability} & p^0(n\tau^+)\\
        \frac{\mu^n\cal{T}_L^+}{(1-\mu^n)\cal{T}_L^-+\mu^n\cal{T}_L^+} & \text{ with probability}&  p^1(n\tau^+) \\
        1 & \text{ with probability}& p^2(n\tau^+) \\
    \end{array}\right. .
\end{equation}
Its average also remains constant and a recursion shows once again that after some time, the sequence stabilizes at a constant value $\mu_n = 0$ or $\mu_n = 1$ depending on which of the two is reached first.

\subsection{Selective measurement of the total charge}

Let us now consider the total charge $C = n_{\ua}+n_{\da}+n_R$, acting on both QD. We can calculate the four possible projected states:
\begin{equation}
    \left\{ \begin{array}{l}
         \rho^0(\tau^+) = |0\ra\la 0| \\
         \rho^1(\tau^+) = \frac{\rho_{11}|1\ra\la 1| + \rho_{22}|2\ra\la 2| + \rho_{44}|4\ra\la 4| + \rho_{14}|1\ra\la 4| + \rho_{41}|4\ra\la 1|}{\rho_{11} + \rho_{22} + \rho_{44}}\\
         \rho^2(\tau^+) = \frac{\rho_{33}|3\ra\la 3| + \rho_{55}|5\ra\la 5| + \rho_{66}|6\ra\la 6| + \rho_{36}|3\ra\la 6| + \rho_{63}|6\ra\la 3|}{\rho_{33} + \rho_{55} + \rho_{66}}\\
         \rho^3(\tau^+) = |7\ra\la 7|
    \end{array} \right.\ ,
\end{equation}
and the probabilities:
\begin{equation}
    \left\{ \begin{array}{l}
         p^0(\tau^+) = \rho_{00} \\
         p^1(\tau^+) = \rho_{11} + \rho_{22} + \rho_{44}\\
         p^2(\tau^+) = \rho_{33} + \rho_{55} + \rho_{66}\\
         p^3(\tau^+) = \rho_{77}
    \end{array} \right.\ .
\end{equation}
This time, off-diagonal terms are present in the projected states! And we can now easily see that the average energy increment $\overline{\Delta E} = p^0\Delta E^0 + p^1\Delta E^1 + p^2\Delta E^2 + p^3\Delta E^3$ vanishes! In this case, we cannot expect the measurement to energize the system on average because it does not produce the necessary separation.

\section{\label{SI5} Entropy for the various measurement protocols}

\subsection{Second-order correction}

As we have showed in SI Note. 2, the perturbative kernel approach solution we have derived cannot yield a second-order correction because the system is not invertible. Nonetheless, it is still possible to find such a correction using another trick. Indeed, one should remember that the density matrix is a positive hermitian matrix, but the solution we have found up to first order may not verify this positivity condition whenever $\mu = 0$ or $\mu = 1$. We shall thus try to find a second-order correction that does not change the eigenvalues of the density matrix up to second order. 

Let us first renumerate the basis and set:
\begin{equation}
\begin{array}{c}
     |\tilde{0}\ra \equiv |0\ra = |00\ra,\ |\tilde{1}\ra \equiv |1\ra = |0\da\ra,\ |\tilde{2}\ra \equiv |4\ra = |\da 0\ra,\ |\tilde{3}\ra \equiv |3\ra = |\ua\da\ra,\\ |\tilde{4}\ra \equiv |6\ra = |20\ra,\ |\tilde{5}\ra \equiv |5\ra = |\da\da\ra,\ |\tilde{6}\ra \equiv |2\ra \equiv |\ua 0\ra,\ |\tilde{7}\ra \equiv |7\ra = |2\da\ra\ 
\end{array} ,
\end{equation}
such as the perturbative steady state matrix now reads in this basis
\begin{equation}
    \rho = \begin{pmatrix}
        \rho_{00} & 0 & 0 & 0 & 0 & 0 & 0 & 0 \\
        0 & \rho_{11} & \rho_{14} & 0 & 0 & 0 & 0 & 0 \\
        0 & \rho_{14}^* & \rho_{44} & 0 & 0 & 0 & 0 & 0 \\
        0 & 0 & 0 & \rho_{33} & \rho_{36} & 0 & 0 & 0 \\
        0 & 0 & 0 & \rho_{36}^* & \rho_{66} & 0 & 0 & 0 \\
        0 & 0 & 0 & 0 & 0 & \rho_{55} & 0 & 0 \\
        0 & 0 & 0 & 0 & 0 & 0 & \rho_{22} & 0 \\
        0 & 0 & 0 & 0 & 0 & 0 & 0 & \rho_{77}
    \end{pmatrix},
\end{equation}
so that it be block-diagonal.

Let us then focus on a block $\begin{pmatrix} p_+ & ra \\ ra^* & p_- \end{pmatrix} $ with $p_+, p_- >0$ and let us suppose without loss of generality that $p_+>p_-$ (as we are only interested in the eigenvalues of such a matrix, one can still permute the basis vectors and rename $p_+$ and $p_-$ to satisfy this condition). The eigenvalues $\lambda_+$ and $\lambda_-$ then read:
\begin{equation}
    \lambda_{\pm} = \frac{p_++p_-}{2} \pm \sqrt{\Big(\frac{p_+-p_-}{2}\Big)^2 + r^2|a|^2} \ ,
\end{equation}
which expands into:
\begin{equation}
    \lambda_{\pm} = p_{\pm} \pm \frac{r^2|a|^2}{p_+-p_-} + o(r^2)\ .
\end{equation}
Therefore, if we want to kill the second-order term in these eigenvalues, we may set $p_{\pm} \equiv p_{\pm} \mp \frac{r^2|a|^2}{p_+-p_-}$.

We should now check that this correction to the diagonal indeed does not change the eigenvalues. Let us then consider the block $\begin{pmatrix} p_+ - r^2\lambda & ra \\ ra^* & p_- + r^2\lambda \end{pmatrix} $ where $\lambda \equiv \frac{|a|^2}{p_+-p_-}$. Then the eigenvalues now read:
\begin{equation}
    \mu_{\pm} = \frac{p_++p_-}{2} \pm \sqrt{\Big(\frac{p_+-p_-}{2}\Big)^2 + r^4\lambda^2} = p_{\pm} + o(r^2)\ ,
\end{equation}
so the correction indeed does not change the eigenvalue up to the second order.

This demonstration allows us to build a second-order correction term $\rho^{(2)}$:
\begin{equation}
    \rho^{(2)} = \frac{|\rho_{14}|^2}{\rho_{11}-\rho_{44}}(|4\ra\la 4|-|1\ra\la 1|) + \frac{|\rho_{36}|^2}{\rho_{33}-\rho_{66}}(|6\ra\la 6|-|3\ra\la 3|)\ .
\end{equation}

\subsection{\label{5A} Thermodynamic quantities}

\subsubsection{\label{5Aa} Entropy for unselective quantum measurements}

Let us study the evolution of the Von-Neumann entropy of the system at each step of the cycle. Before the measurement, for any observable separating the state we consider, the entropy of the thermalized state is given by 
\begin{equation}
    S(n\tau^-) = S(\rho) = -\rm{Tr}[\rho\,\rm{ln}\, \rho] \ .
\end{equation}
Then, the final entropy after the measurement of the observable $n_R$ will depend on the measurement outcome. When unselective quantum measurements breaking the entanglement are performed, we have showed in the main text as well as in SI Note. 3 that the cycle is the same regardless of the considered observable since $\rho(n\tau^-) = \rho$ and $\rho(n\tau^+) = \sum_{i=0}^7 \rho_{ii}|i\ra\la i|$. Therefore, the first term is trivial because $\rho(\tau^+)$ is diagonal and leads to
\begin{equation}
    S(\rho(\tau^+)) = -\sum_{i=0}^7 \rho_{ii}\rm{ln}\, \rho_{ii}.
\end{equation}
Then the second term may be evaluated using the second order we just derived. By diagonalizing $\rho$, we obtain
$S(\rho) =  S(\rho) + o(\gamma^2)$ since the second order term has been chosen exactly so that the eigenvalues of $\rho$ calculated at the second order remain the same as the first and zero order. In the end, we get $\Delta S = o(\gamma^2)$ and we shall not try to extend this calculation to the third order but simply consider that the measurement process is nearly isentropic, which is sufficient for the purpose of this study.

From the previous calculation, we notice that the corrections we found for $\rho$ vanish in the entropy, only resulting in null corrections up to the third-order of this entropy. We can thus neglect the entropy change as long as we stay at low temperature and in the perturbative regime. The only statement we can make up to this point is on the sign of $\Delta S$ which should be positive because of the thermodynamic effect of unselective quantum measurements \cite{nielsen_quantum_2002, kammerlander_coherence_2016}.

This means that the measurement is reorganizing the two-QD system so that it creates localized information from a delocalized form.

\subsubsection{\label{4C} Free Energy}

The previous derivations allow us to define the maximum extractable work during the thermalization process. This quantity $W_{th}$ is defined by the difference between the average free energy of the initial state at the beginning of a cycle at time $n\tau^+$ and the free energy of the final state at the end of the thermalizing stroke at time $(n+1)\tau^-$. So we obtain directly:
\begin{equation}
    W_{th} = -\Delta E + T\Delta S \ .
\end{equation}
We can thus expect to extract energy at finite temperature whenever $W_{th} < 0$. It leads to a critical temperature $T_c \equiv \Delta E / \Delta S$ above which this engine cannot possibly work.

\subsection{\label{4D} Efficiency}

We may define the engine efficiency $\eta$ as the ratio of the electronic work $W_{el}$ obtained during the thermalization process to the total average energy provided by the quantum measurement:
\begin{equation}
    \eta \equiv \frac{W_{el}}{\Delta E} \ .
\end{equation}
By definition, this quantity is less than unity. Indeed, according to the first law of thermodynamics which should hold during the time-dependent evolution of the thermalizing stroke according to Kumar and Stafford \cite{kumar_first_nodate}, we may write
\begin{equation}
    \Delta E= W_{el} + Q
\end{equation}
with $\Delta E \le W_{th} \le W_{el} \le 0$ as we expect $Q \le 0$. This means that heat should be dissipated to the reservoir as the system thermalizes with the environment (a negative $Q$, with the same sign as $W_{el}$ means that the heat has been transferred from the system to the environment). Using this inequality, we should therefore have
\begin{equation}
    \eta \le \frac{W_{th}}{\Delta E} \le 1 - \frac{T}{T_c}\ ,
\end{equation}
so, as expected, the engine efficiency is still bounded by a form of the Carnot efficiency.
As a final remark on this quantity, we should state that it does not have a very practical importance in the design of the engine as it can have in classical or other semi-classical quantum engines. Indeed, as our engine relies on the energy provided by autonomous quantum measurements performed by the environment, the fuel we are harvesting is present in infinite quantity in the self-sustained bath we are exploiting. Therefore, even a poor efficiency can be of interest given the limitless, constantly refuelling amount of energy we are trying to harvest. This energy might come from the local breaking of the second law of thermodynamics during the measurement processes \cite{hormoz_quantum_2013, lloyd_use_1989, dabramo_peculiar_2012}, from a minuscule temperature gradient between the measurement bath considered as hot and the opposite electrode bath considered as cold, or from other non-thermal resource such as squeezing \cite{manzano_entropy_2016, klaers_squeezed_2017, xiao_thermodynamics_nodate}. We note that our formalism does not take into account the thermodynamic cost of information erasure as the spinterface electronically interacts with the ferromagnetic electrode acting as an entropy sink \cite{chowrira_quantum_2022}. We suppose here that the overall energy balance will be favorable, and will address the thermodynamic cost of interfacial spin accumulation in a future paper.

\subsubsection{\label{4E} Power Output}

The power $P$ follows instantly from the previous section. It is defined as
\begin{equation}
    P \equiv \frac{W_{el}}{\tau} \le \frac{W_{th}}{\tau}\ \equiv P_{th}.
\end{equation}
It is inversely proportional to the duration of a cycle. This means that devices with fast measurement frequencies can maximize this quantity.

\subsection{Entropy for selective quantum measurements}

\subsubsection{Selective measurement of the occupation of the right quantum dot}

From the previous calculation, the entropy of the thermalized state reduces to the entropy of the diagonal terms of $\rho$:
\begin{equation}
    S(\rho) = -\rho_{00}\rm{ln}\,\rho_{00}-\rho_{11}\rm{ln}\,\rho_{11}-\rho_{22}\rm{ln}\,\rho_{22}-\rho_{33}\rm{ln}\,\rho_{33} + o(\gamma^2)\ .
\end{equation}

Let us now study how the entropy changes after the selective measurement. The final entropy after the measurement of the observable $n_R$ will depend on the measurement outcome. This leads to the definition of two different entropies $S^1 \equiv S(\rho^1(\tau^+))$ and $S^0 \equiv S(\rho^0(\tau^+))$, which are calculated simply because the projected states are diagonal:
\begin{equation}
\left\{\begin{array}{l}
    S^1 = \frac{-\rho_{11}\rm{ln}\,\rho_{11}-\rho_{33}\rm{ln}\,\rho_{33}+(\rho_{11}+\rho_{33})\rm{ln}(\rho_{11}+\rho_{33})}{\rho_{11}+\rho_{33}}\\
    S^0 = \frac{-\rho_{00}\rm{ln}\,\rho_{00}-\rho_{22}\rm{ln}\,\rho_{22}+(\rho_{00}+\rho_{22})\rm{ln}(\rho_{00}+\rho_{22})}{\rho_{00}+\rho_{22}}
\end{array}\right. \ .
\end{equation}
This leads us to define the expected value of the entropy of the projected state $\overline{S} = p^1S^1 + p^0S^0$, which reads
\begin{multline}
    \overline{S} = -\rho_{11}\rm{ln}\,\rho_{11}-\rho_{33}\rm{ln}\,\rho_{33}+(\rho_{11}+\rho_{33})\rm{ln}(\rho_{11}+\rho_{33})\\ -\rho_{00}\rm{ln}\,\rho_{00}-\rho_{22}\rm{ln}\,\rho_{22}+(\rho_{00}+\rho_{22})\rm{ln}(\rho_{00}+\rho_{22}) \ .
\end{multline}
So the expectancy of the entropy increase defined as $\overline{\Delta S} \equiv \overline{S} - S(\rho)$, reads
\begin{equation}
    \overline{\Delta S} = (\rho_{00}+\rho_{22})\rm{ln}(\rho_{00}+\rho_{22}) + (\rho_{11}+\rho_{33})\rm{ln}(\rho_{11}+\rho_{33}) \ .
\end{equation}
We witness that the sign of $\overline{\Delta S}$ is negative. This means that the reading of the measurement outcome reduces the mixture of states through the elimination of the components coding for the unmeasured states. Contrary to the case of unselective measurements, we thus find that reading the measurement outcome is reducing the entropy of the state.

This entropy difference can be minimized and we find that $\overline{\Delta S} \ge -\rm{ln}\, 2$, reaching equality whenever
\begin{equation}
     \rho_{00}+\rho_{22}=\frac{1}{2} \text{ and } \rho_{11}+\rho_{33}=\frac{1}{2}\ .
\end{equation}
Indeed the measurement can yield two possible outcomes so the information extracted from from the system should not exceed $\ln \,2$ as expected.

\subsubsection{Measurement of the charge of the left quantum dot}

The derivation of the entropies of each outcome of the measurement of the charge on the left QD is again straightforward and leads to:
\begin{equation}
    \left\{ \begin{array}{l}
         S^0 = \frac{1}{\rho_{00}+\rho_{33}}[-\rho_{00}\rm{ln}\,\rho_{00}-\rho_{33}\rm{ln}\,\rho_{33}+(\rho_{00}+\rho_{33})\rm{ln}(\rho_{00}+\rho_{33})] \\
         S^1 = 0\\
         S^{-1} = 0
    \end{array} \right. \ .
\end{equation}
This leads to the average projected entropy $\overline{\Delta S} = p^0S^0 + p^1S^1 + p^{-1}s^{-1}$ : 
\begin{equation}
    \la S \ra = -\rho_{00}\rm{ln}\,\rho_{00}-\rho_{33}\rm{ln}\,\rho_{33}+(\rho_{00}+\rho_{33})\rm{ln}(\rho_{00}+\rho_{33}) \ ,
\end{equation}
and to the average entropy increase:
\begin{equation}
    \overline{\Delta S} = \rho_{11}\rm{ln}\,\rho_{11} + \rho_{22}\rm{ln}\,\rho_{22} + (\rho_{00}+\rho_{33})\rm{ln}(\rho_{00}+\rho_{33}) < 0\ .
\end{equation}
This time, the minimum is different as we have $\overline{\Delta S} \ge -\rm{ln}\, 3$, with an equality when 
\begin{equation}
     \rho_{11} = \frac{1}{3},\ \rho_{22}=\frac{1}{3} \text{ and } \rho_{00}+\rho_{33}=\frac{1}{2}\ .
\end{equation}
Again, this result is consistent with a measurement that can yield three possible values, leading to a decrease in entropy capped by $\rm{ln}\,3$.

\subsubsection{Measurement of the spin of the left quantum dot}

In the case of the spin on the left QD, we have again:
\begin{equation}
    \left\{ \begin{array}{l}
         S^0 = 0 \\
         S^1 = \frac{1}{\rho_{11}+\rho_{22}}[-\rho_{11}\rm{ln}\,\rho_{11}-\rho_{22}\rm{ln}\,\rho_{22}+(\rho_{11}+\rho_{22})\rm{ln}(\rho_{11}+\rho_{22})]\\
         S^{-1} = 0
    \end{array} \right. \ ,
\end{equation}
leading to 
\begin{equation}
    \overline{S} = -\rho_{11}\rm{ln}\,\rho_{11}-\rho_{22}\rm{ln}\,\rho_{22}+(\rho_{11}+\rho_{22})\rm{ln}(\rho_{11}+\rho_{22}) \ ,
\end{equation}
and 
\begin{equation}
    \overline{\Delta S} = \rho_{00}\rm{ln}\,\rho_{00} + \rho_{33}\rm{ln}\,\rho_{33} + (\rho_{11}+\rho_{22})\rm{ln}(\rho_{11}+\rho_{22}) < 0\ .
\end{equation}
In this case, the minimum is the same $ \overline{\Delta S} \ge -\rm{ln}\,3$, but the equality is achieved for different population
\begin{equation}
     \rho_{00} = \frac{1}{3},\ \rho_{33}=\frac{1}{3} \text{ and } \rho_{11}+\rho_{22}=\frac{1}{2}\ .
\end{equation}

\section{\label{SI6} Engine Operational Simulations}

\subsection{Unselective measurements}

\subsubsection{Measurement of right qubit}

\begin{figure*}[b]
    \includegraphics[width=0.98\textwidth]{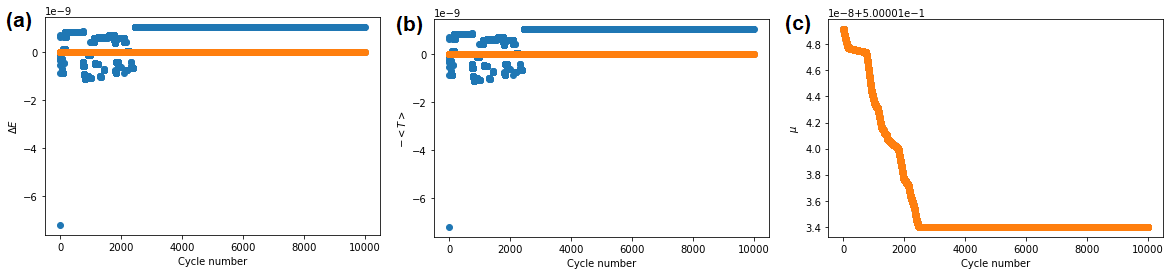}
    \caption{Simulation results of $\Delta E$ in (a) and $-\la T \ra$ in (b) and $\mu$ in (c) for $10^5$ cycles when measuring $n_R$. The parameters used are $\epsilon_{\da} = 8,\ \epsilon_{\ua} = -3,\ \epsilon_R = 1,\ J = 8,\ U = 100,\ \gamma = 0.01,\ \cal{T}_L^+ = 5,\ \cal{T}_R^+=\cal{T}_L^- = \cal{T}_R^-=1$ and with the initial condition $\rho_0 = \frac{1}{2}|\ua\da\ra\la\ua\da| +  \frac{1}{2}|\da\da\ra\la\da\da|$.}
    \label{E1_1e5}
\end{figure*}

In Fig.~\ref{E1_1e5}, we show additional information corresponding to Fig.3 in the main text. When starting with the sub-optimal mixed state $\rho_0 = \frac{1}{2}|\ua\da\ra\la\ua\da| +  \frac{1}{2}|\da\da\ra\la\da\da|$, we first witness a perfect agreement between $\Delta E$ and $-\la T \ra$, as predicted in the main text. This concordance between the two values has been tested for all the runs we did involving unselective measurements, which confirms the validity of our analysis. Second, compared to the main text in which we show if Fig.3(c) and (d) the dataset obtained starting from cycle number $10^5$, we see that in this case, the system presents a transitory regime lasting for about 3000 cycles during which $\mu=1/2+\d\mu$, $\d\mu \approx 4.10^{-8}$, is steadily decreasing while $\Delta E$ is presenting an erratic behaviour around 0. Then, a stationary regime is reached where $\d\mu \approx 3.4.10^{-8}$ and the energy increment $\Delta E$ is alternating between two values $E_0 \approx 10^{-9}$ and $E_1 \approx 0$, thus leading to a positive average value $\overline{\Delta E} \approx 8.10^{-10}$. It would thus seem that the system may indeed act as an active device since $\overline{\Delta E} > 0$, albeit returning a very low power output. A final peculiar pojnt we should raise regarding this experiment is that, we do observe a convergence towards $\mu = \mu_c = 1/2$ for these parameters, as predicted by the perturbation model, with a singular difference though, revealing that $\Delta E > 0$ while $\mu > \mu_c$, which should be prohibited in the first order solution. This shows that the perturbation solution we developed can be relevant for predicting the operational steady-state for the engine, but it fails to render the right behavior of the system in the neighborhood of this point. 

\subsubsection{Measurement of left qubit}

As we saw in SI Note. 3, the behavior of the engine might be different depending on the observable we choose for the measurement. We showed previously that changing the observable does not change the energy increment one can hope to harvest, and in fact, in the case of the unselective measurements, we argue that this choice will not change the time evolution of the engine in any way. Indeed, based on previous calculations, it is straightforward to see that the projective channel $\rho \to \sum_k \Pi_k\rho\Pi_k$ is exactly the same channel if we decide to measure $n_R$, $Q$ or $S$, as they all have the same effect of deleting the off-diagonal terms in $\rho$, while keeping its diagonal unchanged. This argument shows that the engine is completly independant from the choice of the measurement basis, provided that it separates the two QDs, thereby killing the entanglement terms between them. 

However, we should emphasize that each observable will lead a different behavior when studying selective measurement protocols because of the non-linear probability rescaling that is applied in this case. Hence, the evolution of the linear system is independent on the local measurement basis chosen to measure the system, the situation becomes less trivial when the measurement result relative to a specific local basis is read as this result starts to condition an erratic evolution of the system. Further details is given below regarding the choice of observable for selective measurements.

\subsection{Selective measurement}

\subsubsection{Measurement of the right qubit}

\begin{figure*}[b]
    \includegraphics[width=0.98\textwidth]{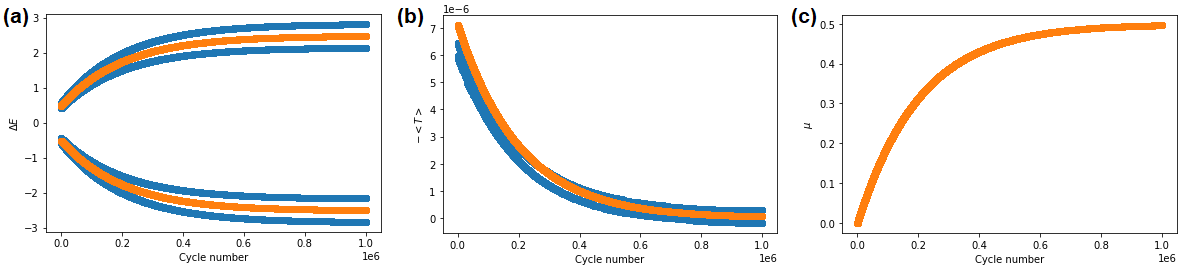}
    \caption{Simulation results of $\Delta E$ in (a) and $-\la T \ra$ in (b) and $\mu$ in (c) for $10^6$ cycles when measuring $n_R$. The parameters used are $\epsilon_{\da} = 8,\ \epsilon_{\ua} = -3,\ \epsilon_R = 1,\ J = 8,\ U = 100,\ \gamma = 0.01,\ \cal{T}_L^+ = \cal{T}_R^+=\cal{T}_L^- = \cal{T}_R^-=1$ and with the initial condition $\rho_0 = |\da\da\ra\la\da\da|$.}
    \label{sel_E3}
\end{figure*}

\begin{figure*}[t]
    \includegraphics[width=0.98\textwidth]{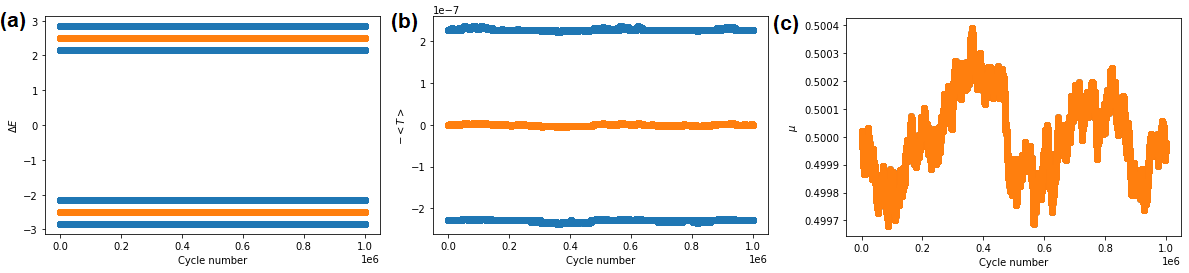}
    \caption{Simulation results of $\Delta E$ in (a) and $-\la T \ra$ in (b) and $\mu$ in (c) for $10^6$ cycles when measuring $n_R$. The parameters used are $\epsilon_{\da} = 8,\ \epsilon_{\ua} = -3,\ \epsilon_R = 1,\ J = 8,\ U = 100,\ \gamma = 0.01,\ \cal{T}_L^+ = \cal{T}_R^+=\cal{T}_L^- = \cal{T}_R^-=1$ and with the initial condition $\rho_0 = \frac{1}{2}|\ua\da\ra\la\ua\da| +  \frac{1}{2}|\da\da\ra\la\da\da|$.}
    \label{sel_E1}
\end{figure*}

As comparison, we show in Fig.~\ref{sel_E3} and Fig.~\ref{sel_E1} the results of numerical simulations obtained for the same set of parameters used in Fig.3 in the main text, but for selective measurements. Compared to the unselective case, when starting with the sate $\rho_0 = |\ua\da\ra\la\ua\da|$ we first notice a striking difference, which is the appearance of a negative branch in the line shapes of $\Delta E$ and $-\la T \ra$. This new possible state emerges from the two measurement possibilities or $n_R$. One result will energize the system, the other will lower its energy, while on average keeping the average energy increment positive. Such alternation is not present in the corresponding unselective measurement because it projects linearly the system onto one mixed state, so the behaviour is more continuous and remains positive. Nevertheless, we emphasize that we should not need to provide any form of work during the thermalization process when the measurement decreased the energy of the system because the quantum dots will just equilibrate themselves with the baths without the need for an external energy input:  heat alone will suffice to reset the state.

Nonetheless, in Fig.3(c) and (d) in the main text, we see that unselective measurements can result in discontinuous alternating behavior. This stochastic process is not due to the randomness of a measurement result but to a sensitivity on the initial conditions. We can compare this experiment to its equivalent for selective measurements, presented in Fig.~\ref{sel_E1}. Compared to the unselective case where $\mu=  1/2 + \d \mu$ is alternating fast, within a clear uptrend, in the selective case, $\mu$ looks like a random walk around $\mu = 1/2$. It leads in the unselective case to a stable and sharp bimodal distribution for $\Delta E$ and $\la T \ra$ which contrasts with the plots for the selective case: $\la T \ra$ is also bimodal but its distribution is not positive and its dispersion is larger around the modes. This negative feature comes from the two measurement outcomes leading to energy-dissipating projections, and the dispersion comes from a stronger sensitivity to the initial conditions, induced by the non-linear projective measurement; and $\Delta E$ now has four modes! Two of them originates from the two measurement outcomes and, each of them splits into two because of the sensitivity to the initial conditions, mainly represented by the value of $\mu$.

These experiments show that reading the measurement result can be detrimental to the engine's performance as it generates power-dissipating cycles, which are counter-balanced by energizing cycles of larger amplitudes, thus resulting in a device with stronger power fluctuations. The non-linear random walk issued by the selection of the measurement is in this case mostly undesirable because of the stroboscopic chaotic behavior it entails: indeed, the projection onto a specific eigenstate drastically changes the initial conditions of the next cycle, which can place it in a unfavorable domain ($\mu > \mu_c$), where the system will be trapped in a power-dissipating phase. Indeed, we calculate the temporal averages over the cycles and found for this unselective case run $-\mathbb{E}[\la T \ra] = -4.10^{-11}$ and $\mathbb{E}[ \Delta E ]= -2.10^{-4}$, which compare with the value found for the selective case $\mathbb{E}[\Delta E ]=\mathbb{E}[-\la T \ra] = 8.10^{-10}$. We thus notice that the fundamental equality $\Delta E =-\la T \ra$ is broken in the selective case and that, for this run, the device is on average passive. We believe that convergence between these two quantities to a positive value corresponding to the unselective case may happen for at infinity, for a larger duration, due to ergoticity as we will describe below. It thus looks like selecting the measurement forces the system to stay in a non-thermal behavior, stabilizing its quantum behavior via a Zeno effect which keep the initial conditions in memory for a longer time, such as it can behave only statistically when studied for a larger time-scale; while unselective measurements allow for a faster convergence towards a cyclic steady-state which is closer to a statistical thermal state in which the history of the system becomes insignificant.

While we have seen that selective measurement may be detrimental to the engine efficiency due to the quantum-trajectory like behavior they allow to stabilize for a larger timeframe, we shall keep in mind that selective measurement may very well be more adapted to other kinds of Maxwell demons which could better use this stroboscopic and negentropic quantum feature to its advantage, eventually through a feedback that could counter the larger fluctuations induced by selection.

\subsubsection{Measurement of the left qubit}

As detailed in SI Note. 3, measuring the charge or the spin of the left qubit leads to a qualitatively very different behaviour for the engine with selective measurements. Indeed, the measurement of $n_R$ leaves the initial parameter $\mu = \la n_{\da}\ra$ unchanged up to the first order, therefore the system is slowly drifting from the initial condition $\mu_0$ because the components that encode the population $n_{\da}$ in the density matrix are progressively mixing due to higher order terms ; contrary to a measurement on the left quantum dot which is stabilizing the population $\mu$, forcing it to remain close to 0 or 1.  

\begin{figure*}[t]
    \includegraphics[width=0.98\textwidth]{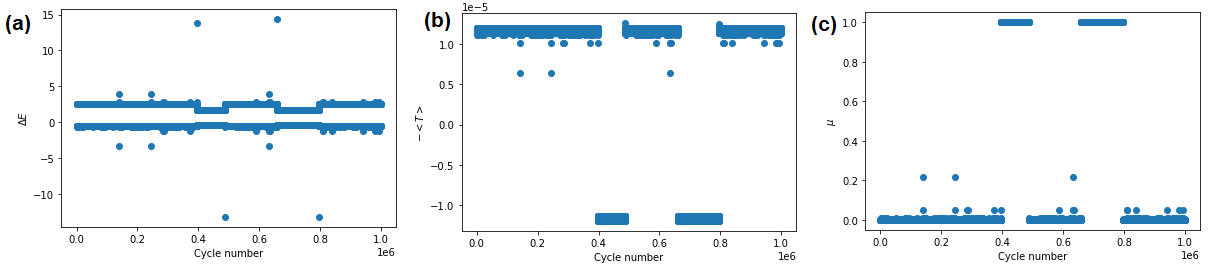}
    \caption{Simulation results of $\Delta E$ in (a) and $-\la T \ra$ in (b) and $\mu$ in (c) for $10^6$ cycles when measuring the charge $Q$ of the left QD. The parameters used are $\epsilon_{\da} = 8,\ \epsilon_{\ua} = -3,\ \epsilon_R = 1,\ J = 8,\ U = 1,\ \gamma = 0.01,\ \cal{T}_L^+ = 5,\ \cal{T}_R^+=\cal{T}_L^- = \cal{T}_R^-=1$ and with the initial condition $\rho_0 = |\ua\da\ra\la\ua\da|$.}
    \label{mu0}
\end{figure*}

In Fig.~\ref{mu0}, we show a test run of $10^6$ cycles of the engine when we measure the spin of the left qubit. We observe in Fig.~\ref{mu0}(c) a stabilization of $\mu$ close to 0 or 1, and that the system indeed stays longer in the state 0 than in the state 1, which should ensure that the measurement is energizing the system on average.

The data of Fig.~\ref{mu0}(a) and (b) confirms this point by yielding a positive value for the temporal averages $\mathbb{E}[{\Delta E}] \approx 18\  \mu$eV and $\mathbb{E}[-\la T \ra] \approx 1.7\ \mu$eV. Nonetheless, we notice a discrepancy between the two values, which differs from the proven result showing the equality between the measurement average $\Delta E$ and $-\la T\ra$ for each cycle. Therefore, averaging over time steps is not equivalent to averaging over the measurement outcomes. The reason we may get different results for these two averages is that the information of the initial state can persist over time and is somewhat transferred from cycle to cycle due to the incomplete thermalization. It thus remains unclear at this point whether it is possible to harvest quantum fluctuation energy with this system after a long time, as the temporal average $\mathbb{E}[{\Delta E}]$ remains close to 0, and with a fluctuating sign depending on the parameters and initial conditions. Indeed, numerical experiments (not shown) that are identical to the present ones revealed the strong dependence of its sign on the parameters chosen. A more detailed analysis of the relation between the initial density of state, the temporal average and the measurement average is beyond the scope of the current paper and will be the subject of a separated study. Here, we shall only state that ergodicity will be of use in order to study the temporal transfer of information, and will ultimately link the temporal and measurement averages to some measure of $\la T \ra$ over the space of initial density matrices. This followup study should provide a definitive answer to the question of harvesting quantum fluctuations with this system after a long time, and may lead to profound connections between thermodynamics, quantum measurement and ergodicity.

\subsubsection{Proof of energy generation}

At the end of each cycle, while the first-order solution showed that after some time, the system would get stuck in one of the two values $\mu_n = 0$ or $\mu_n = 1$, the numerical experiment in Fig.\ref{mu0} showed that there is still a small probability to jump back to $\mu_{n+1} = 1-\mu_n$. This means that the charge of the left quantum dot - driven by the occupation of the down spin energy level - may change during the operation of a cycle.

As shown in the main text, the system may behave as an engine during the cycle $n$ whenever $\mu_n < 0$. Therefore, we can hope to extract energy continuously from this system when the expectation value $\overline{\mu}_n$ is negative. This should be the case if the probability $p \equiv \mathbb{P}(\mu_{n+1}=1|\mu_n=0)$ to jump from $\mu_n = 0$ to $\mu_{n+1}=1$ is lower than the symmetric probability $q \equiv \mathbb{P}(\mu_{n+1}=0|\mu_n=1)$ to jump from $\mu_n=1$ to $\mu_{n+1}=0$; in which case it would guarantee a longer lifetime in the state $\mu = 0$ than in the state $\mu = 1$. Intuitively, one could hope to achieve such asymmetry whenever $U$ is sufficiently small. Indeed, a small $U$ would allow for excessive charging on the left site, hence it would favor a state with a charge of 0 or 2 which would then relax into the thermalized state with a charge of 1, eventually releasing more power if measuring 2 is more probable than measuring 1.

To prove this physical intuition, we begin with the second order correction we have derived in SI Note. 4, which allows us to refine the sequence $\mu_n$. We have:
\begin{equation}
    \mu_{n+1} = \left\{\begin{array}{lllll}
        0 & \text{if} & \mu_n = 0 & \text{with probability} & 1-p \\
        1 & \text{if} & \mu_n = 0 & \text{with probability} & p \\
        0 & \text{if} & \mu_n = 1 & \text{with probability} & q \\
        1 & \text{if} & \mu_n = 1 & \text{with probability} & 1-q 
    \end{array}\right.,
\end{equation}
where 
\begin{equation}
\left\{\begin{array}{l}
    p = \rho_{66}^0+\rho_{77}^0 = \frac{|\rho_{63}^0|^2}{\rho_{33}^0-\rho_{66}^0} = \frac{|\gamma|^2\alpha^2}{|\rm{det}\,B|^2}\cal{T}_L^-\cal{T}_R^-[(s+r)^2+(\Delta+U)^2]\\
    q = \rho_{00}^1+\rho_{11}^1 = \frac{|\rho_{14}^1|^2}{\rho_{44}^1-\rho_{11}^1} = \frac{|\gamma|^2\alpha^2}{|\rm{det}\,B|^2}\cal{T}_L^+\cal{T}_R^+[(s+r)^2+\Delta^2]
\end{array}\right..
\end{equation}
Using the total probability formula, we can then write:
\begin{equation}
\left\{\begin{array}{l}
    \mathbb{P}(\mu_{n+1}=0) = \mathbb{P}(\mu_{n+1}=0|\mu_n = 0)\mathbb{P}(\mu_n=0) + \mathbb{P}(\mu_{n+1}=0|\mu_n = 1)\mathbb{P}(\mu_n = 1) = (1-p)\mathbb{P}(\mu_n=0) + q\mathbb{P}(\mu_n = 1)\\
    \mathbb{P}(\mu_{n+1}=1) = \mathbb{P}(\mu_{n+1}=1|\mu_n = 0)\mathbb{P}(\mu_n=0) + \mathbb{P}(\mu_{n+1}=1|\mu_n = 1)\mathbb{P}(\mu_n = 1) = p\mathbb{P}(\mu_n=0) + (1-q)\mathbb{P}(\mu_n = 1)
\end{array}\right.,
\end{equation}
which reads in matrix format $\mathbb{P}(\mu_{n+1}) = M\mathbb{P}(\mu_{n})$ with $M = \begin{pmatrix} 1-p & q \\ p & 1-q \end{pmatrix}$. This matrix $M$ thus describes the transition of a Markov chain, and in order to study the behaviour or $\mu_n$ at long times, we shall study the stationary distribution. The diagonalization of $M$ is straightforward and yields:
\begin{equation}
M = \frac{1}{p(p+q)}\begin{pmatrix}
    q & -p \\ p & p
\end{pmatrix} \begin{pmatrix}
    1 & 0 \\ 0 & 1-p-q
\end{pmatrix} \begin{pmatrix}
    p & p \\ -p & q
\end{pmatrix},
\end{equation}
which shows that after some time:
\begin{equation}
M^n \underset{n\to +\infty}{\longrightarrow} \frac{1}{p+q}\begin{pmatrix}
    q & q \\ p & p
\end{pmatrix} .
\end{equation}
Therefore,
\begin{equation}
\overline{\mu_{n}} = \mathbb{P}(\mu_n = 1) \underset{n\to +\infty}{\longrightarrow} \overline{\mu_{\infty}} \equiv \frac{1}{1+\frac{q}{p}} = \frac{1}{1+\frac{\cal{T}_L^+\cal{T}_R^+}{\cal{T}_L^-\cal{T}_R^-}\frac{(s+r)^2+\Delta^2}{(s+r)^2+(\Delta+U)^2}},
\end{equation}
when starting with $P(\mu_0=0) = 1$ and $P(\mu_0=1) = 0$. This leads to the condition:
\begin{equation}\label{mu_vs_muc}
    \overline{\mu_{\infty}} < \mu^c \eq p < \frac{\cal{T}_R^-}{\cal{T}_R^+}q \eq 1 \le \frac{(s+r)^2+(\Delta+U)^2}{(s+r)^2+\Delta^2} < \frac{\cal{T}_L^+}{\cal{T}_R^-}.
\end{equation}

This result partially confirms our intuition by showing two points. First, our conjecture was almost correct, meaning that $p < q$ may not be sufficient to allow for energy generation, but we shall instead require $p < \frac{\cal{T}_R^-}{\cal{T}_R^+}q$ to guarantee it. This shows that a higher $\frac{\cal{T}_R^-}{\cal{T}_R^+}$ ratio would increase the value of $\mu^c$ and increase the range of parameters that allow for energy generation.\\

Second, $U$ must indeed be small enough in order to fulfill the condition \ref{mu_vs_muc}. This inequality additionally shows another critical point, which is the condition $\cal{T}_L^- < \cal{T}_L^+$. In physical terms, it means that the injection of electrons on the left side must be stronger than the injection of holes, and the opposite should be true on the right side to maximise $\mu_c$. Intuitively, this should favor a flux of electrons going from left to right, so that it would create a current going above the built-in potential ladder imposed by the placement of the energy levels. This exactly describes exactly the behavior of an active device.

\newpage

\section{\label{SI7} Sampling experiments}

\subsection{Comparison between the perturbative and the numerical solutions}

\begin{wrapfigure}{l}{0.5\textwidth}
    \centering
    \includegraphics[width=0.43\textwidth]{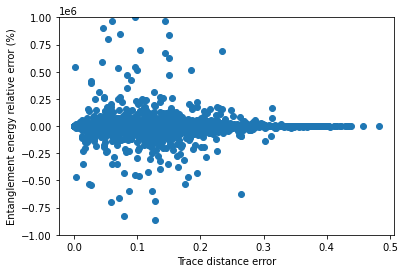}
    \caption{Distribution of the trace distance error and the entanglement energy error between the perturbative and the numerical solution.}
    \label{p_vs_num}
\end{wrapfigure}

Our perturbative solution holds only shortly after each thermalization stroke, yet we expect the exact final state to be reached only after many cycles, and partial thermalization can occur. To evaluate the robustness of this approximation, we now perform simulations using QuTip. We calculated the density matrix at time $t = 1$ meV$^{-1} \approx 4$ ps, and $\Delta E $, starting from different pure states and with different parameters taken randomly within a range that is compatible with the perturbative assumption. More precisely, we calculated the perturbative and the numerical solution for a set of $10^6$ corpus of parameters taken uniformly within a physically reasonable range that preserves the relative positions of the energy levels. For this experiment, we have chosen a uniform sampling of the parameters such as $\epsilon_{\ua} \in \lb -100, 99 \rb$,  $\epsilon_{\da} \in \lb \epsilon_{\ua}+1, 100\rb$, $\epsilon_R \in \lb \epsilon_{\ua}, \epsilon_{\da}\rb$, $J \in \lb 1, 100\rb$, $U \in \lb 1, 1000\rb$, $10^4\gamma \in \lb 10, 1000\rb$, $\cal{T}_L^+,\ \cal{T}_L^-,\ \cal{T}_R^-,\ \cal{T}_R^+ \in \lb 1,100\rb$ and $\rho_0$ a random 8$\times$8 density matrix. 

Results of this experiment are presented in Fig~SI.\ref{p_vs_num} and show a sample of $10^6$ trials, the trace distance between the perturbative solution $\rho$ and the calculated solution $\sigma$ defined by $T(\rho,\sigma) = \frac{1}{2}\rm{Tr}\,\sqrt{(\rho-\sigma)(\rho-\sigma)^{\dagger}}$ is lower than $0.5$ and the corresponding error on $\Delta E$ can reach up to $10^8$. The statistics show that 98\% of the runs lead to an error on $\Delta E$ higher than 1\%. This shows that the derived perturbative solution is clearly unable to describe the entanglement energy created through partial thermalization.

Although this first experiment clearly shows the limits of the perturbative approach to accurately describe the state at the end of the thermalization step, it may still be usable to give qualitative interpretations of the operation of the engine and orient the search for optimal parameters. In the main text, we also display one special case where the perturbative solution remains pretty close to the numerical solution at the level of a single cycle.

\subsection{Comparison between the partially thermalized and the steady-state solutions}

\begin{wrapfigure}{l}{0.5\textwidth}
    \centering
    \includegraphics[width=0.43\textwidth]{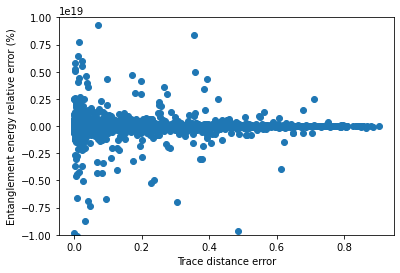}
    \caption{Distribution of the trace distance error and the entanglement energy error between the perturbative and the numerical solution.}
    \label{th_vs_ss}
\end{wrapfigure}

Ideally, we would want to achieve full thermalization during the thermalizing stroke, meaning that the system would reach the steady-state of the master equation, independent from time. Getting close to full thermalization would indeed completely wipe out the memory of the initial condition, hence making the cycle more reliable and easier to study. We therefore wanted to test this assertion by comparing the partially thermalized solution after $\tau = 4$ ps with the numerically calculated steady-state solution for a large sample of the set of parameters, using the same method as above. 

Using the same ranges of parameters used in the previous experiments, we therefore calculated the trace distance error and the entanglement energy error between the two numerically calculated states for a sample size of $10^6$. The results presented in Fig~\ref{th_vs_ss} once again show a large discrepancy between the two solutions, with a trace distance error reaching up to 0.9 and an entanglement energy error of up to $10^21$. This experiment thus shows that the cycle time of 4 ps is largely insufficient to fully thermalize the system with the baths. Nonetheless, this partial thermalization may not be critical to the efficiency of the engine. In fact, it may even be beneficial! Indeed, as we will show in SI. Note. 9, the entanglement energy that is obtained after a duration of 4 ps is much larger and smoother than the energy that can be achieved with full thermalization. This means that the energy provided by the quantum measurement to the system may be greater by up to ~20 orders of magnitude in the case of partial thermalization than when the steady-state is reached. This can be seen by witnessing the scale of the vertical axis in Fig~\ref{th_vs_ss}. And even though a larger magnitude of the entanglement energy can be detrimental to the power output, in the end, due to increasing fluctuations (see SI Note 10), it seems that partial thermalization would still be valuable to enhance the power of the device given the very low magnitude of $|\la T \ra|$ and the sign variability obtained for the steady-state (see SI Note. 9).

\section{\label{SI8} On the applicability of the relation between the energy increment given by the measurement and the entanglement energy}

Referring to SI Note. 1, the master equation reads:
\begin{equation}
    \frac{\d\rho_S}{\d t} = -i[H_S,\rho_S] + \cal{T}^-_L\cal{D}[c_{\ua}^{\dagger}](\rho_S) + \cal{T}^+_L\cal{D}[c_{\ua}](\rho_S) + \cal{T}^-_R\cal{D}[c_{R}^{\dagger}](\rho_S) + \cal{T}^+_R\cal{D}[c_{R}](\rho_S)
\end{equation}
In the basis chosen in SI Note. 2, the very same argument used in this section shows that a set of 12 equations are independent of the 52 others. We can thus decompose the density matrix $\rho_S$ taken as a vector such as $\vec{\rho}_S = \vec{\rho}_D\otimes \vec{\rho}_F$ with 
\begin{equation}
    \frac{\d \vec{\rho}_D}{\d t} = D\vec{\rho}_D \text{ and } \frac{\d \vec{\rho}_F}{\d t} = F\vec{\rho}_F,
\end{equation}
where $D$ and $F$ are two matrices. The first equation with $D$ is describing the evolution of the 12 coefficients we are interested in, in particular the 8 diagonal terms ; while the second equation in $F$ dictates the evolution of the rest of the coefficients. It thus appears that when the device is initialized in a pure state, a thermal state, or any state with no coherence, then $\vec{\rho}_F(0) = 0$, which would lead to $\vec{\rho}_F(t) = e^{tF}\vec{\rho}_F(0) = 0$. This - not too restrictive - initial condition would thus imply that the 52 coefficients of $\rho_S$ describing $\vec{\rho}_F$ remain null during the whole time-dependent evolution of the thermalizing stroke. Furthermore, the measurement is also not changing these coefficients as it is projecting the density matrix onto a diagonal state, therefore the measurement only has an effect on the coefficients described by $\vec{\rho}_D$. This thus shows that $\vec{\rho}_F = 0$ at any time, for all cycles, provided that we start with $\vec{\rho}_F(0) = 0$.

This reduction down to the 12 coefficients is documented in  SI Notes. 1, 2 and 3, and reveals that the essential relation $\Delta E = - \la T \ra$ remains valid at any time, for all cycles, since the calculus that led to it remains the same given the shape of the density matrix.

\section{\label{SI9} On the maximization of the entanglement energy}

\subsection{Regime where $\gamma \gg U \sim \epsilon \sim \cal{T}$}

Gaining insights into the influence of the parameters on the device is necessary to guide a potential optimization for a physical implementation. To this end, we performed several numerical experiments in which we plotted $-\la T \ra$ as a function of different pairs of parameters, while keeping the other fixed, for different regimes.

\begin{figure}
    \centering
    \includegraphics[width=0.98\textwidth]{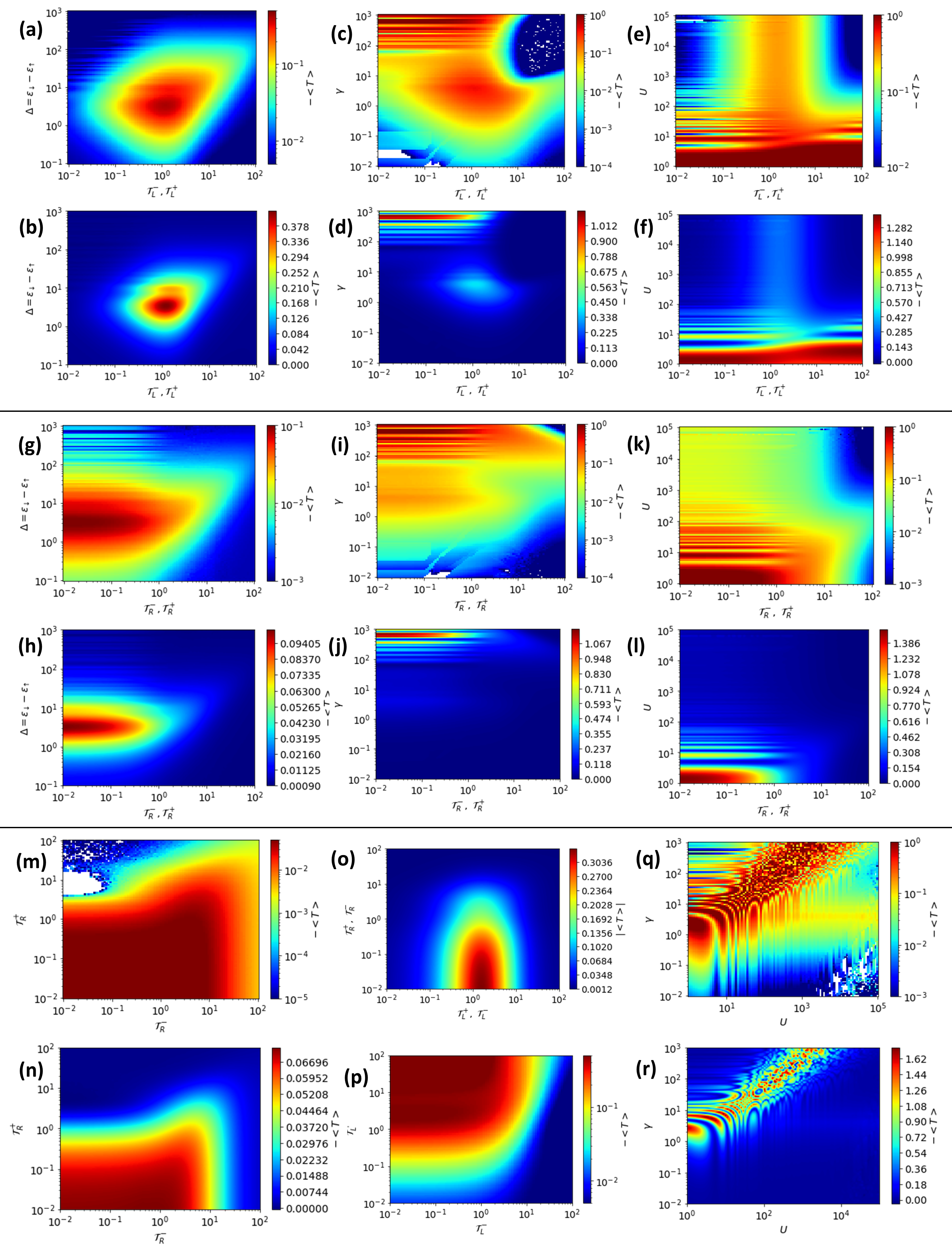}
    \caption{Color plots of the entanglement energy $-\la T \ra$ calculated after 4 ps as a function of various pairs of parameters. For all figures, the fixed parameters are set to $\epsilon_{\da} = 8,\ \epsilon_{\ua} = -3,\ \epsilon_R = 1,\
J = 8,\ U = 10,\ \gamma = 1000,\ \cal{T}_L^+ = 1,\ \cal{T}_L^- = 1,\ \cal{T}_R^+ =5,\ \cal{T}_R^- = 5$}
    \label{SI_Tmax_1}
\end{figure}

In the regime where $\gamma \gg U \sim \epsilon \sim \cal{T}$, the data is represented in Fig~\ref{SI_Tmax_1}. Figures (a) to (f) show the dependence of $\Delta = \epsilon_{\da}-\epsilon_{\ua}$, $\gamma$ and $U$ with respect to $\cal{T}_L^+=\cal{T}_L^-$. We notice that, in this regime, maximizing $\Delta$ is beneficial to the entanglement energy, but $\gamma$ and $U$ present a sweet spot around $\gamma \approx 1-10$ and $U \approx 10^3-10^4$. Intuitively, maximizing $\Delta$ could indeed lead to a higher energizing of the system as this difference in energy between the two levels is strongly linked to the built-in potential difference in the device. We also notice that when $\cal{T}_L$ is too strong, then the entanglement energy decreases. The sweet spot for $\gamma$ and $U$ may be explained by arguing that tunnelling may be impaired when these two parameters become either too strong or too weak.

Then, Figures (g) to (l) display the same dependence of $\Delta,\ \gamma$ and $U$ but with respect to $\cal{T}_R^+ = \cal{T}_R^-$. The same tendencies can be observed: $-\la T \ra$ is maximal when $\Delta$ is maximal, and when $\gamma \approx 10$ and $U \approx 10^3$. The interesting feature in this case is the appearance of a chaotic phase at high $\cal{T}_R$. Here, the entanglement energy almost vanishes and its sign is subject to strong fluctuations that depend on small variations of the parameters, as we can see in the log plots of Fig~\ref{SI_Tmax_1}(g), (i) and (k). This chaotic phase can also be observed in panels (m) and (o). We therefore ascribe its origin to an interaction with the right electrode that overcomes the tunneling interaction between the two QDs, thereby killing the entanglement energy between the two sites. We understand the influence of $\cal{T}_L$ to be less significant than the influence of $\cal{T}_R$ because the left electrode is not directly linked by a tunnel interaction to the two spin energy levels of interest. Indeed, in our hypothesis, only the $\ua$ level is connected to the lead while tunnelling between the dots couple the $\da$ level of the left site with the right site.

This first experiment thus leads us to the following regimes in which we can hope to maximize $-\la T \ra$: $\cal{T} \ll \epsilon$, then $\gamma \sim \epsilon$ or $\gamma \sim U \gg \epsilon$. Let us study these two regimes corresponding to the two branches we can identify in Fig~\ref{SI_Tmax_1}(q) and (r).

\subsection{Regime where $\gamma \gg U \sim \epsilon \gg \cal{T}$.}

\begin{figure}
    \centering
    \includegraphics[width=0.98\textwidth]{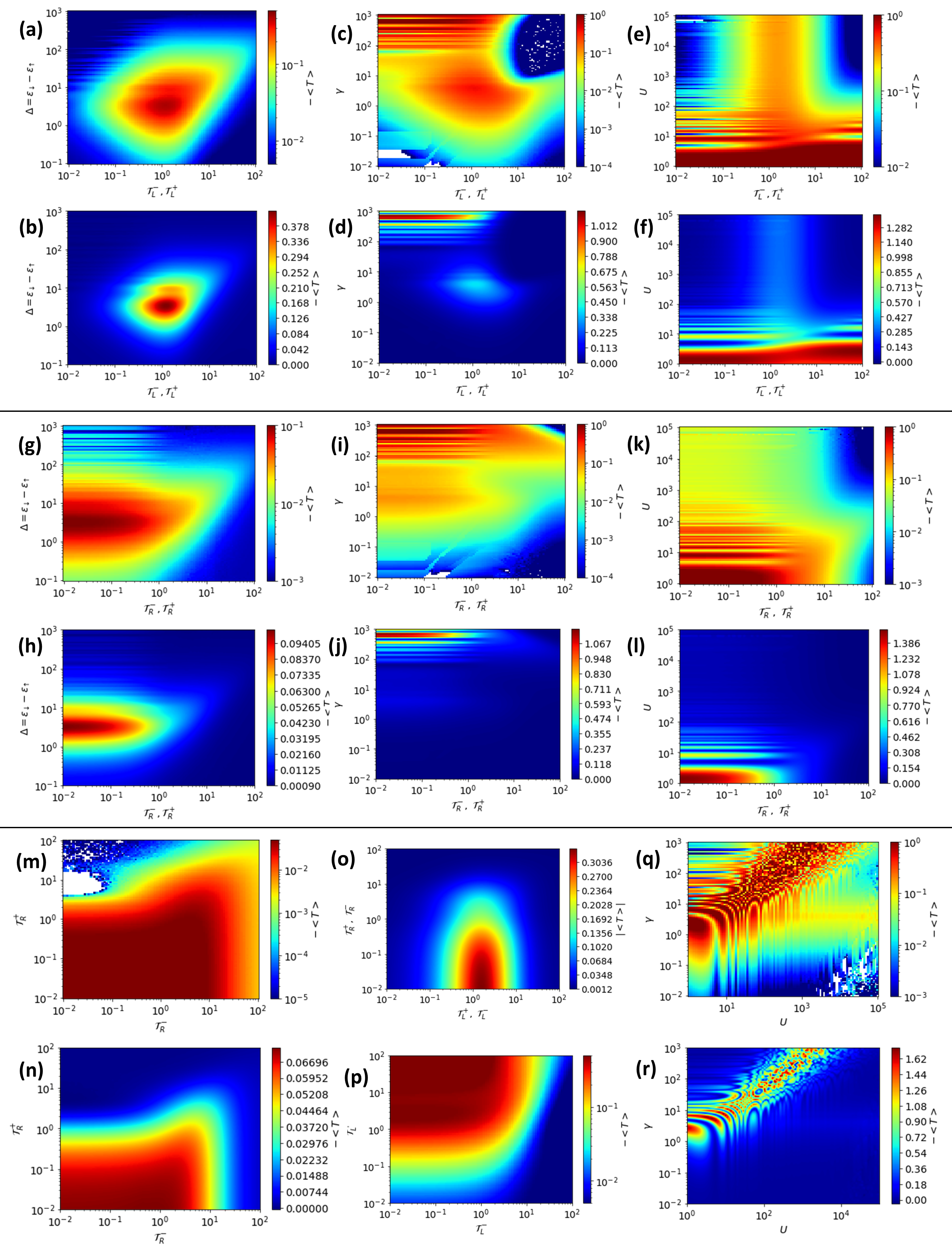}
    \caption{Color plots of the entanglement energy $-\la T \ra$ calculated after 4 ps as a function of various pairs of parameters. For all figures, the fixed parameters are set to $\epsilon_{\da} = 8,\ \epsilon_{\ua} = -3,\ \epsilon_R = 1,\
J = 8,\ U = 3000,\ \gamma = 2,\ \cal{T}_L^+ = 0.1,\ \cal{T}_L^- = 0.1,\ \cal{T}_R^+ =0.1,\ \cal{T}_R^- = 0.1$}
    \label{SI_Tmax_2}
\end{figure}

We therefore repeated the previous simulation with different parameters corresponding first to the branch where $\gamma \gg U \sim \epsilon$. The results presented in Fig~\ref{SI_Tmax_2} are ordered in the same manner as in Fig~\ref{SI_Tmax_1} such that we will just comment on the differences. Contrary to the previous case, we witness this time that there is a sweet spot for $\Delta$ for both the $\cal{T}_L$ and the $\cal{T}_R$ dependence and we should have $\Delta \sim U$ to maximize $-\la T \ra$. This can be explained by arguing that in this case $U$ is very large, so increasing $\Delta$ even more should kill the eventual flow of electrons because they are not be able to overcome both the potential barrier imposed by the placement of the energy levels and the repulsive Coulombic energy. Fig~\ref{SI_Tmax_2}(a)-(f) also shows that there is also a sweet spot for $\cal{T}_L$ such as $\cal{T}_L \sim \epsilon$. This can be explained through the fact that a stronger coupling to the left is now needed to overcome the on-site Coulomb interaction, but this is not the case for $\cal{T}_R$ which could be minimized, and this should be due to the stronger link between the interdot-coupling and the right electrode which is sufficient to overcome this repulsion even in the weak coupling regime. All these analyses remain valid in the data of Fig~\ref{SI_Tmax_2}(o), which shows the asymmetry between left and right and reveals the sweet spot for $\cal{T}_L$ and the minimization of $\cal{T}_R$.

In Fig~\ref{SI_Tmax_2}(m) and (n), we also notice the presence of the both the dissipative and the chaotic phases depending on the parameter $\cal{T}_R$, but the parameter extent of these phases has been considerably reduced as $\cal{T}_R$ appears to become less relevant given the magnitude of $U$. Finally, Fig~\ref{SI_Tmax_2}(q) and (r) once again feature these two distinct branches $\gamma \sim U$ and $\gamma \sim 1$ with the particularities that the branch $\gamma \sim U$ now leads to a clearly superior entanglement energy and that strong vertical and horizontal fluctuations can now be observed. It seem that as $-\la T \ra$ becomes stronger, the fluctuations with respect to the different parameters also become stronger.

\subsection{Regime where $\gamma \sim U \gg \epsilon \gg \cal{T}$.}

In this last simulation, we study the regime of the higher branch in the $\gamma$-$U$ plane where $\gamma \sim U$. The first striking feature we notice is the presence of strong fluctuations around  zones with a higher entanglement energy in all plots, which can reach up to $-\la T \ra \sim 10^3$. We interpret this as evidence that we are scanning quite close to the global maximum of $-\la T \ra$ in this range of parameters.

\begin{figure}
    \centering
    \includegraphics[width=0.98\textwidth]{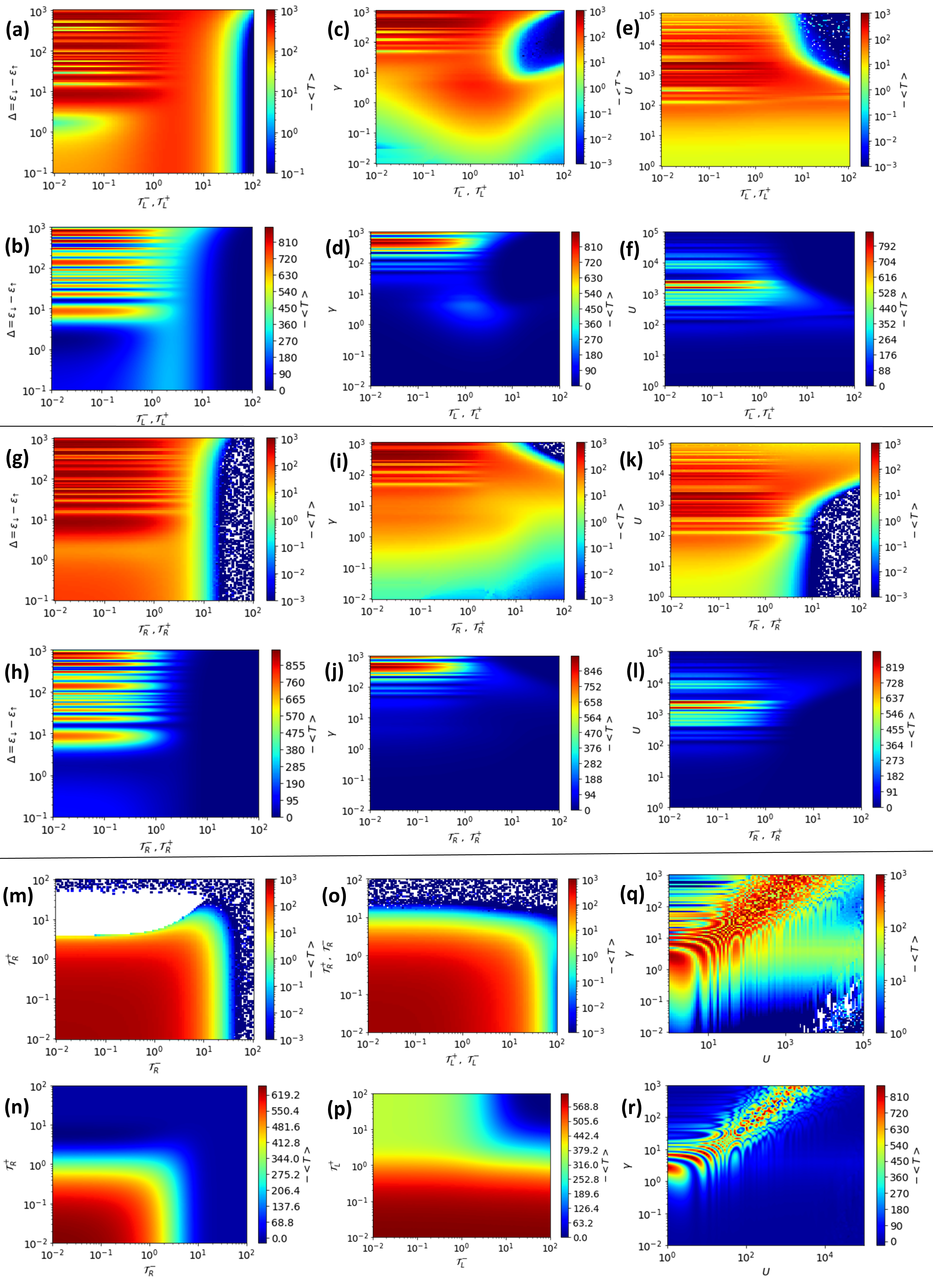}
    \caption{Color plots of the entanglement energy $-\la T \ra$ calculated after 4 ps as a function of various pairs of parameters. For all figures, the fixed parameters are set to $\epsilon_{\da} = 8,\ \epsilon_{\ua} = -3,\ \epsilon_R = 1,\
J = 8,\ U = 1000,\ \gamma = 1000,\ \cal{T}_L^+ = 0.1,\ \cal{T}_L^- = 0.1,\ \cal{T}_R^+ =0.1,\ \cal{T}_R^- = 0.1$}
    \label{SI_Tmax_3}
\end{figure}

The following observations follow from Fig~\ref{SI_Tmax_3}.(a)-(l): the dependence with respect to $\Delta$ is unclear and $-\la T \ra$ can be maximized as long as $\Delta \gtrsim \epsilon$ ; both $\cal{T}_R$ and $\cal{T}_L$ should not be too large and we witness the emergence of a chaotic phase for $\cal{T} \gtrsim \epsilon$, this time also for $\cal{T}_L$. This phase transition may now be observed for $\cal{T}_L$ in this range as this parameter now becomes more relevant in the considered scale ; and finally, we confirm that the maximum is indeed approached when $\gamma \sim U \sim 10^3\epsilon$.

Then, Fig~\ref{SI_Tmax_3}(m) to (r) look quite similar to the previous case with a larger range for $\cal{T}_L$, with a particularly notable difference that, in this case, we should maximize $\cal{T}_L^-$ over $\cal{T}_L^+$, which could mean that a current going in the reverse direction in the vicinity of the left electrode is favored. A satisfying qualitative explanation for this current inversion remains to be found. 

\newpage


\end{document}